\documentclass[twocolumn]{aastex63}

\usepackage{amsmath}
\received{\today}
\usepackage{graphicx}
\usepackage[caption=false]{subfig}

\usepackage{hyperref}
\usepackage{amssymb}
\usepackage{pifont}

\newcommand{\xmark}{\ding{55}}
\newcommand{\hi}{\ion{H}{1}}

\usepackage{array,multirow}

\newcommand{\STAB}[1]{\begin{tabular}{@{}c@{}}#1\end{tabular}}

\newcommand*{\figuretitle}[1]{%
    {\centering%   <--------  will only affect the title because of the grouping (by the
    \textbf{#1}%              braces before \centering and behind \medskip). If you remove
    \par\medskip}%            these braces the whole body of a {figure} env will be centered.
}
\submitjournal{ApJ}

\shorttitle{Scylla I}
\shortauthors{Murray et al.}
\graphicspath{{./}{figures/}}
\begin{document}

\title{Scylla I: A pure-parallel, multi-wavelength imaging survey of the ULLYSES fields in the LMC and SMC}

\correspondingauthor{C.\,E.\,M}
\email{cmurray1@stsci.edu}

\author[0000-0002-7743-8129]{Claire E. Murray}
\affil{Space Telescope Science Institute, 
3700 San Martin Drive, 
Baltimore, MD 21218, USA}
\affil{The William H. Miller III Department of Physics \& Astronomy, Bloomberg Center for Physics and Astronomy, Johns Hopkins University, 3400 N. Charles Street, Baltimore, MD 21218, USA}

\author[0000-0003-0588-7360]{Christina W. Lindberg}
\affil{The William H. Miller III Department of Physics \& Astronomy, Bloomberg Center for Physics and Astronomy, Johns Hopkins University, 3400 N. Charles Street, Baltimore, MD 21218, USA}
\affil{Space Telescope Science Institute, 
3700 San Martin Drive, 
Baltimore, MD 21218, USA}

\author[0000-0002-9912-6046]{Petia Yanchulova Merica-Jones}
\affil{Space Telescope Science Institute, 
3700 San Martin Drive, 
Baltimore, MD 21218, USA}

\author[0000-0002-7502-0597]{Benjamin F. Williams}
\affil{Department of Astronomy, University of Washington, Box 351580, U. W., Seattle, WA 98195-1580, USA}

\author[0000-0002-2970-7435]{Roger E. Cohen}
\affil{Department of Physics and Astronomy, Rutgers the State University of New Jersey, 136 Frelinghuysen Rd., Piscataway, NJ, 08854, USA}

\author[0000-0001-5340-6774]{Karl D.\ Gordon}
\affil{Space Telescope Science Institute, 3700 San Martin Drive, Baltimore, MD 21218, USA}

\author[0000-0001-5538-2614]{Kristen B. W. McQuinn}
\affil{Department of Physics and Astronomy, Rutgers the State University of New Jersey, 136 Frelinghuysen Rd., Piscataway, NJ, 08854, USA}
\affil{Space Telescope Science Institute, 3700 San Martin Drive, 
Baltimore, MD 21218, USA}

\author[0000-0003-1680-1884]{Yumi Choi}
\affil{NSF National Optical-Infrared Astronomy Research Laboratory, 950 N. Cherry Avenue, Tucson, AZ 85719 USA}

\author[0009-0005-0339-015X]{Clare Burhenne}
\affil{Department of Physics and Astronomy, Rutgers the State University of New Jersey, 136 Frelinghuysen Rd., Piscataway, NJ, 08854, USA}

\author[0000-0002-4378-8534]{Karin M. Sandstrom}
\affil{Department of Astronomy \& Astrophysics, University of California San Diego, 9500 Gilman Drive, La Jolla, CA 92093, USA}

\author[0000-0001-6118-2985]{Caroline Bot}
\affil{Observatoire Astronomique de Strasbourg, Universit\'e de Strasbourg, UMR 7550, 11 rue de l'Universit\'e, F-67000 Strasbourg, France}

\author[0000-0001-6421-0953]{L. Clifton Johnson}
\affil{Center for Interdisciplinary Exploration and Research in Astrophysics (CIERA) and Department of Physics and Astronomy, Northwestern University, 1800
Sherman Avenue, Evanston, IL 60201, USA}

\author[0000-0002-8937-3844]{Steven R. Goldman}
\affil{Space Telescope Science Institute, 3700 San Martin Drive, Baltimore, MD 21218, USA}

\author[0000-0001-7959-4902]{Christopher J. R. Clark}
\affil{AURA for the European Space Agency, Space Telescope Science Institute, 3700 San Martin Drive, Baltimore, MD 21218, USA}

\author[0000-0001-6326-7069]{Julia C. Roman-Duval}
\affil{Space Telescope Science Institute, 3700 San Martin Drive, Baltimore, MD 21218, USA}

\author[0000-0003-0394-8377]{Karoline M. Gilbert}
\affil{Space Telescope Science Institute, 3700 San Martin Drive, Baltimore, MD 21218, USA}
\affil{The William H. Miller III Department of Physics \& Astronomy, Bloomberg Center for Physics and Astronomy, Johns Hopkins University, 3400 N. Charles Street, Baltimore, MD 21218, USA}

\author[0000-0003-4797-7030]{J. E. G. Peek}
\affil{Space Telescope Science Institute, 3700 San Martin Drive, Baltimore, MD 21218, USA}
\affil{The William H. Miller III Department of Physics \& Astronomy, Bloomberg Center for Physics and Astronomy, Johns Hopkins University, 3400 N. Charles Street, Baltimore, MD 21218, USA}

\author[0000-0002-2954-8622]{Alec S. Hirschauer}
\affil{Space Telescope Science Institute, 3700 San Martin Drive, Baltimore, MD 21218, USA}

\author[0000-0003-4850-9589]{Martha L. Boyer}
\affil{Space Telescope Science Institute, 3700 San Martin Drive, Baltimore, MD 21218, USA}

\author[0000-0001-8416-4093]{Andrew E. Dolphin}
\affil{Raytheon Company, 1151 E. Hermans Road, Tucson, AZ 85756, USA}

\begin{abstract}
Scylla is a deep \emph{Hubble Space Telescope} survey of the stellar populations, interstellar medium and star formation in the LMC and SMC. As a pure-parallel complement to the Ultraviolet Legacy Library of Young Stars as Essential Standards (ULLYSES) survey, Scylla obtained 342 orbits of ultraviolet (UV) through near-infrared (IR) imaging of the LMC and SMC with Wide Field Camera 3. In this paper, we describe the science objectives, observing strategy, data reduction procedure, and initial results from our photometric analysis of 96 observed fields. Although our observations were constrained by ULYSSES primary exposures, we imaged all fields in at least two filters (F475W and F814W), and $64\%$ of fields in at least three and as many as seven WFC3 filters spanning the UV to IR. Overall, we reach average $50\%$ completeness of $m_{\rm F225W}=26.0$, $m_{\rm F275W}=26.2$, $m_{\rm F336W}=26.9$, $m_{\rm F475W}=27.8$, $m_{\rm F814W}=25.5$, $m_{\rm F110W}=24.7$, and $m_{\rm F160W}=24.0$ Vega magnitudes in our photometric catalogs, which is faintward of the ancient main sequence turnoff in all filters. The primary science goals of Scylla include characterizing the structure and properties of dust in the MCs, as well as their spatially-resolved star formation and chemical enrichment histories. Our images and photometric catalogs, which represent the widest-area coverage of MCs with HST photometry to date, are available as a high-level science product at the Barbara A. Mikulski Archive for Space Telescopes (MAST).
\end{abstract}


\section{Introduction} 
\label{sec:intro}

\begin{figure*}[t!]
 \centering
    \includegraphics[width=\textwidth]{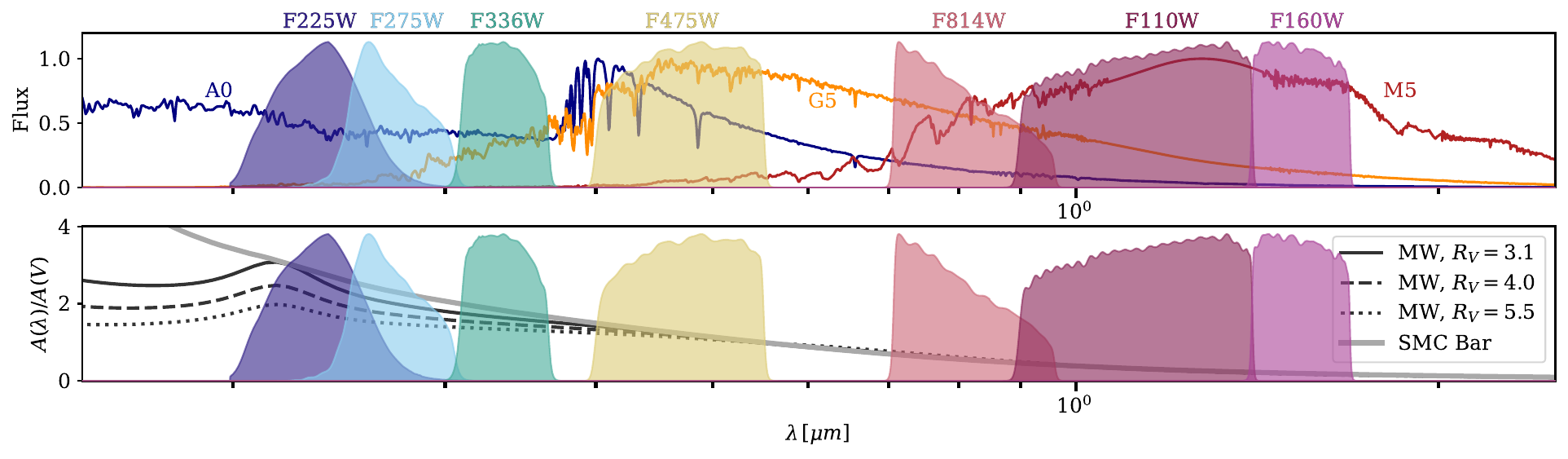}
	\caption{Top: Representative stellar spectra ($F_{\lambda}(\lambda)$) normalized to have the same peak flux, from three stars \citep[A0, G5, M5 types;][]{pickles1998} compared with the transmission curves for six HST/WFC3 filters spanning the UV (F275W, F336W), optical (F475W, F184W) and near-IR (F110W, F160W), to illustrate how distinguishing between intrinsic stellar properties (i.e., variation in temperature between A0 and M5 stars) and the effects of dust reddening requires broad wavelength coverage. Bottom: Dust extinction curves of varying $R_V$ and curve shape.}
	\label{f:filters}
\end{figure*}

The LMC and SMC (combined MCs) are the most massive dwarf satellites of the Milky Way (MW).  Located at distances of $\sim 50-60\rm\,kpc$ \citep{scowcroft2016,pietrzynski2019}, individual stars in the LMC and SMC can be resolved, as well as the individual structures in the interstellar medium (ISM) from which they formed (e.g., molecular clouds) and into which they feed back (e.g., \ion{H}{2} regions, supernova remnants). Furthermore, the MCs feature a diverse range of interstellar conditions, including metallicity  \citep[$Z<0.5\rm\,Z_{\odot}$;][]{russell1992}, gas density \citep[e.g.,][]{stanimirovic1999, kim1999, wong2011, jameson2016, pingel2022}, and radiation field \citep[e.g.,][]{galliano2011, chastenet2017, chastenet2019}. These systems therefore provide excellent laboratories for studying the detailed properties of stellar populations and the ISM in low metallicity environments, which are markedly different from those found in the Milky Way (MW). 

\subsection{Open questions on the MCs}

Despite the proximity of the MCs, fundamental questions regarding the nature of their ISM and stellar populations remain open due to strong observational limitations. Both systems are so rich in neutral hydrogen (\hi) that emission at 21 cm can suffer from strong optical depth effects \citep[e.g.,][]{dempsey2022}. Molecular hydrogen (H$_2$), the most abundant molecule in the ISM, is not observable in cold, star-forming conditions in the MCs due to its lack of excitation at low temperature. Carbon monoxide (CO), a popular H$_2$ tracer, cannot self-shield (unlike H$_2$), and therefore fails to trace the total molecular gas reservoir, especially in the low-$Z$ MCs, where shielding by dust is minimized, and also within strong radiation fields, where molecule destruction is enhanced \citep{bolatto2011}. As a result, we still lack a comprehensive census of the total ISM mass of the MCs and how it is distributed as a function of phase (i.e., density and temperature), which severely limits our ability to diagnose the star formation efficiency and baryon cycle of these systems.

Fortunately, dust grains are distributed throughout the ISM and thus provide a convenient tracer of all phases. However, converting between dust emission in the IR -- a ubiquitous tracer of dust at all redshifts -- and mass, is complicated. First of all, the physical properties of dust grains, including their size, shape, and composition, are known to vary \emph{significantly} within and between galaxies \citep[e.g.,][]{pei1992, gordon2003, clayton2015, schlafly2017}. These properties determine the manner in which dust will absorb, scatter and emit light as a function of wavelength \citep[e.g.][]{draine2003}, which is crucial for interpreting dust observations. In addition, key ingredients of dust models are wildly uncertain. For example, the value of the dust mass absorption coefficient --- required to estimate dust mass from dust emission --- varies by an order of magnitude in the literature \citep{clark2019}.

\begin{figure*}[t!]
 	\centering
\includegraphics[width=\textwidth]{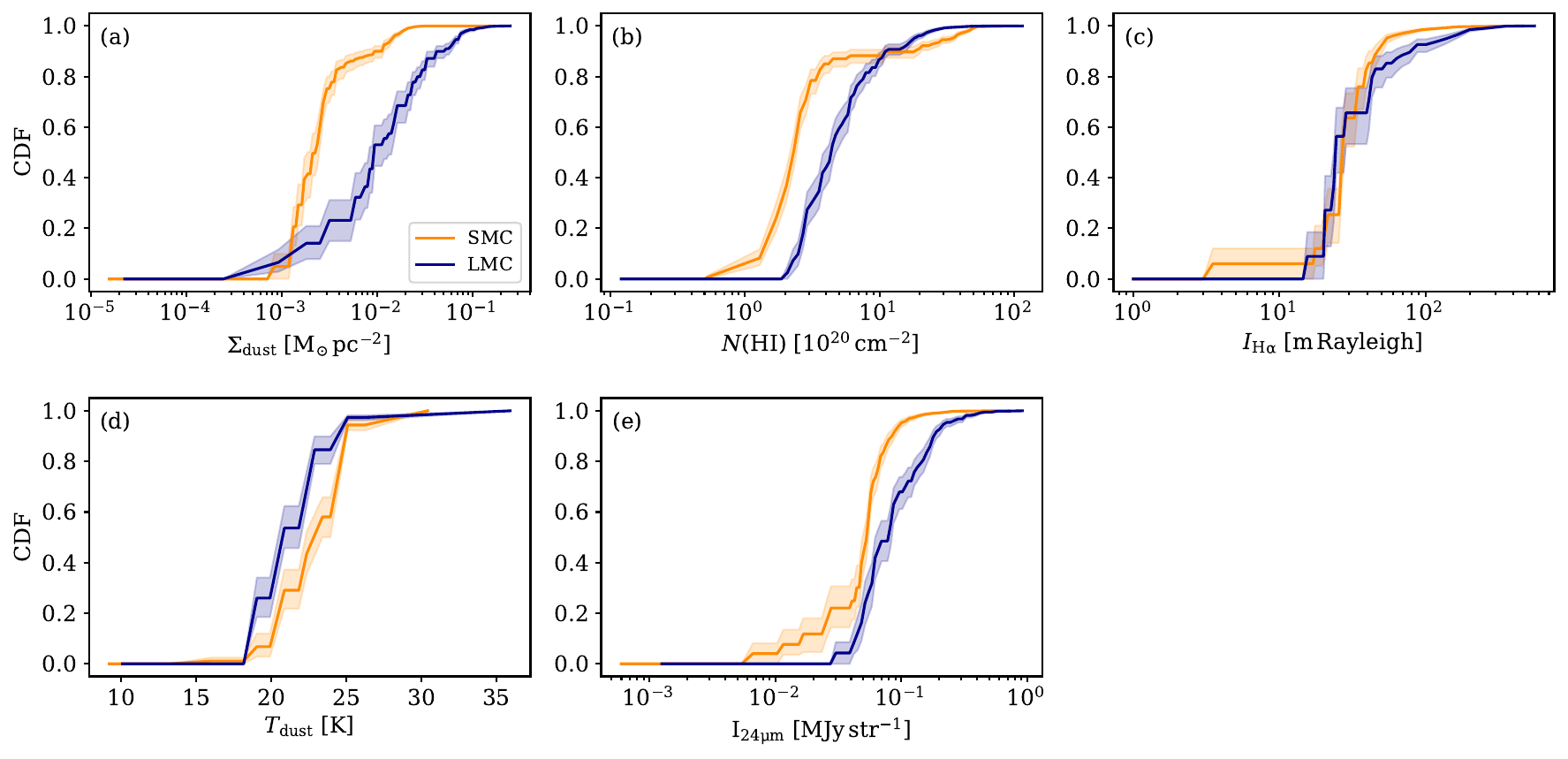}
	\caption{Cumulative distribution functions of ancillary tracers of the ISM at the postion of Scylla fields in the SMC (orange) and LMC (blue). (a)  Dust surface density \citep[$\Sigma_{\rm dust}$;][]{clark2023}; (b) \hi\ column density \citep[$N({\rm HI})$;][]{kim1999, pingel2022}; (c) H$\alpha$ emission \citep[$I_{H\alpha}$;][]{gaustad2001}; (d) Dust temperature \citep[$T_{\rm dust}$;][]{clark2023}; (e) $24\mu$m intensity \citep[$I_{\rm 24\mu m}$;][]{gordon2011}. } 
	\label{fig:envs}
\end{figure*}

Even with an accurate census of the total ISM budget in the MCs, we still lack an observational view of how the ISM moves in these nearby systems. How and where stellar feedback carves holes into the surrounding ISM is fundamental for understanding mass and radiation escape into the ISM. In addition, the momentum injected by starbursts can trigger secondary generations of star formation at the intersections of colliding gas flows. Disentangling the expansion {versus} collapse of individual ISM structures on 10s of pc scales, as well as large-scale outflows and inflows on kpc scales is crucial for understanding how these systems form stars, enrich the ISM and feed back into their circumgalactic environments. The crucial missing piece of information is the distance along the line of sight to individual ISM structures, which cannot be directly constrained by standard (emission, absorption) observations. 

In addition, the physical mechanisms responsible for converting the dusty ISM into stars and planets remain uncertain \citep{kennicutt2012}. Standard empirical approximations for the star formation rate (SFR) and its evolution, used widely to infer how stars form, evolve and feedback, are coarsely calibrated in time ($\Delta t \sim 100\rm\,Myr$) and derived from global scales \citep[$\gtrsim \rm kpc$;][]{kennicutt2012, leroy2013}. Not only are these models known to break down on the scales of individual clouds \citep{schruba2010, kruijssen2014}, they are insufficient to describe the star formation histories (SFH) of local galaxies \citep{mcquinn2010}, MCs included.

\begin{figure*}
 	\centering
\includegraphics[width=\textwidth]{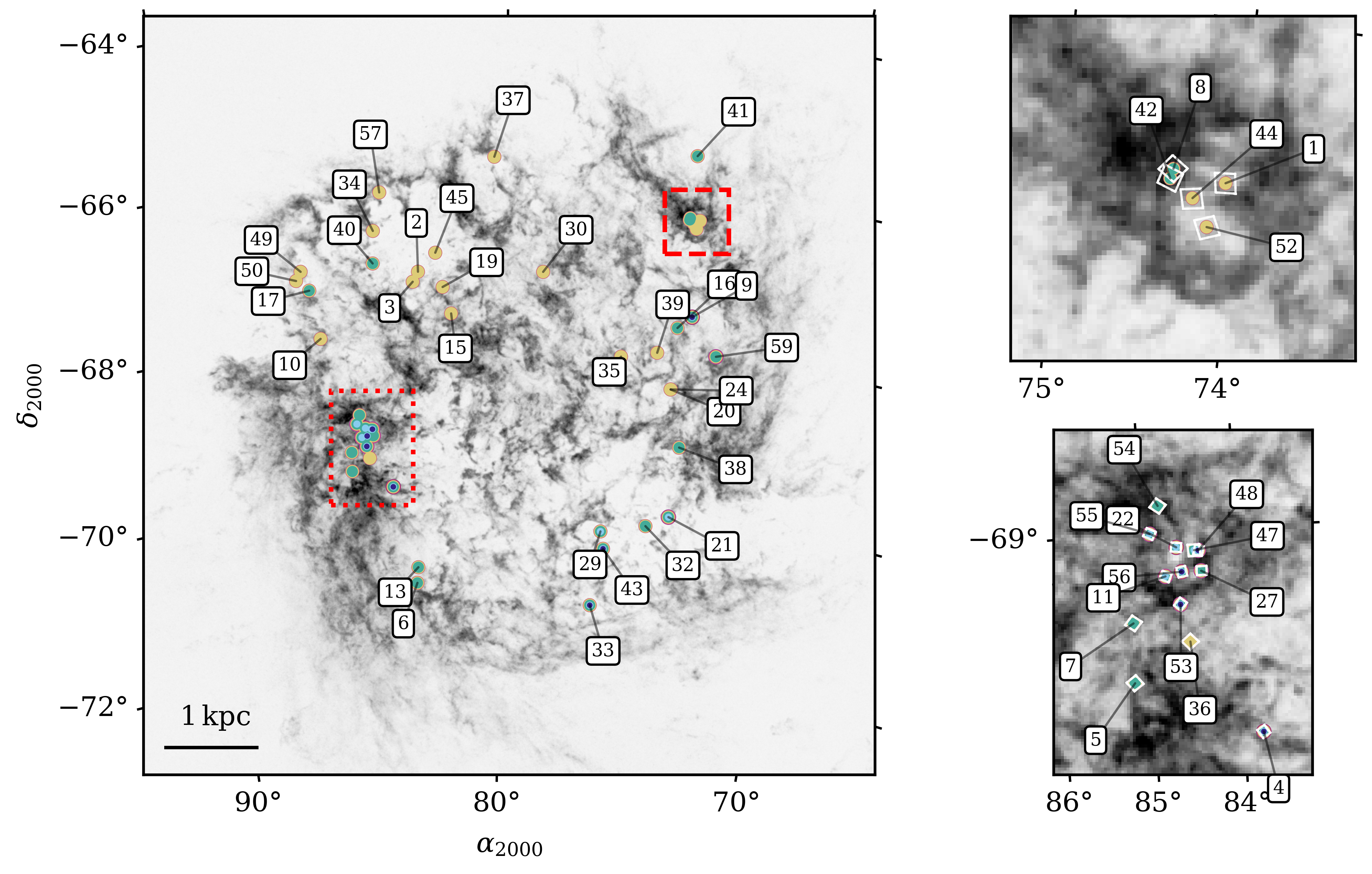}
\includegraphics[width=\textwidth]{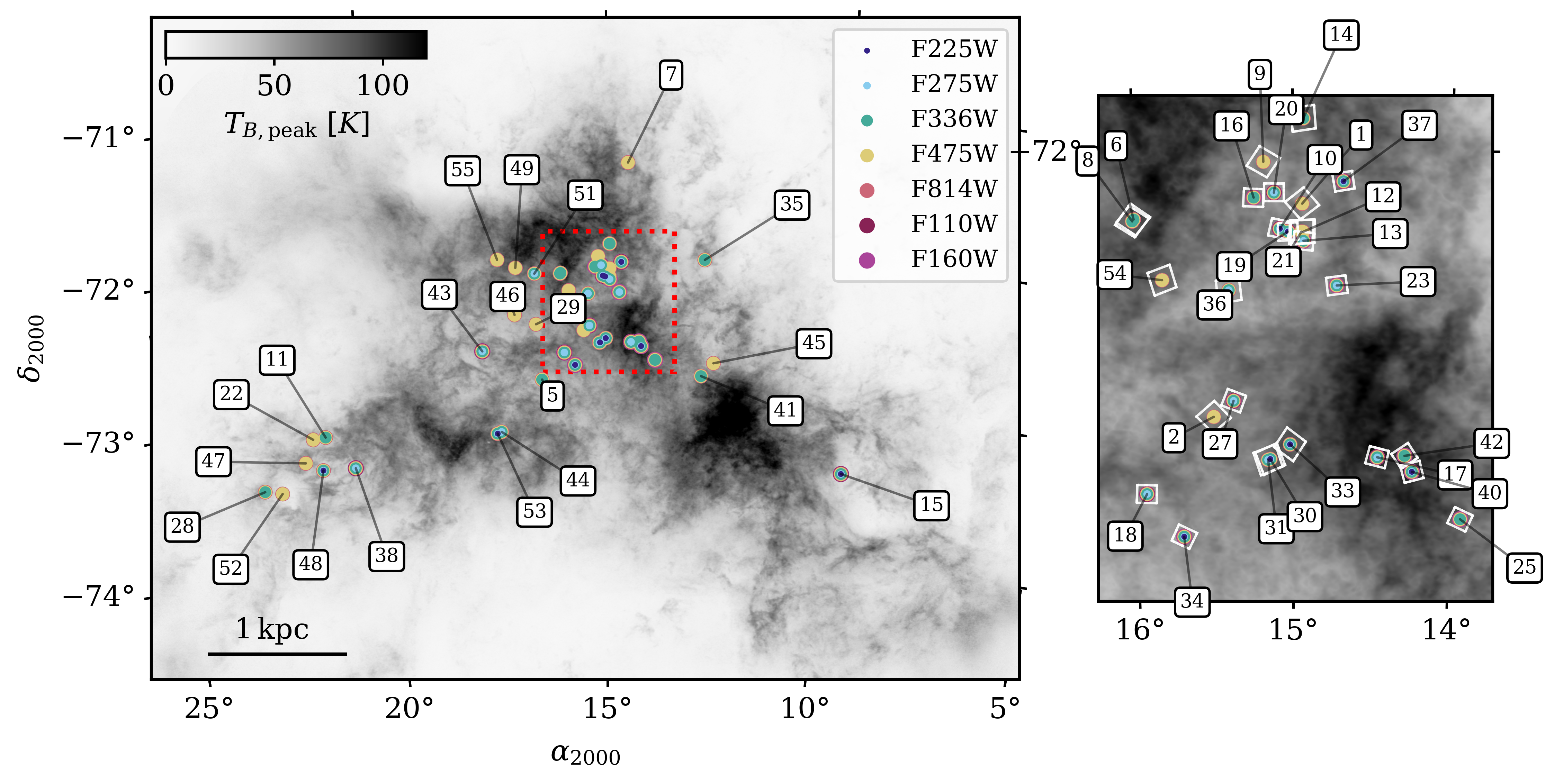}
	\caption{Maps of the peak brightness temperature of $21\rm\,cm$ emission in the LMC \citep{kim1999} and the SMC \citep{pingel2022}, overlaid with observed Scylla fields. In each panel, regions where fields overlap are identified by dashed red bounding boxes and expanded at right. There are 96 fields as part of the release described in this paper, 48 in the SMC and 48 in the LMC. The number of filters used to observe each field is illustrated by the number of concentric circles overlaid at each position. Each field is tagged by its ``short" name number (see Table~\ref{tab:summary_obs}). In the expanded panels, the size of each footprint is plotted in white outline.} 
	\label{f:targets}
\end{figure*}

For the MCs in particular, the influence of their rich history of dynamical interactions on their SFH is significant. How and whether the MCs have collided with each other in the past, and where they are on their orbit through the MW halo are key open questions. Beyond the unique characteristics of the LMC/SMC system, these galaxies provide a high-resolution window into the effects of dwarf galaxy interactions on SFHs in a broader context. Resolving precisely how interactions trigger bursts of star formation is crucial for testing theories of how galaxy interactions and feedback drive inside-out {versus} outside-in star formation, which need detailed, radial SFHs at the oldest ages to test \citep[e.g.,][]{stinson2009, El-Badry2016}. 

\subsection{How Scylla will solve MC mysteries}

A powerful technique for deciphering the intrinsic properties of stars in the MCs, as well as the nature of the intervening ISM, is to observe resolved stellar populations across a wide wavelength range. The ability to resolve individual sources is crucial to distinguish between the degenerate effects of the relative geometry between stars and dust, dust radiative transfer, and SFH on stellar observations. Wide wavelength coverage is also important -- it is well known that optical photometry alone is insufficient to disentangle the effects of dust extinction and stellar temperature on observed stellar properties \citep{gordon2016}. In Figure~\ref{f:filters}, we illustrate this point by comparing synthetic stellar spectra (top) and dust extinction curves (bottom; e.g., parameterized by $R_V$, the total-to-selective extinction ratio) across ultraviolet (UV) through infrared wavelengths (IR). It is clear that in order to simultaneously and precisely distinguish between stars of different stellar types, as well as the amount and nature of dust extinction, observations spanning UV through near-IR wavelengths are required. {While UV coverage is critical for measuring dust extinction, the IR allows us to observe stars in dustier regions, improving our stellar completeness and range of extinction we can probe \citep[for more details on the filter selection justification, see Section 3.2 in][]{dalcanton2012}.}

Fortunately, the \emph{Hubble Space Telescope} (HST) is perfectly suited to this task. At the distances of the MCs, individual stars are resolvable by HST far below the oldest main sequence turn-off (oMSTO) in relatively short exposure times \citep[e.g.,][]{weisz2013, romanduval2019}. HST is also equipped with imaging detectors sensitive to UV-IR wavelengths. However, a fully-resolved survey of the two systems is unfeasible given the small observing field of view of HST. This underscores the power of ground-based surveys, and future wide-area facilities (e.g., the Nancy Grace Roman Space Telescope) to cover the full extents of the MCs \citep[e.g.,][]{cioni2011, nidever2017}. Existing HST observations of the LMC and SMC therefore tend to target prominent, individual regions, such as the 30 Doradus star-forming region in the LMC \citep{sabbi2016} and the Southwest Bar of the SMC \citep{ymj2017}, small numbers of deep pointings \citep{weisz2013, cignoni2013} or regions parallel to targeted observations \citep[METAL;][]{romanduval2019}. 

In this paper, we introduce Scylla, a pure-parallel HST imaging survey which operated alongside the Ultraviolet (UV) Legacy Library of Young Stars as Essential Standards (ULLYSES) survey \citep{romanduval2020}. Named after the multi-headed monster from the myth of Ulysses, Scylla imaged fields parallel to spectroscopic ULLYSES targets, obtaining maximum photometric coverage from the near-UV to the near-IR with Wide Field Camera 3 (WFC3). In this paper, we describe the first 96 fields imaged over 342 orbits during Cycles 27-31 (2020-2023). In a future release, we will include all survey data (27 additional fields). 

Although Scylla does not cover large, contiguous fields, our spatial coverage of the MCs is unprecedented as the 48 fields in the LMC are distributed over the central $\sim$4.5 $\times$ $\sim$5 kpc, and the 48 fields in the SMC are distributed over the central $\sim$4 $\times$ $\sim$2 kpc. As a result, Scylla samples diverse environments in terms of metallicity, gas column densities, radiation fields and SFHs across kpc scales. 

In Figure~\ref{fig:envs} we plot cumulative distribution functions of several ISM tracers: dust surface density and temperature \citep{clark2023}, \hi\ column density \citep{kim1999, pingel2022}, H$\alpha$ emission \citep{gaustad2001}, and $24\mu$m emission \citep{gordon2011}. These results were computed by identifying the pixels associated with each Scylla footprint in each of the sampled maps and bootstrapping the distributions with replacement to compute uncertainties. We find that Scylla observations in the LMC and SMC probe different parameter space in each tracer, and also a wide range. For example, we probe nearly three orders of magnitude in dust surface density in the LMC, and a similarly wide range of \hi\ column densities in the LMC. The two samples are very similar in ionized hydrogen and dust temperature, and the LMC fields sample regions of enhanced star formation activity relative to the SMC (traced by $24\rm\mu m$). 

Armed with this data set, our science goals are to resolve how dust properties vary with interstellar environment, to constrain the multi-dimensional structure of the ISM of the MCs, and to probe their detailed SFHs. Scylla will probe the total dust column density ($A_V$) at $\sim$parsec-scale resolution within the MCs independently of other ISM tracers. By comparing our results with $A_V$ derived from far-IR emission, we will probe variations in the opacity and emissivity of dust throughout the MCs. In addition, we will generate maps of the SFH and chemical enrichment history of the MCs with the highest resolution to date, and resolve how the SFH evolves as a function of position and time (Cohen et al. 2024a, b).

\begin{deluxetable*}{c|ccccccc|cc} 
\tablecaption{\label{tab:filter_combos} Filter Combinations}
\tablehead{\colhead{Name}  & \colhead{F225W} & \colhead{F275W} & \colhead{F336W} & \colhead{F475W} & \colhead{F814W} & \colhead{F110W} & \colhead{F160W} & \colhead{$N_{\rm LMC}$} & \colhead{$N_{\rm SMC}$}}
\startdata
A &  \xmark & \xmark & \xmark & \xmark & \xmark & \xmark & \xmark &  3 & 4 \\ 
B &   & \xmark & \xmark & \xmark & \xmark & \xmark & \xmark &  1 &  6 \\ 
C &  \xmark & \xmark & \xmark & \xmark & \xmark &  & \xmark &  2 &  2 \\ 
D &  & \xmark & \xmark & \xmark & \xmark &  & \xmark &  3 &  1\\ 
E &  \xmark & \xmark & \xmark & \xmark & \xmark &  &  &  2 &  4\\ 
F &   &  & \xmark & \xmark & \xmark &  & \xmark & 2 & 3 \\ 
G &   & \xmark & \xmark & \xmark & \xmark &  &  &  2 &  5 \\ 
H &   &  & \xmark & \xmark & \xmark &  &  &  13 & 8 \\ 
I &   &  &  & \xmark & \xmark &  &  &  20 & 15 
\enddata  
\tablecomments{This table summarizes the ten unique filter combinations for the Scylla survey. The number of fields with each combination in the LMC ($N_{\rm LMC}$) and SMC ($N_{\rm SMC}$) is indicated in the right-hand columns.}   
\end{deluxetable*}

In subsequent publications, we will produce a catalog of stellar and dust parameters by fitting the multi-band photometry of each star using the Bayesian Extinction and Stellar Tool \citep[BEAST;][]{gordon2016}, a probabilistic method for modeling spectral energy distributions (SEDs) for surveys of resolved stellar populations. Our catalogs will describe the age, initial mass, metallicity, distance, $A_V$, average grain size ($R_V$), and the mixture coefficient between the Milky Way $R_V$-dependent dust extinction and the SMC Bar dust extinction \citep{gordon2003} of the observed sources in the MCs.

This paper describes the observing strategy (Section~\ref{sec:observations}), and data reduction procedure (Sections~\ref{sec:data_reduction} and \ref{sec:asts}) for the Scylla survey. We present color-magnitude diagrams (CMDs) and discuss the results of analyzing the CMDs in the context of previous observations (Section~\ref{sec:results}). Finally, we present concluding remarks (Section~\ref{sec:conclusions}).

\section{Observations} 
\label{sec:observations}

Following the strategy of other large HST photometric surveys of nearby galaxies, including the Panchromatic Hubble Andromeda Treasury \citep[PHAT;][]{dalcanton2012, williams2014}, the SMC Investigation of Dust and Gas Evolution \citep[SMIDGE;][]{ymj2017, ymj2021}, and Metal, Evolution, Transport and Abundance in the LMC \citep[METAL;][]{romanduval2019} surveys, we observed each ULLYSES parallel field with Wide Field Camera 3 (WFC3)\footnote{We note that although the previous surveys also used the Advanced Camera for Surveys to image regions in optical/IR filters, our parallel program status led us to choose using only WFC3 for the full survey.} in filters spanning UV, optical and near-IR wavelengths. 

As a pure-parallel complement to the ULLYSES survey, we were constrained by the requirements of the primary COS and STIS spectroscopic observations. Our fields are located near the primary targets ($\sim5^{\prime}$ away, based on the footprint of WFC3 relative to COS/STIS), and were positioned according to the roll angle of the telescope at the time of observation. The inability to set the roll angle means that we were unable to guarantee follow-up imaging of the same fields in the event of observing failures (e.g., guide star acquisition failures).

Based on the arrangement of buffer dump overheads imposed by the primary observations, we arranged exposures in as many of the WFC3/UVIS (F225W, F275W, F336W, F475W, F814W) and WFC3/IR (F110W and F160W) filters shown in Figure~\ref{f:filters} as possible. We selected filters for each field based on the following hierarchy: (1) F475W and F814W, (2) F336W, F275W (3) F110W, F160W, (4) F225W. We prioritized observations of optical filters (F475W, F814W) to maximize the number of observed sources and to optimize the constraining power of the SFH across a large variety of stellar types. Bluer and UV observations (F336W, F275W, F225W) give us additional leverage when measuring hotter stars, extinction, and dust grain size distribution. Lastly, depending on the constraints imposed by the primary observing setup, we obtained imaging in the IR (F110W, F160W) which allows us to measure cooler stars (e.g., asymptotic giant branch stars) and stars embedded in dustier regions. When possible, we also included one or more exposures in F225W to constrain the shape of the $2175\textup{~\AA}$ extinction feature \citep[e.g.,][]{romanduval2019}. Overall, we used nine unique combinations of filters across the Scylla observations. These filter combinations are summarized in Table~\ref{tab:filter_combos}.

As the telescope pointing is fully determined by the primary ULLYSES observations, we were unable to dither our exposures. For WFC3/UVIS, we included multiple exposures of as close to equal length as possible in each visit to enable cosmic ray rejection. For WFC3/IR, this is done via up-the-ramp data reduction. In many cases, the ULLYSES overheads mandated that we obtain long exposures in a small number of filters (1-2) per orbit, which hampered our ability to reliably detect bright stars (which saturate in long exposures). Therefore, where possible we inserted short (3-5 second, also known as ``guard") exposures in F475W or F814W.

In Figure~\ref{f:targets}, we summarize the positions and wavelength coverage of Scylla imaging from this release (observations through February 2023) in the LMC and SMC. Scylla increases the total spatial coverage of UV-IR imaging from HST \citep[e.g.,][]{weisz2013, cignoni2013, sabbi2016, ymj2017, romanduval2019} by factors of two and eight in the LMC and SMC, opening unexplored parameter space in their ISM and SFHs. {Assuming average distances to the LMC and SMC of $50$ and $62$ kpc respectively, and a field of view of WFC3 UVIS of $160^{\prime \prime}$, each field covers an area of $\sim 39\,(48)\rm\,pc$ on a side in the LMC (SMC)}.

\startlongtable
\begin{deluxetable*}{llccccc} 
\tablecaption{\label{tab:intro} Field Summary}
\tablehead{
 \colhead{Field Name} &  \colhead{Field Name} & \colhead{RA} & \colhead{Dec} & \colhead{$N_{\rm filters}$} & \colhead{$N_{\rm orbits}$} & \colhead{Combination}  \\
 \colhead{(short)} & \colhead{(long)} &  \colhead{($^{\circ}$)} & \colhead{($^{\circ}$)} & \colhead{} & \colhead{} &  \colhead{} \\
\colhead{(1)} & \colhead{(2)} & \colhead{(3)} & \colhead{(4)} & \colhead{(5)} & \colhead{(6)}  & \colhead{(7)} }
\startdata  
SMC$\_ \rm $1  &  15891$\_ \rm$SMC-3192ne-8290  &  14.939  &  -72.1032  &  2  &  2  &  I  \\ 
SMC$\_ \rm $2  &  15891$\_ \rm$SMC-2750ne-8567  &  15.5059  &  -72.5123  &  2  &  2  &  I  \\ 
SMC$\_ \rm $5  &  15891$\_ \rm$SMC-3514se-8584  &  16.4523  &  -72.835  &  3  &  3  &  H  \\ 
SMC$\_ \rm $6  &  15891$\_ \rm$SMC-3956ne-9632  &  15.9954  &  -72.1305  &  3  &  2  &  H  \\ 
SMC$\_ \rm $7  &  15891$\_ \rm$SMC-5278ne-9802  &  14.5464  &  -71.4007  &  2  &  2  &  I  \\ 
SMC$\_ \rm $8  &  15891$\_ \rm$SMC-3955ne-9818  &  16.0039  &  -72.1341  &  3  &  2  &  H  \\ 
SMC$\_ \rm $9  &  15891$\_ \rm$SMC-3587ne-10112  &  15.1806  &  -72.0226  &  2  &  2  &  I  \\ 
SMC$\_ \rm $10  &  15891$\_ \rm$SMC-3149ne-12269  &  15.0793  &  -72.1515  &  7  &  4  &  A  \\ 
SMC$\_ \rm $11  &  15891$\_ \rm$SMC-8743se-11371  &  21.4361  &  -73.1212  &  3  &  2  &  H  \\ 
SMC$\_ \rm $12  &  15891$\_ \rm$SMC-3025ne-13499  &  14.9344  &  -72.1575  &  2  &  2  &  I  \\ 
SMC$\_ \rm $13  &  15891$\_ \rm$SMC-2983ne-12972  &  14.9313  &  -72.1748  &  5  &  2  &  D  \\ 
SMC$\_ \rm $14  &  15891$\_ \rm$SMC-3669ne-13972  &  14.9314  &  -71.9391  &  3  &  2  &  H  \\ 
SMC$\_ \rm $15  &  15891$\_ \rm$SMC-4292sw-13841  &  9.562  &  -73.4013  &  7  &  3  &  A  \\ 
SMC$\_ \rm $16  &  15891$\_ \rm$SMC-3435ne-13258  &  15.2425  &  -72.0912  &  4  &  2  &  F  \\ 
SMC$\_ \rm $17  &  15891$\_ \rm$SMC-1588ne-12105  &  14.4594  &  -72.5908  &  6  &  3  &  B  \\ 
SMC$\_ \rm $18  &  15891$\_ \rm$SMC-3029ne-13288  &  15.9413  &  -72.6589  &  6  &  3  &  B  \\ 
SMC$\_ \rm $19  &  15891$\_ \rm$SMC-3032ne-13306  &  14.944  &  -72.1573  &  2  &  2  &  I  \\ 
SMC$\_ \rm $20  &  15891$\_ \rm$SMC-3370ne-13459  &  15.1156  &  -72.082  &  6  &  3  &  B  \\ 
SMC$\_ \rm $21  &  15891$\_ \rm$SMC-3104ne-13781  &  15.0284  &  -72.1569  &  7  &  4  &  A  \\ 
SMC$\_ \rm $22  &  15891$\_ \rm$SMC-9034se-13316  &  21.723  &  -73.127  &  2  &  2  &  I  \\ 
SMC$\_ \rm $23  &  15891$\_ \rm$SMC-2584ne-14274  &  14.7218  &  -72.2607  &  5  &  2  &  D  \\ 
SMC$\_ \rm $25  &  16235$\_ \rm$SMC-879ne-11082  &  13.9209  &  -72.7082  &  4  &  2  &  F  \\ 
SMC$\_ \rm $27  &  16235$\_ \rm$SMC-2668ne-11415  &  15.379  &  -72.4818  &  6  &  4  &  B  \\ 
SMC$\_ \rm $28  &  16235$\_ \rm$SMC-10336se-14099  &  22.9725  &  -73.4318  &  3  &  2  &  H  \\ 
SMC$\_ \rm $29  &  16235$\_ \rm$SMC-3870ne-14647  &  16.5543  &  -72.4672  &  2  &  2  &  I  \\ 
SMC$\_ \rm $30  &  16235$\_ \rm$SMC-2259ne-15609  &  15.1462  &  -72.5944  &  5  &  3  &  E  \\ 
SMC$\_ \rm $31  &  16235$\_ \rm$SMC-2272ne-15308  &  15.1603  &  -72.5957  &  4  &  2  &  G  \\ 
SMC$\_ \rm $33  &  16235$\_ \rm$SMC-2167ne-18821  &  15.0188  &  -72.5662  &  5  &  3  &  E  \\ 
SMC$\_ \rm $34  &  16235$\_ \rm$SMC-2728ne-28918  &  15.706  &  -72.7421  &  6  &  4  &  C  \\ 
SMC$\_ \rm $35  &  16235$\_ \rm$SMC-2773nw-32334  &  12.8842  &  -72.0375  &  3  &  3  &  H  \\ 
SMC$\_ \rm $36  &  16235$\_ \rm$SMC-3127ne-32138  &  15.403  &  -72.269  &  4  &  2  &  G  \\ 
SMC$\_ \rm $37  &  16235$\_ \rm$SMC-3154ne-32442  &  14.6806  &  -72.0602  &  7  &  5  &  A  \\ 
SMC$\_ \rm $38  &  16235$\_ \rm$SMC-8151se-32530  &  20.82  &  -73.3444  &  6  &  3  &  B  \\ 
SMC$\_ \rm $40  &  16235$\_ \rm$SMC-1339ne-33009  &  14.2361  &  -72.6181  &  6  &  3  &  C  \\ 
SMC$\_ \rm $41  &  16235$\_ \rm$SMC-286sw-34349  &  12.8849  &  -72.809  &  3  &  3  &  H  \\ 
SMC$\_ \rm $42  &  16235$\_ \rm$SMC-1443ne-34945  &  14.2842  &  -72.5876  &  4  &  3  &  F  \\ 
SMC$\_ \rm $43  &  16235$\_ \rm$SMC-4996ne-34726  &  17.7638  &  -72.633  &  6  &  3  &  B  \\ 
SMC$\_ \rm $44  &  16786$\_ \rm$SMC-4646se-5833  &  17.4098  &  -73.1726  &  4  &  2  &  G  \\ 
SMC$\_ \rm $45  &  15891$\_ \rm$SMC-641nw-12753  &  12.6171  &  -72.721  &  2  &  2  &  I  \\ 
SMC$\_ \rm $46  &  16235$\_ \rm$SMC-4450ne-32733  &  17.0188  &  -72.4006  &  2  &  2  &  I  \\ 
SMC$\_ \rm $47  &  16786$\_ \rm$SMC-9277se-14900  &  21.9543  &  -73.2765  &  2  &  2  &  I  \\ 
SMC$\_ \rm $48  &  16786$\_ \rm$SMC-8904se-15007  &  21.5707  &  -73.3374  &  5  &  3  &  E  \\ 
SMC$\_ \rm $49  &  16786$\_ \rm$SMC-4926ne-15573  &  16.9656  &  -72.089  &  2  &  2  &  I  \\ 
SMC$\_ \rm $51  &  16786$\_ \rm$SMC-4451ne-16362  &  16.5598  &  -72.1312  &  4  &  2  &  G  \\ 
SMC$\_ \rm $52  &  16786$\_ \rm$SMC-9946se-16175  &  22.5766  &  -73.4592  &  2  &  2  &  I  \\ 
SMC$\_ \rm $53  &  16786$\_ \rm$SMC-4745se-7610  &  17.5022  &  -73.1831  &  5  &  3  &  E  \\ 
SMC$\_ \rm $54  &  16786$\_ \rm$SMC-3529ne-15172  &  15.8228  &  -72.2482  &  2  &  2  &  I  \\ 
SMC$\_ \rm $55  &  16786$\_ \rm$SMC-5409ne-15524  &  17.3518  &  -72.0303  &  2  &  2  &  I  \\ 
LMC$\_ \rm $1  &  15891$\_ \rm$LMC-15629nw-4948  &  74.0669  &  -66.3823  &  2  &  2  &  I  \\ 
LMC$\_ \rm $2  &  15891$\_ \rm$LMC-9617ne-5147  &  82.6984  &  -67.1787  &  2  &  2  &  I  \\ 
LMC$\_ \rm $3  &  15891$\_ \rm$LMC-9256ne-6744  &  82.8737  &  -67.3005  &  2  &  2  &  I  \\ 
LMC$\_ \rm $4  &  15891$\_ \rm$LMC-3610se-7920  &  83.7896  &  -69.8  &  7  &  3  &  A  \\ 
LMC$\_ \rm $5  &  15891$\_ \rm$LMC-5442ne-8000  &  85.1885  &  -69.5809  &  3  &  2  &  H  \\ 
LMC$\_ \rm $6  &  15891$\_ \rm$LMC-5082se-6540  &  83.06  &  -70.9753  &  3  &  2  &  H  \\ 
LMC$\_ \rm $7  &  15891$\_ \rm$LMC-5619ne-9411  &  85.1607  &  -69.3493  &  3  &  2  &  H  \\ 
LMC$\_ \rm $8  &  15891$\_ \rm$LMC-15413nw-9621  &  74.3669  &  -66.3639  &  3  &  2  &  H  \\ 
LMC$\_ \rm $9  &  15891$\_ \rm$LMC-12315nw-11221  &  73.9739  &  -67.58  &  7  &  3  &  A  \\ 
LMC$\_ \rm $10  &  15891$\_ \rm$LMC-9446ne-10765  &  85.9116  &  -67.9299  &  2  &  2  &  I  \\ 
LMC$\_ \rm $11  &  15891$\_ \rm$LMC-5389ne-11134  &  84.7783  &  -69.1768  &  5  &  2  &  D  \\ 
LMC$\_ \rm $13  &  15891$\_ \rm$LMC-4489se-10451  &  83.0047  &  -70.7874  &  3  &  2  &  H  \\ 
LMC$\_ \rm $15  &  15891$\_ \rm$LMC-7454ne-11865  &  81.675  &  -67.7037  &  2  &  2  &  I  \\ 
LMC$\_ \rm $16  &  15891$\_ \rm$LMC-11456nw-12627  &  74.4019  &  -67.735  &  3  &  3  &  H  \\ 
LMC$\_ \rm $17  &  15891$\_ \rm$LMC-11384ne-12295  &  86.1452  &  -67.325  &  3  &  3  &  H  \\ 
LMC$\_ \rm $19  &  15891$\_ \rm$LMC-8680ne-12405  &  81.9348  &  -67.3754  &  2  &  2  &  I  \\ 
LMC$\_ \rm $20  &  15891$\_ \rm$LMC-9679nw-13399  &  74.4149  &  -68.4953  &  2  &  2  &  I  \\ 
LMC$\_ \rm $21  &  15891$\_ \rm$LMC-8532sw-13647  &  74.0085  &  -70.0449  &  5  &  3  &  D  \\ 
LMC$\_ \rm $22  &  15891$\_ \rm$LMC-5421ne-12728  &  84.6425  &  -69.0671  &  6  &  3  &  B  \\ 
LMC$\_ \rm $24  &  15891$\_ \rm$LMC-9690nw-13623  &  74.4057  &  -68.4953  &  2  &  2  &  I  \\ 
LMC$\_ \rm $27  &  16235$\_ \rm$LMC-4958ne-31479  &  84.3877  &  -69.1643  &  4  &  2  &  F  \\ 
LMC$\_ \rm $29  &  16235$\_ \rm$LMC-5812sw-7744  &  76.3845  &  -70.3046  &  4  &  2  &  G  \\ 
LMC$\_ \rm $30  &  16235$\_ \rm$LMC-9740nw-7508  &  78.7752  &  -67.1761  &  2  &  2  &  I  \\ 
LMC$\_ \rm $32  &  16235$\_ \rm$LMC-7623sw-22524  &  74.7905  &  -70.1869  &  3  &  3  &  H  \\ 
LMC$\_ \rm $33  &  16235$\_ \rm$LMC-7234sw-22225  &  76.5819  &  -71.2035  &  5  &  3  &  E  \\ 
LMC$\_ \rm $34  &  16235$\_ \rm$LMC-12057ne-22332  &  84.0445  &  -66.6476  &  2  &  2  &  I  \\ 
LMC$\_ \rm $35  &  16235$\_ \rm$LMC-8599nw-23221  &  76.1349  &  -68.1583  &  2  &  2  &  I  \\ 
LMC$\_ \rm $36  &  16235$\_ \rm$LMC-4763ne-26440  &  84.55  &  -69.4356  &  2  &  2  &  I  \\ 
LMC$\_ \rm $37  &  16235$\_ \rm$LMC-14421nw-26822  &  80.3377  &  -65.7595  &  2  &  2  &  I  \\ 
LMC$\_ \rm $38  &  16235$\_ \rm$LMC-9173nw-28313  &  73.9133  &  -69.1883  &  3  &  3  &  H  \\ 
LMC$\_ \rm $39  &  16786$\_ \rm$LMC-10028nw-33586  &  74.9718  &  -68.067  &  2  &  2  &  I  \\ 
LMC$\_ \rm $40  &  16235$\_ \rm$LMC-10728ne-8437  &  84.1026  &  -67.0508  &  3  &  3  &  H  \\ 
LMC$\_ \rm $41  &  16235$\_ \rm$LMC-17892nw-9532  &  74.3489  &  -65.5843  &  3  &  3  &  H  \\ 
LMC$\_ \rm $42  &  16786$\_ \rm$LMC-15342nw-2460  &  74.3789  &  -66.3868  &  3  &  3  &  H  \\ 
LMC$\_ \rm $43  &  16786$\_ \rm$LMC-6222sw-15490  &  76.2446  &  -70.5113  &  5  &  3  &  E  \\ 
LMC$\_ \rm $44  &  16786$\_ \rm$LMC-15331nw-5447  &  74.2426  &  -66.4259  &  2  &  2  &  I  \\ 
LMC$\_ \rm $45  &  16786$\_ \rm$LMC-10253ne-6545  &  82.1444  &  -66.9497  &  2  &  2  &  I  \\ 
LMC$\_ \rm $47  &  16786$\_ \rm$LMC-5187ne-4440  &  84.4514  &  -69.0828  &  4  &  2  &  G  \\ 
LMC$\_ \rm $48  &  16786$\_ \rm$LMC-5127ne-3118  &  84.4053  &  -69.0844  &  6  &  3  &  C  \\ 
LMC$\_ \rm $49  &  16786$\_ \rm$LMC-12311ne-5715  &  86.3622  &  -67.0819  &  2  &  2  &  I  \\ 
LMC$\_ \rm $50  &  16786$\_ \rm$LMC-12141ne-5771  &  86.5299  &  -67.1893  &  2  &  2  &  I  \\ 
LMC$\_ \rm $52  &  16786$\_ \rm$LMC-15254nw-24541  &  74.1457  &  -66.4882  &  2  &  2  &  I  \\ 
LMC$\_ \rm $53  &  16786$\_ \rm$LMC-5045ne-18484  &  84.6333  &  -69.2891  &  6  &  3  &  C  \\ 
LMC$\_ \rm $54  &  16786$\_ \rm$LMC-5943ne-9430  &  84.8145  &  -68.9016  &  3  &  3  &  H  \\ 
LMC$\_ \rm $55  &  16786$\_ \rm$LMC-5850ne-10777  &  84.9219  &  -69.0085  &  5  &  3  &  D  \\ 
LMC$\_ \rm $56  &  16786$\_ \rm$LMC-5199ne-23482  &  84.5993  &  -69.1631  &  7  &  4  &  A  \\ 
LMC$\_ \rm $57  &  16786$\_ \rm$LMC-13556ne-15380  &  83.793  &  -66.1738  &  2  &  2  &  I  \\ 
LMC$\_ \rm $59  &  16786$\_ \rm$LMC-12269nw-24827  &  73.0619  &  -68.0239  &  4  &  3  &  F  \\ 
\enddata  
\tablecomments{The complete table will be available in the online journal. (1): Short field name; (2): Long field name; (3): Average RA per field; (4): Average Dec per field; (5): Number of filters used in the field's observations; (6): Number of orbits per field; (7): The filter combination (see Table~\ref{tab:filter_combos}). }  
\end{deluxetable*}

\label{tab:intro}

In Table~\ref{tab:intro}, we summarize the field names, positions and observing parameters. Each field obtains two names: one short and one long. The short names are formatted as ``\texttt{L(S)MC}{\textunderscore}$n$'', where $n$ is a number corresponding to the order in which the field was observed in the survey (e.g. \texttt{SMC{\textunderscore}15}). All values of $n$ are not present in the table, as some observations overlapped with other fields and therefore those exposures were processed together. In addition, thirteen fields suffered from guide star failures or other imaging artifacts which made them unusable for photometry. {Overall, 109 fields were observed through Februrary 2023, and the 96 without observing failures are presented here. The rest of the Scylla survey, comprising observations {of 27 fields} between February and August 2023, will be presented in a future release}. The long field names are formatted as ``[\texttt{PID}]{\textunderscore}[\texttt{galaxy}]$-$[\texttt{distance from galaxy center in arcsec}][\texttt{cardinal direction}]$-$[\texttt{telescope rotation angle in deg (V3PA)}]'', where \texttt{PID} is the HST program ID under which the observation was taken (e.g. \texttt{15891{\textunderscore}SMC-4292sw-13841}). The long-form names are used for a qualitative assessment of the field location within each galaxy, and therefore the lack of precision in the name (i.e., using an integer value for the rotation angle) is not a significant issue.

In Appendix~\ref{a:exposures}, we summarize the observing parameters for all exposures in the Scylla survey to date, including filters, field centers, position angles, exposure times, and post-flash exposures to reduce charge-transfer efficiency (CTE) effects. The complete table is available in the online journal.

\begin{figure*}[!ht]
  \subfloat[][SMC 21 with extended emission.]{\includegraphics[width=.49\textwidth]{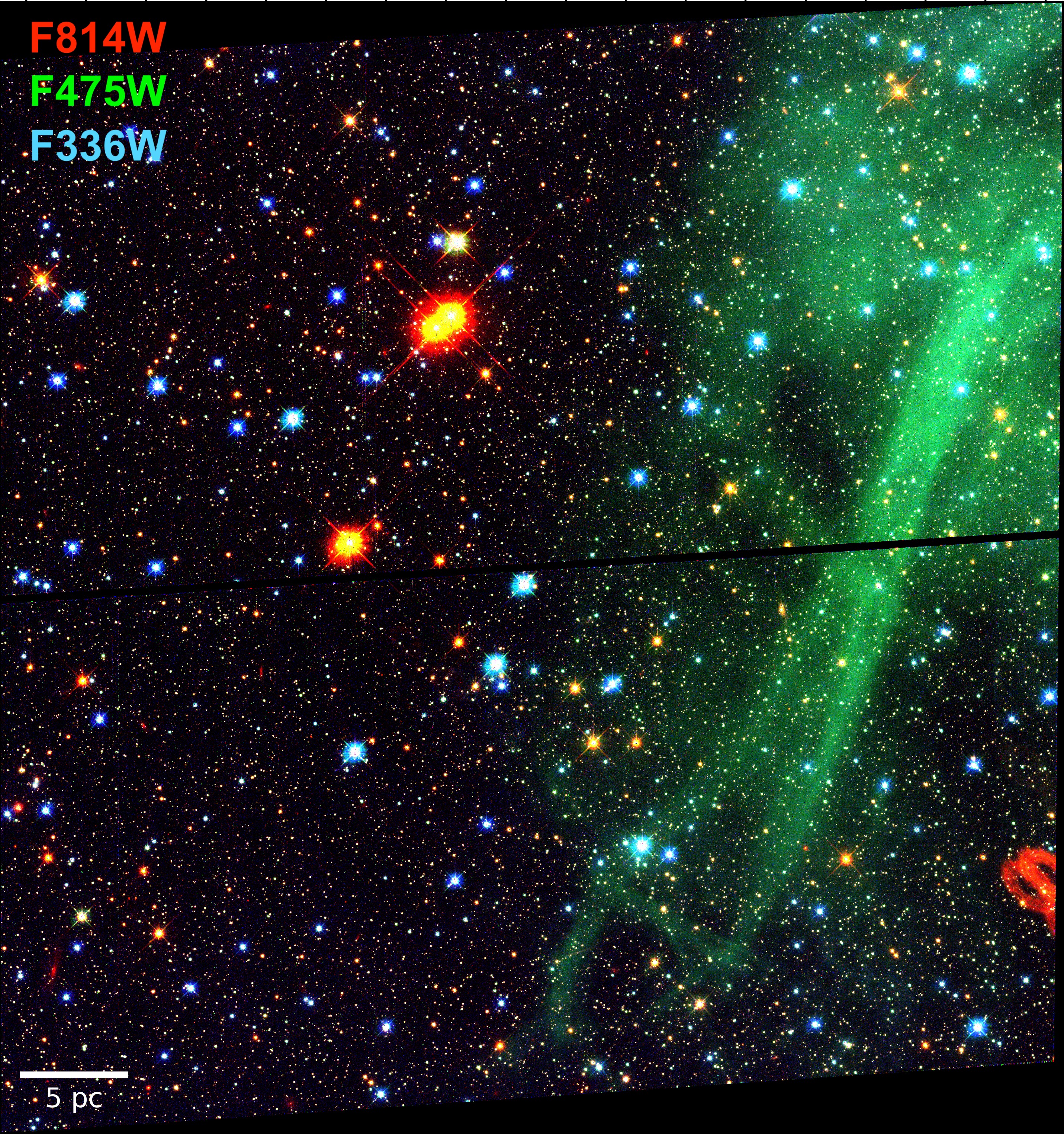}}\quad
  \subfloat[][SMC 15 with no emission.]{\includegraphics[width=.49\textwidth]{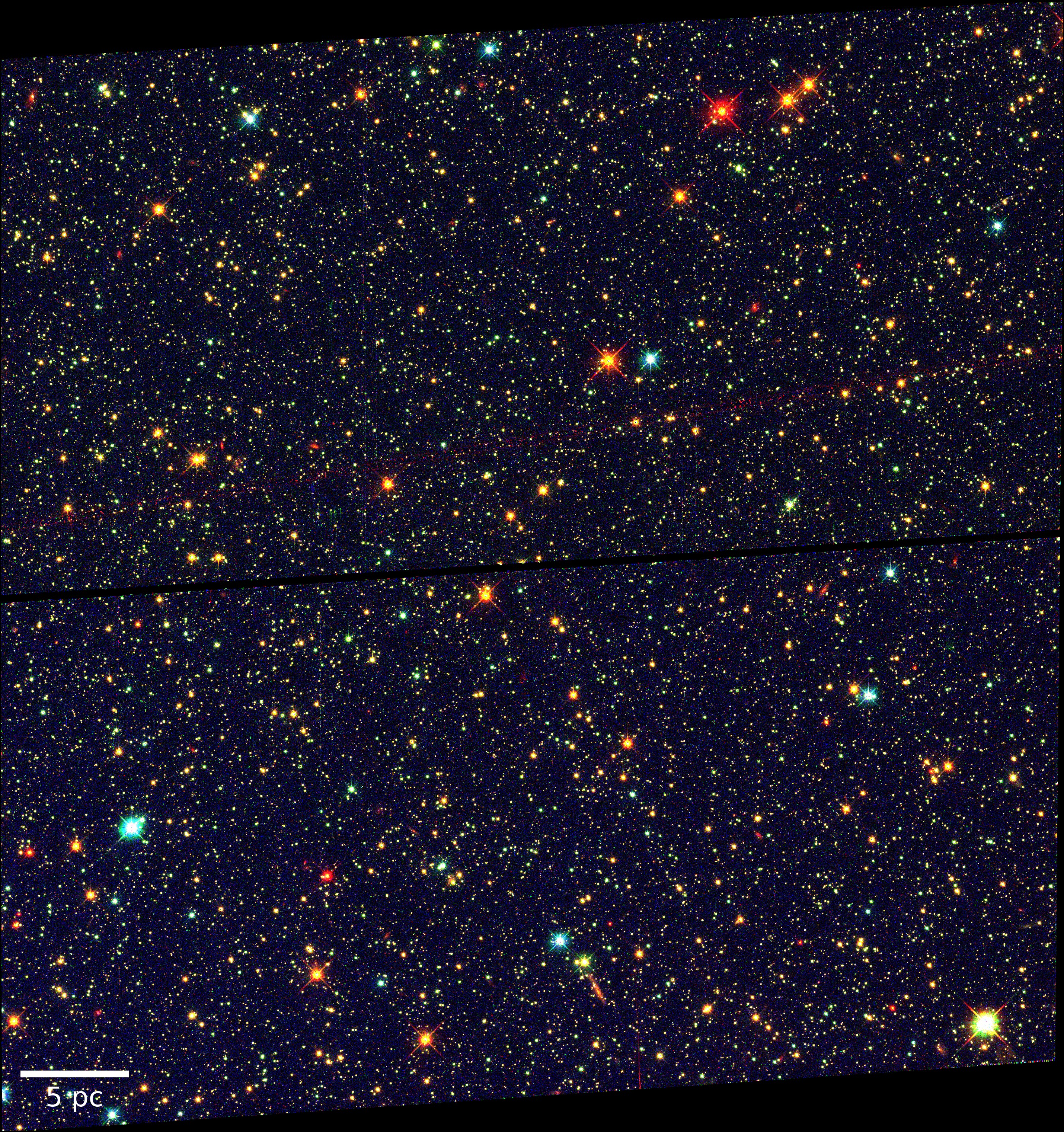}}\\
  \centering
  \subfloat[][LMC 11 with extended emission.]{\includegraphics[width=.49\textwidth]{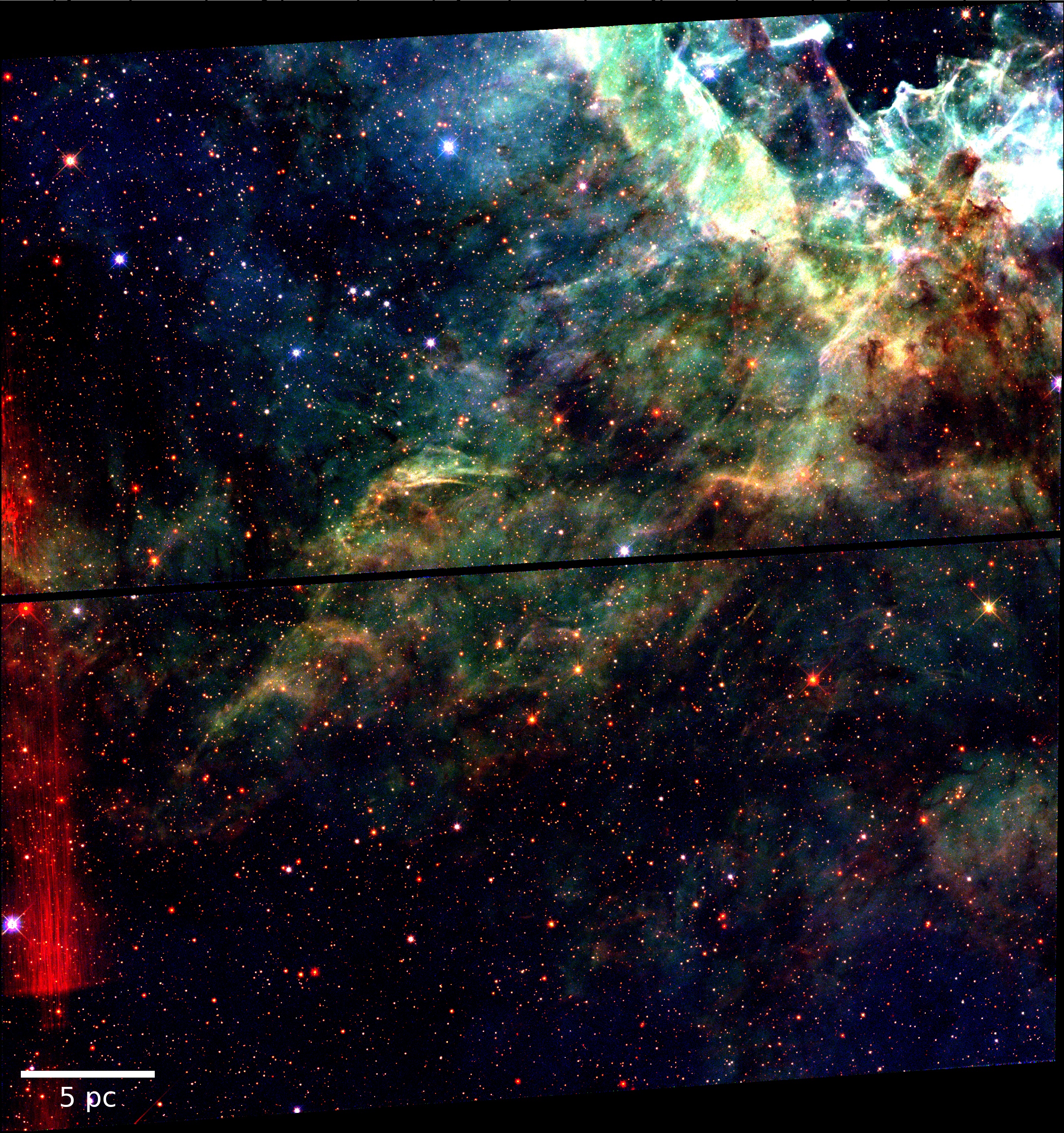}}\quad
  \subfloat[][LMC 16 with no emission.]{\includegraphics[width=.49\textwidth]{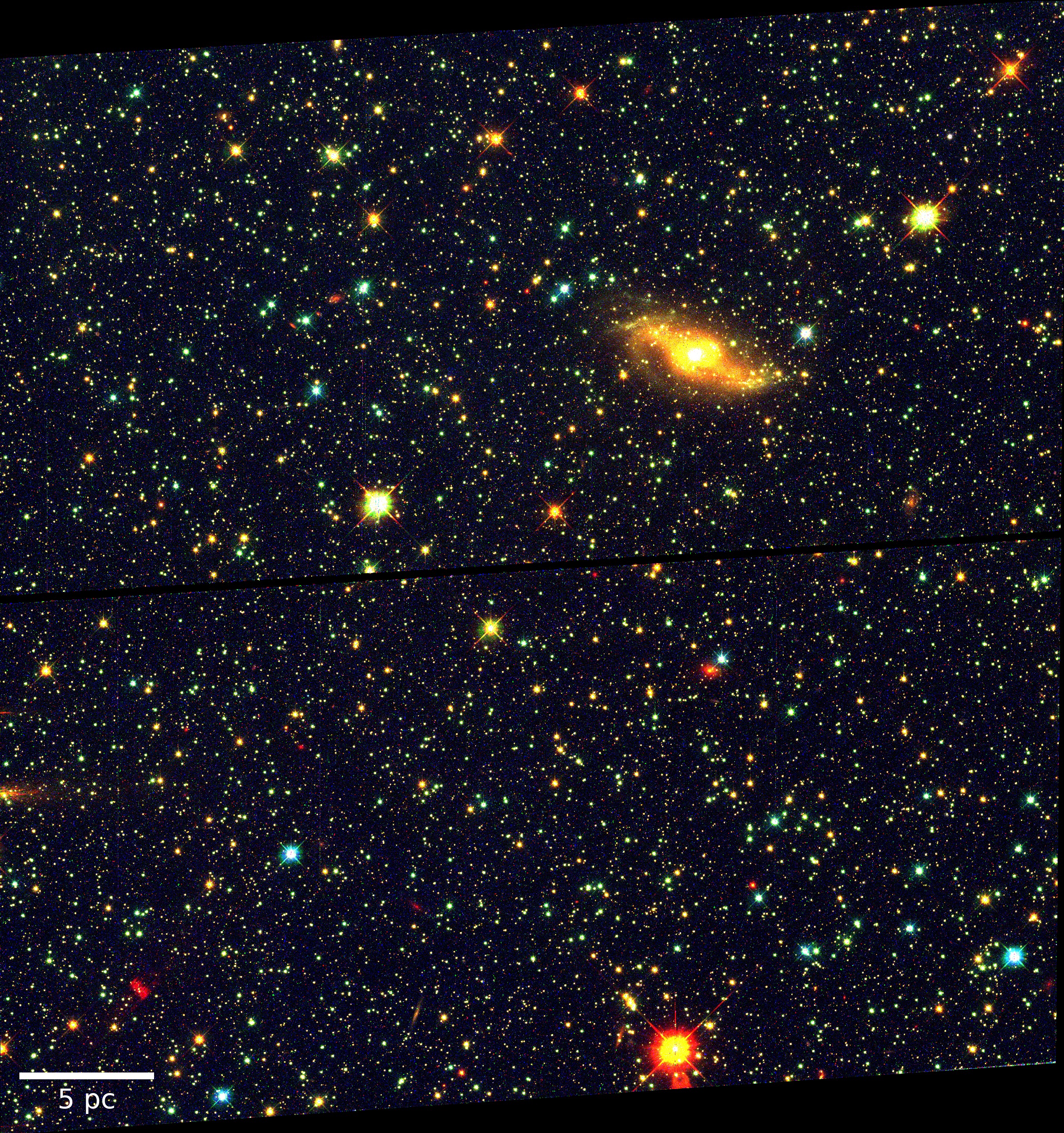}}
  \caption{Example 3-color RGB images of fields in the SMC (top) and LMC (bottom) with (left) and without (right) extended emission. Images were created using \texttt{aplpy} by normalizing the observed flux range from the F814W (red),  F475W (green), and F336W (blue) filters from WFC3. Three-color (or two-color where F336W is not available) images for all fields are available as part of our data release.}
  \label{fig:3color}
\end{figure*}

In Figure~\ref{fig:3color}, we show example three-color images (red=F814W, green=F475W, blue=F336W) for a selection of fields with these filters available. In the images on the left (a, c), we see evidence of rich extended emission, likely from [OIII] $\lambda5007$, [OIII] $\lambda4959$, H$\beta$, H$\gamma$ (F475W), {as well as nebular [S III] $\lambda9069$ (F814W), and nebular continuum, Balmer continuum, and He II $\lambda3203$ (F336W)} from the ISM \citep[as explained by][]{romanduval2019}. While beautiful, these features confuse sources and increase photometric uncertainties, making it harder to resolve individual stars. 

In SMC$\_$21 (Figure \ref{fig:3color}a), we see a smattering of bright blue sources from the open cluster NGC 346, associated with the nearby star-forming region N66 \citep{rubio2018}. We also see several saturated red sources with diffraction spikes, two of which are blended in the upper left quadrant. This field also contains imaging artifacts in the lower right corner, in the shape of a figure-8\footnote{See \hyperlink{https://www.stsci.edu/files/live/sites/www/files/home/hst/instrumentation/wfc3/documentation/instrument-science-reports-isrs/_documents/2022/WFC3-ISR-2022-03.pdf}{Instrument Science Report WFC3 2022-03} for more details.}. These artifacts appear when light from bright objects in one quadrant is reflected on the WFC3/UVIS CCD, creating ghosts in the diagonal quadrant. In this case, the two blended sources with diffraction spikes are likely the source of these imaging artifacts. However, these artifacts were not identified as sources in our catalogs, meaning we do not attempt to remove them from our images.

In LMC$\_$11 (Figure \ref{fig:3color}c), we also see extended emission structure, however, this emission is present in all optical bands we use to construct the three-color images. Since this field is near the star-forming region 30 Doradus, this emission is likely correlated with ionized gas and/or stellar light scattered by dust. In addition, we see evidence for extinction towards some of the regions that show prominent diffuse emission.

In SMC$\_$15 and LMC$\_$16 (Figures \ref{fig:3color}b and d), we find no significant evidence of extended emission. However, in LMC\textunderscore16, we do see a background galaxy.

\section{Data Reduction}
\label{sec:data_reduction}

In the following section, we describe the data reduction process, including image reduction, photometry, and post-processing (quality cuts). All HST data used in this paper can be found in MAST: \dataset[https://doi.org/10.17909/8ads-wn75]{https://doi.org/10.17909/8ads-wn75}.

\subsection{Image Reduction}

The calibrated images for the Scylla program (multi-cycle PIDs 15891, 16235, 16786) were downloaded from MAST during observations spanning April 2020 and February 2023\footnote{Additional Scylla observations were conducted between February 2023 and August 2023, and will be described as part of a future data release}. The images were processed using CalWF3 versions 3.5.0, 3.5.1, 3.5.2, 3.6.1 or 3.6.2 (versions spanning October 2018 through April 2021)\footnote{As part of a future data release, we will re-reduce all images using the same pipeline version}. For UVIS exposures, we downloaded the calibrated and CTE-corrected images (flc), and for the IR, we downloaded the calibrated images (flt) as WFC3/IR data does not require CTE correction.

We ran all of our images through the task {\tt astrodrizzle} from the Drizzlepac package\footnote{https://hst-docs.stsci.edu/drizzpac} \citep{avila2012, avila2015, hoffman2021}, which produced combined, rectified, resampled images in each band.  This process also uses a median filter to flag cosmic ray affected pixels on each of the input exposures by updating the data quality FITS file extensions. In addition, the quality of the combined images provides a visual check on the astrometric alignment of the individual exposures in each band. We processed our fields in batches as they arrived between Cycles 27 and 30, and therefore not all images are aligned in the same way. At least $70\%$ of fields were aligned to \emph{Gaia}, and the remaining fields used an a priori solution based on the observed guide stars. In a future data release, we will employ the same astrometric alignment on all fields.

\begin{figure*}[ht!]
 	\centering
\figuretitle{(a) SMC$\_$15: Quality cuts for typical Scylla field}
\includegraphics[width=\textwidth]{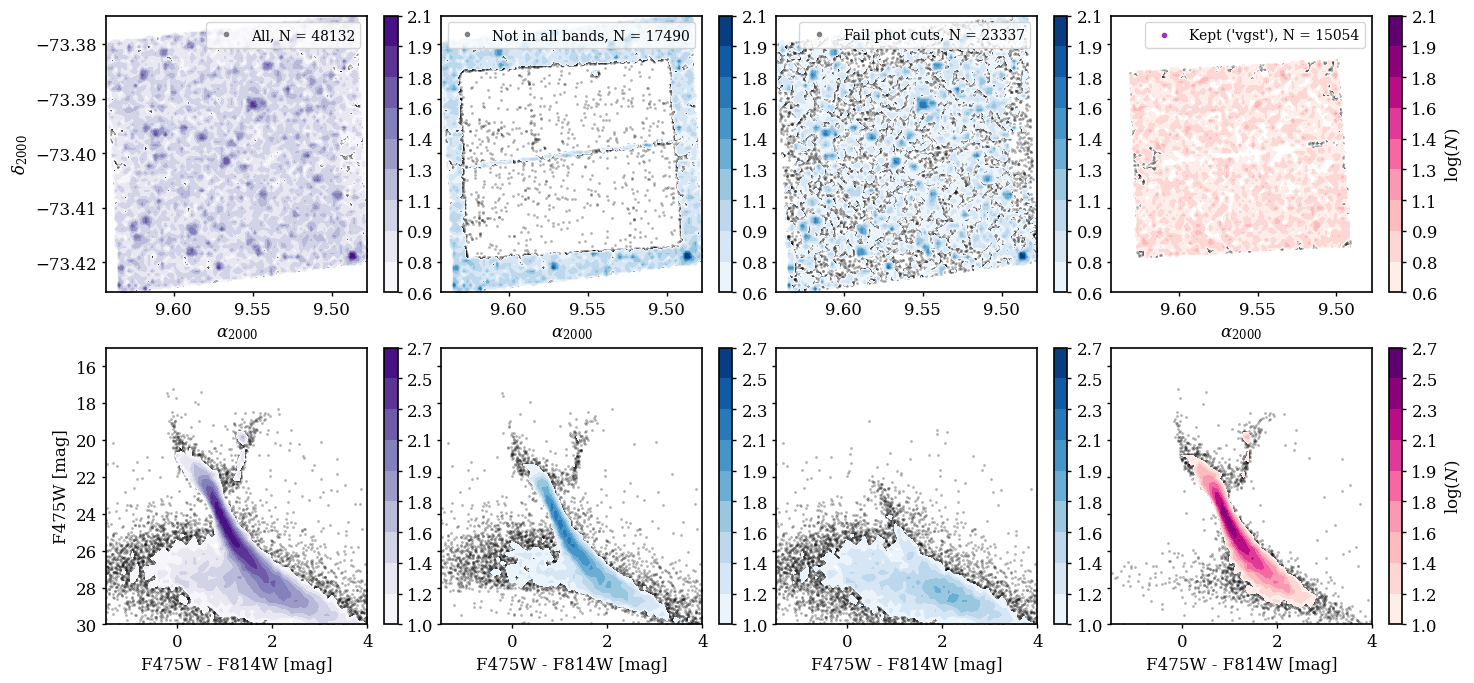}
\vspace{0.5cm}
\\
\figuretitle{(b) LMC$\_$11: Quality cuts for a high-background Scylla field }
\includegraphics[width=\textwidth]{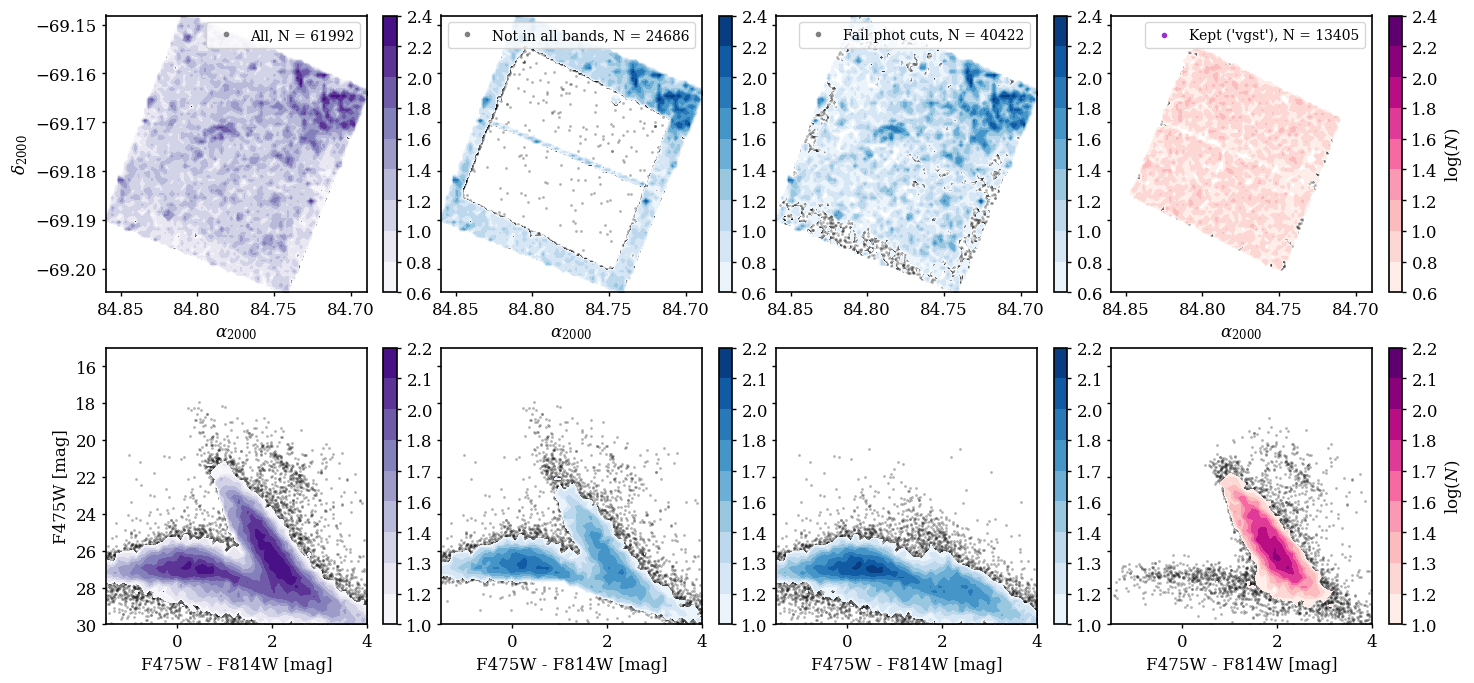}
\caption{\textbf{Illustration of the photometric quality cuts in two example fields}. (a) A typical field (SMC Field 15, Figure~\ref{fig:3color}b) and (b) a field with strong background emission (LMC Field 11, Figure~\ref{fig:3color}c). Top rows: spatial distribution of sources; Bottom row: Optical color-magnitude diagram of the same sources. Left: Full photometric catalog (.st.fits); Middle Left: Sources removed due to lack of detection in all bands; Middle Right: Sources removed after failing quality cuts; Right: Final catalog (.vgst.fits). } 
	\label{f:quality_cuts}
\end{figure*}

With the data quality extensions updated, and the combined images available for astrometric reference, we then prepared the images for crowded-field point spread function (PSF) photometry, as described in the following section.

\subsection{Photometry}

We measured the resolved photometry of all detected sources on all flt/flc exposures using the same photometry pipeline as the PHAT \citep{williams2014}, PHATTER \citep{williams2021}, METAL \citep{romanduval2019} and SMIDGE \citep{ymj2017} surveys. We independently analyzed the data from each field of the survey, providing a resolved star catalog for each location. Since detailed descriptions of the pipeline are available in those papers, here we provide an overview.    

Our photometry pipeline uses the software package {\tt DOLPHOT} \citep{dolphin2002, dolphin2016}. After running astrodrizzle on the images to flag pixels affected by cosmic rays, we applied the tools for masking and pixel area correction ({\tt wfc3mask}) to each of the flc/flt images. We then separated each of the individual CCD reads from the multi-extension fits files, and we applied the tool for generating a sky image for each CCD readout.  All processed images covering a survey pointing were then put into the {\tt DOLPHOT} task, along with a drizzled reference frame, which was the deepest single-band combined drizzled image. The {\tt DOLPHOT} task then calculated the alignment of a subset of stars in each individual image to locations of stars in the reference image, applied small corrections if necessary, and returned quality statistics on the alignment. All of the Scylla data had excellent (better than 0.1 pixel standard deviation across all alignment stars) alignment statistics, which is typical for sets of exposures all taken in a single HST visit, especially when the fields are well-populated.

Because we supplied all flt/flc exposures to {\tt DOLPHOT}, it optimized sensitivity to detect stars by stacking all pixels at a given on-sky position, i.e., after applying distortion corrections to the individual flc/flt images, covering every location in the field to look for any location with an elevated number of counts.  Then all of these locations were fitted with the appropriate PSF to measure the brightness of the candidate point source. Each source was measured iteratively, taking into account neighboring sources with each iteration. The result was an output catalog that contains combined brightness measurements in each band as well as measurements in each exposure for every candidate point source that was able to be fitted with the PSF. These initial catalogs are very inclusive to provide measurements of the faintest potential sources and allow thorough de-blending.  They are thus likely to contain contaminants and poor measurements, especially at the faint end, along with many very well-measured stars.  To help separate the unreliable measurements, {\tt DOLPHOT} provides many quality metrics for each measurement to allow users to filter the catalog in whatever way works best for their science applications. We describe our filtering below.

\subsection{Photometric Quality Cuts}

\begin{deluxetable*}{clcc|cc}
\tablecaption{Photometric Quality Cuts \label{tab:cuts}}
\tablehead{\colhead{Type} & \colhead{Parameter} & \colhead{Band} & \colhead{Value} & \colhead{$N_{\rm SMC}$} & \colhead{$N_{\rm LMC}$}}  
\startdata
\multirow{1}{*}{\STAB{\texttt{st}}}
 & Signal to noise & $^*\_$\texttt{SNR} (in any one band) & $> 4$ & \multirow{1}{*}{\STAB{4,117,542}} & \multirow{1}{*}{\STAB{3,395,349}} \\ 
  \hline \hline
\multirow{5}{*}{\STAB{\texttt{vgst}}}
  & Sharpness$^{2,a}$ & \texttt{F475W, F814W} & $\leq 0.15$ & \multirow{5}{*}{\STAB{968,217}} & \multirow{5}{*}{\STAB{972,087}} \\
  & Roundness & \texttt{F475W, F814W} & $\leq 0.6$  \\
  & Crowding & \texttt{F475W, F814W} & $\leq 0.2$ \\
  & Quality Flag$^b$ & $^*\_$\texttt{FLAG} (all bands) & $0,\,2$ \\
  & Flux$^c$ & All bands & $!=0$ \\
  \hline
\enddata
\tablecomments{\textbf{The \texttt{vgst} cuts include the \texttt{st} cuts.} $^a$: The sharpness is squared to remove sources with large negative sharpness values. $^b$: If a source is flagged by {\tt DOLPHOT} with anything other than 0 or 2 in any filter, it is cut. $^c$: If a source has a zero flux in any one band (meaning no measurement was made) it is cut, however, fluxes are allowed to be positive or negative.}
\end{deluxetable*}

There are two types of catalogs that we produced for each field: the \texttt{st} (``star") catalog and the \texttt{vgst} (``very good star") catalog. The difference between these two types is that \texttt{vgst} catalogs have more restrictive quality cuts compared to \texttt{st} catalogs {(i.e., the \texttt{vgst} catalogs are subsets of the \texttt{st} catalogs)}. The two catalog types exist to facilitate different science goals --- \texttt{st} for high completeness, \texttt{vgst} for high reliability. In this section, we describe the basic quality cuts we impose to create \texttt{st} catalogs and the additional quality cuts to create \texttt{vgst} catalogs. 

\begin{figure*}
 	\centering
\includegraphics[width=\textwidth]{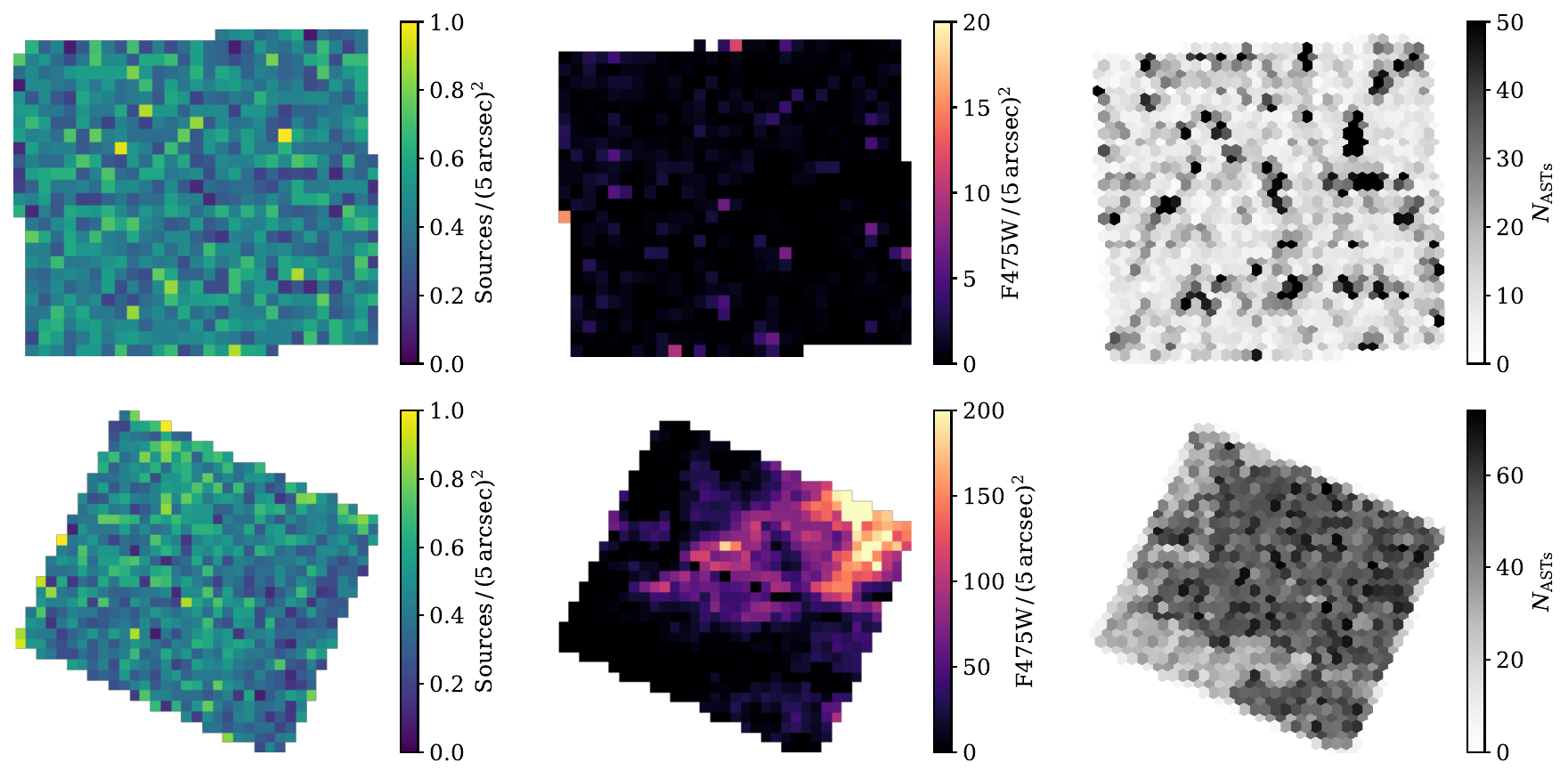}
\caption{Summary of how ASTs are placed for two example fields, one source-density dominated, (SMC Field 15, top), and one background-density dominated (LMC Field 11, bottom). Left: Source density maps; Middle: Background density maps; Right: Positions of ASTs generated before adding supplementary sources.} 
\label{fig:ast_placement}
\end{figure*}

{\tt DOLPHOT} produced \texttt{phot} catalogs that combine the measurements for each source of all the individual CCD chips into one final measurement for each source in each bandpass, including the count rate, rate error, VEGA magnitude and error, $\chi^2$ of the PSF fit, sharpness, roundness, crowding, and signal-to-noise (SNR). For the details of this step see Sec. 2.3 of \cite{williams2014}. The \texttt{st} catalogs are generated from \texttt{phot} including all sources with signal to noise greater than 4 in at least one observed band. 

To construct the \texttt{vgst} catalogs, we imposed additional quality cuts to remove spurious detections from the \texttt{st} catalogs. Specifically, we used the ``sharpness", ``roundness" and ``crowding" parameters from {\tt DOLPHOT}. The sharpness parameter quantifies how centrally-concentrated the source flux is; high values correspond to sources with high values in the central pixels of the PSF relative to the edges (e.g., cosmic rays or hot pixels), and low values correspond to blended sources or galaxies. The roundness parameter quantifies how circular a source is, where a value of zero corresponds to a perfectly round source. The crowding parameter quantifies how neighbors affect the source photometry; large values correspond to sources whose PSF radii intersect with many others. These parameters are described in the {\tt DOLPHOT} documentation\footnote{\url{http://americano.dolphinsim.com/dolphot/}}, as well as discussed and implemented as quality cuts by other surveys \citep[e.g.,][]{williams2014, williams2021}. 

First, we removed any sources not measured in all bands. This step excludes sources around the edges of each field due to the smaller size of the WFC3 IR footprint compared to that of the UVIS camera. In future work, we {plan} analysis which alternately focuses on the HST UV and optical photometric bands to include sources in the UVIS camera footprint, but excludes the HST IR photometric bands.

Next, we applied cuts to exclude diffraction spikes or resolved emission identified as sources by {\tt DOLPHOT}. Inspecting the distribution of sources on the plane of the sky and in color-magnitude space, we found a significant number of sources that are likely spurious detections. These sources are generally bluer and fainter than the main sequence, and are distributed along recognizable imaging artifacts (e.g., edges of the field and diffraction spikes). To remove these sources from the catalogs, we visually identified the contaminant sources by their locations and identified the cuts listed in the Table~\ref{tab:cuts} (see Appendix~\ref{a:spikes} for details). 

Finally, we imposed cuts based on the flags generated by {\tt DOLPHOT} during processing, which have values of 0 (``good") through 8 (``extremely bad"). Per the {\tt DOLPHOT} documentation, sources with flag values of 0-3 are usable. \texttt{FLAG}=1 and \texttt{FLAG}=3 indicate that the photometry aperture extends off the chip, and \texttt{FLAG}=2 indicates too many bad or saturated pixels. For some fields, a significant fraction of sources have \texttt{FLAG}=2 ($\sim25\%$ for SMC-6). Since sources extracted from saturated pixels may be due to bright stars of interest, we kept sources with \texttt{FLAG}=2 in the catalog. {Ultimately, we kept sources with \texttt{FLAG} values of 0 and 2}.

After applying these cuts to a representative sample of fields (i.e., varying depth, filter coverage, and source density), we visually confirmed that they do a reasonable job of removing spurious sources while removing few real sources, in agreement to the performance of similar quality cuts used in other surveys \citep[e.g.,][]{williams2014}. In Figure~\ref{f:quality_cuts}, we compare the full catalog with the kept and cut sources for two fields at representative steps in the process, including sources which are not observed in all photometric bands, and sources which fail our quality cuts (includes {\tt DOLPHOT} and diffraction spike cuts).  

We note that we did not include a cut on SNR or $\chi^2$; additional quality metrics quantifying the goodness-of-fit to the PSF used for cuts by similar surveys \citep[e.g.,][]{williams2014, williams2021}. We made this choice to ensure that even low-SNR sources (which may include rare, interesting stellar populations) are included in the \texttt{vgst} catalogs, which we will characterize with the BEAST as part of future work. Our quality cuts, and the total number of sources retained in each MC, are summarized in Table~\ref{tab:cuts}.

\begin{figure*}
 	\centering
\includegraphics[width=\textwidth]{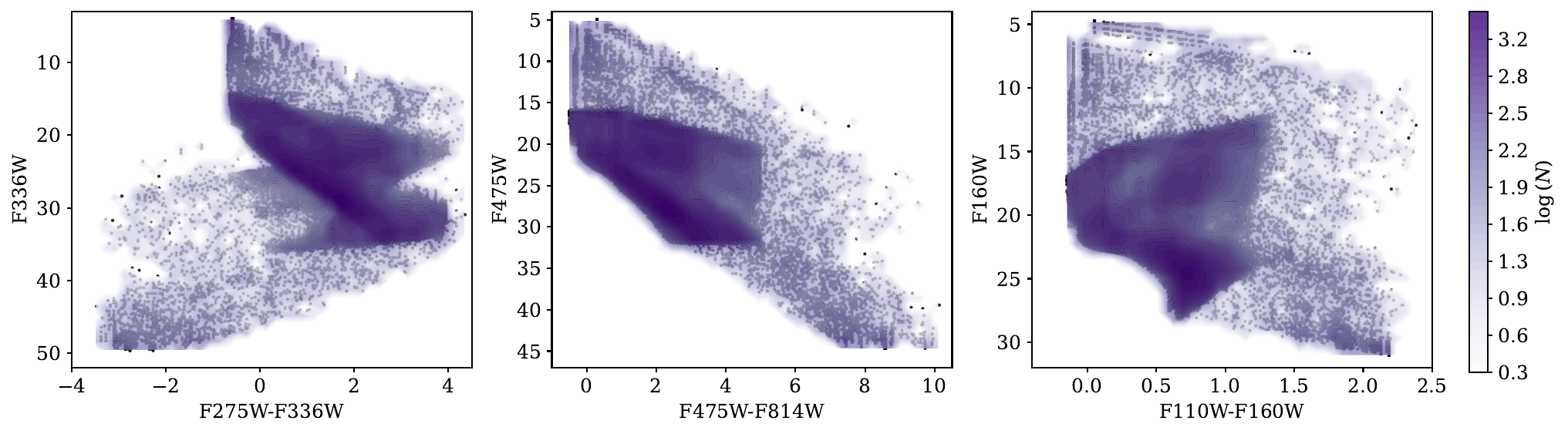}
\caption{CMDs of the input ASTs for background density-dominated Scylla LMC Field 11 (15891$\_$LMC-5389ne-11134), UV (left), optical (center), and IR (right).} 
\label{fig:ast_cmds}
\end{figure*}

\section{Artificial Star Tests}
\label{sec:asts}

To quantify the uncertainty, bias, and completeness in our photometry, we performed {artificial star tests (ASTs)}. This procedure involves placing simulated sources with known parameters within real images, repeating the same photometric extraction procedure, and comparing the derived properties of each fake source with its known intrinsic properties. 

To create the AST inputs, we used the BEAST: a probabilistic method for modeling multi-wavelength SEDs from large photometric surveys. With the BEAST, we generated multi-wavelength SEDs in the Scylla survey bands for sources spanning a representative range of stellar and dust properties for the {MCs}. Using publicly available stellar atmosphere grids \citep{castelli2003, lanz2003, lanz2007} and evolutionary tracks \citep[PARSEC;][]{marigo2008, bressan2012, chen2014, chen2015, marigo2017, pastorelli2019, pastorelli2020}, the BEAST generates the intrinsic spectrum of a star based on its age ($t$), mass (M$_{ini}$), metallicity ($Z$), and distance ($d$). This spectrum is then extinguished according to $A_V$, $R_V$, MW and SMC/LMC dust extinction mixture coefficient \citep[$f_A$; see][]{gordon2016}. Finally, the full extinguished stellar spectrum is converted into SEDs by integrating over the transmission curve of each HST photometric band. We summarize the seven BEAST parameters ($t$, $M_{\rm ini}$, $Z$, $d$, $A_V$, $R_V$ and $f_A$), their ranges, and prior assumptions in Table~\ref{tab:beast}.  

\begin{table*}[t]
\centering
\caption{BEAST Model Parameters Used for AST Generation} \label{tab:beast}
\begin{tabular}{lccccccc}
\hline
\textbf{Parameter} & \textbf{Unit} & \textbf{Description} & \textbf{Min} 
& \textbf{Max} & \textbf{Resolution} & \textbf{Prior} \\
(1) & (2) & (3) & (4) & (5) & (6) & (7) \\
\hline
$\log(t)$ & years & stellar age & 6.0 & 10.13 & 0.1 & flat SFR \\
$\log(M)$ & $M_\odot$ & stellar mass & $-0.8$ & 2.0 & variable & Kroupa IMF \\
$\log(Z/Z_\odot)$ & $\cdots$ & stellar metallicity & $-2.1$ & $- 0.3$ & 0.3 & flat \\
$A_V$ & mag & dust column & 0.01 & 10.0 & 0.05 & flat \\
$R_V$ & $\cdots$ & dust average grain size & 2.0 & 6.0 & 0.5 & peaked at $\sim$3 \\
$f_A$ & $\cdots$ & dust mixture coefficient & 0.0 & 1.0 & 0.2 & peaked at 1 \\
$d$ [LMC] & kpc & distance & 40 & 60 & 2.5 & flat \\
$d$ [SMC] & kpc & distance & 47 & 77 & 2.5 & flat \\
\hline
\end{tabular}
\tablecomments{(1): Parameter name; (2) Unit; (3) Description; (4) Minimum value; (5) Maximum value; (6) Interval; (7) Prior model}
\end{table*}

For each Scylla field, we generated a set of artificial stars using the BEAST model grid. To ensure that the artificial stars cover a wide range of fluxes, we estimated the total flux range in each observed band and split that range into 40 flux bins. We then select BEAST SEDs randomly until we have at least 50 in each bin, which results in $\sim2000$ SEDs per field. 

\subsection{AST placement by source density}

Given the list of artificial stars, the next step is to determine where to place them within the images. Similar HST surveys have established that the noise properties of stars can vary considerably with the local source density, especially for the lower-resolution near-IR bands \citep{gordon2016}.  To incorporate this effect into our uncertainty modeling, we placed artificial stars according to the local source density within the field. We computed the source density across each Scylla field by counting the number of sources with photometric F475W Vega magnitudes between 26 and 15 (the range over which the catalogs are complete based on preliminary completeness tests) within pixels the size of $5^{\prime  \prime}$ on each side. In addition, we eroded the field footprint boundary by $0.5^{\prime \prime}$ on all sides to avoid computing source densities at the very edge of the WFC3 chips. We find that the source density within Scylla fields {follow a roughly log-normal distribution, with most source densities ranging between 0-2 sources per arcsec-squared, peaking at 0.3 and 0.6 sources for the LMC and SMC, respectively. We note that fields in the wing of the SMC (Fields 11, 22, 28, 38, 47, 48, and 52) have extremely low source densities, less than 0.1 sources per arcsec-squared on average. By comparison, other photometric surveys like PHAT observed source densities ranging between $\sim$10-20 in most fields of M31 with similar filters \citep[e.g. Figure 12 in ][]{dalcanton2012}. To reflect the range of source densities observed in the MCs,} we established a set of five source density bins, $(0, 0.27, 0.526, 1.026, 2, >2)$ sources/arcsec$^2$, and placed all $\sim 2000$ artificial stars randomly within the pixels corresponding to each source density bin. In this step, we generated $\sim10000$ artificial stars per field. 

\subsection{AST placement by background density}

\begin{figure*}[h]
 	\centering
        \includegraphics[width=0.45\textwidth]{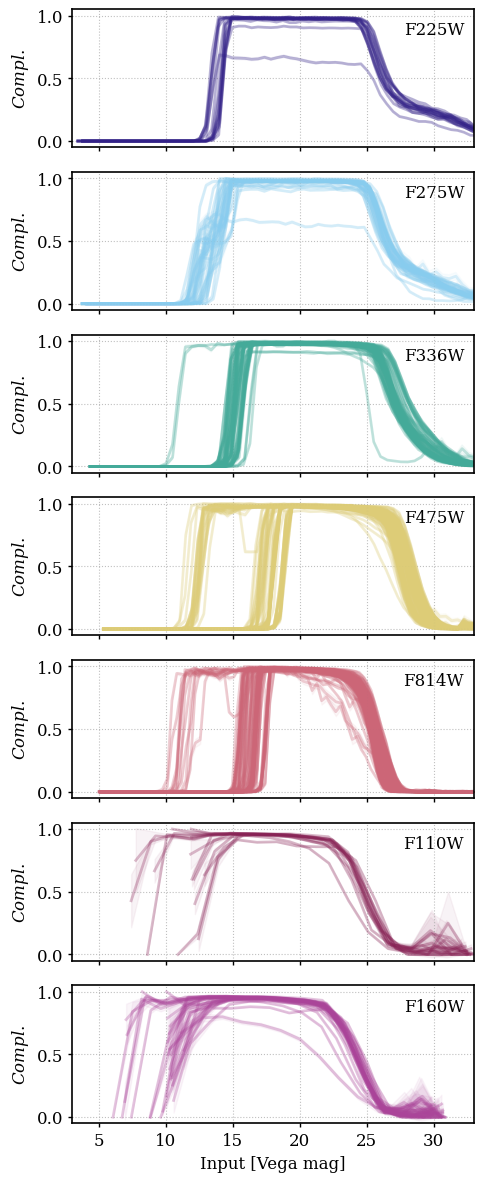}
        \includegraphics[width=0.45\textwidth]{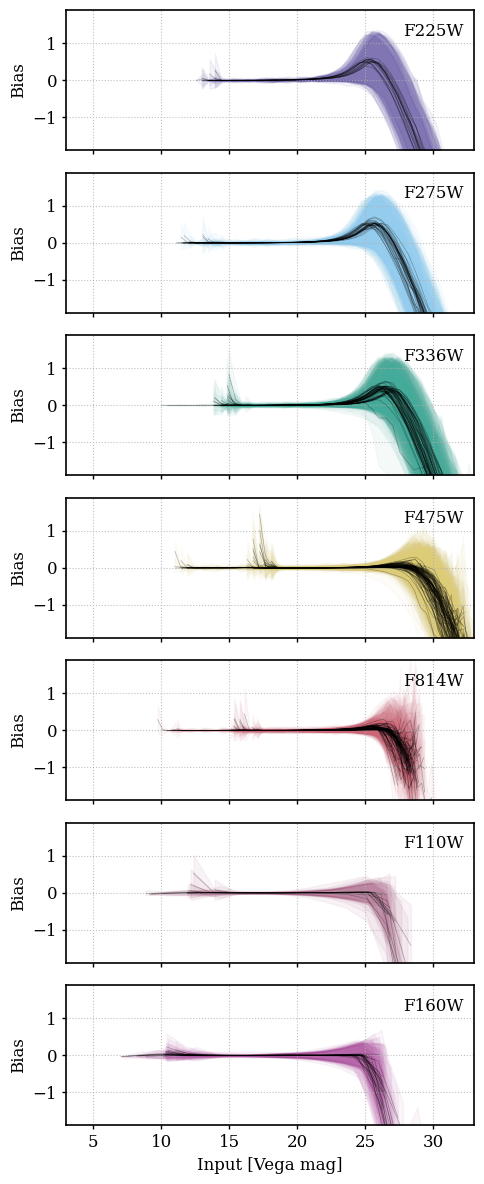}
	\caption{A summary of the completeness, bias and uncertainty ($\sigma$) for all Scylla fields. In the left column, we plot completeness as a function of flux for the Scylla observing bands. In the right column, we plot bias (black lines) with shading corresponding to $\pm 1\sigma$.}
 \label{fig:asts_all}
\end{figure*}

\begin{figure*}[ht!]
 	\centering
        \includegraphics[width=0.95\textwidth]{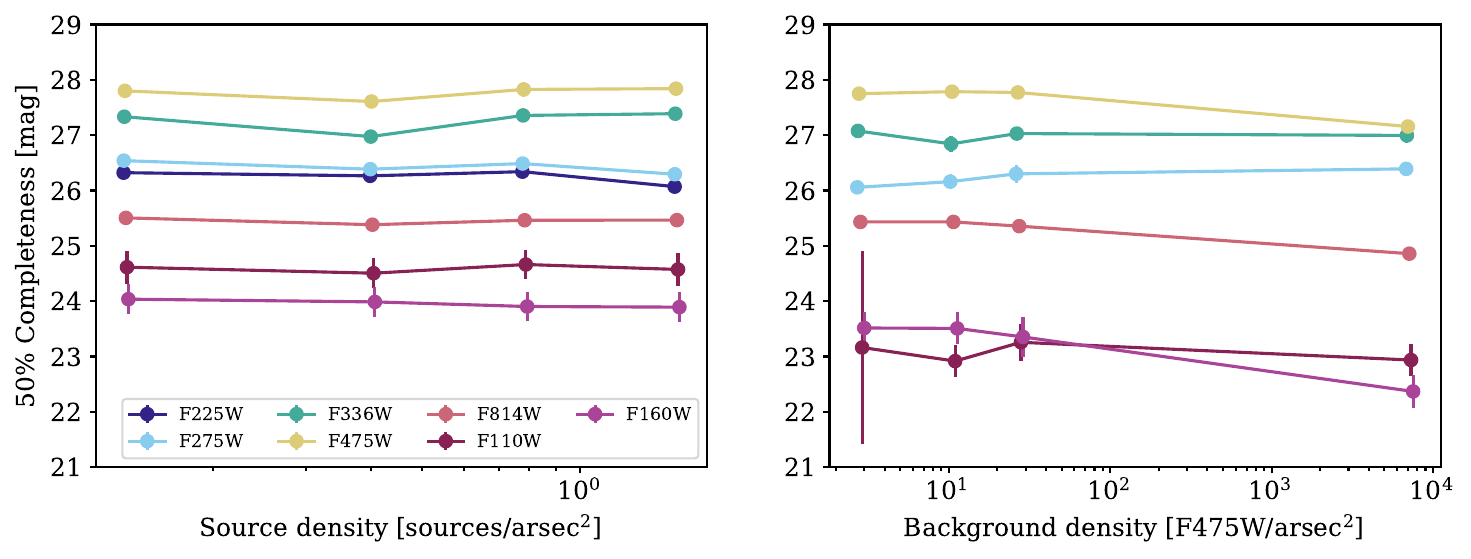}
	\caption{ Left: $50\%$ completeness as a function of source density for each band in fields which are source density-dominated.  Right: $50\%$ completeness as a function of background density for each band in fields which are background-density dominated.  }
 \label{fig:asts_variation}
\end{figure*}

When there is a significant extended emission present in a field (cf.\ Figure~\ref{fig:3color}), the variations in uncertainty, bias, and completeness across the field are no longer dominated by source density. We identified these ``background-dominated" fields by analyzing the flux density in the F475W band (common to all Scylla fields). We first computed the average flux of detected sources in $5\times5$ arcsec$^2$ pixels in each field. We then computed the inter-quartile range (equal to the $84^{\rm th}$ minus $16^{\rm th}$ percentiles) of flux density for all fields. We flagged a field if the inter-quartile range is greater than $17\rm\,mag/arcsec^2$. This cutoff was selected by eye to identify the fields which have significant optical emission. Out of 96 total fields, 12 are background emission-dominated (six in the LMC, six in the SMC). 

For these fields, instead of placing artificial stars by source density, we generated a custom set of background-density bins for each field. Specifically, we selected bin edges to generate five equal-sized (i.e., same number of sources) bins of background density. 

In Figure~\ref{fig:ast_placement}, we plot the source density and background density maps for two representative Scylla fields, one source density-dominated (top; SMC$\_$15) and one background-density dominated (bottom; LMC$\_$11). We also compare the positions of the artificial stars for the two cases. Although the source density-dominated field shows relatively uniform placement, the background density-dominated field's sources are placed where the optical emission is bright.

\subsection{Supplementary ASTs}

In addition to the artificial stars distributed by source density (or background density), we generated supplementary artificial stars in each field to more densely sample the magnitude ranges of our observations. For each field, we generate $>10^5$ SEDs using the BEAST with similar priors as summarized in Table~\ref{tab:beast}, restricted to the observed magnitude ranges (Cohen et al. 2024 a,b). These sources are spatially distributed uniformly within each field. 

In Figure~\ref{fig:ast_cmds}, we plot color-magnitude diagrams of the AST inputs for a single Scylla field (LMC$\_$11). The initial BEAST-generated ASTs span a wide range in color-magnitude space in UV, optical, and IR bands. The supplementary ASTs are clearly identifiable as the high-density range of sources spanning the main observable CMD ranges in each panel. 

\subsection{AST Results}

\begin{figure}[ht!] 
 	\centering
    \includegraphics[width=0.5\textwidth]{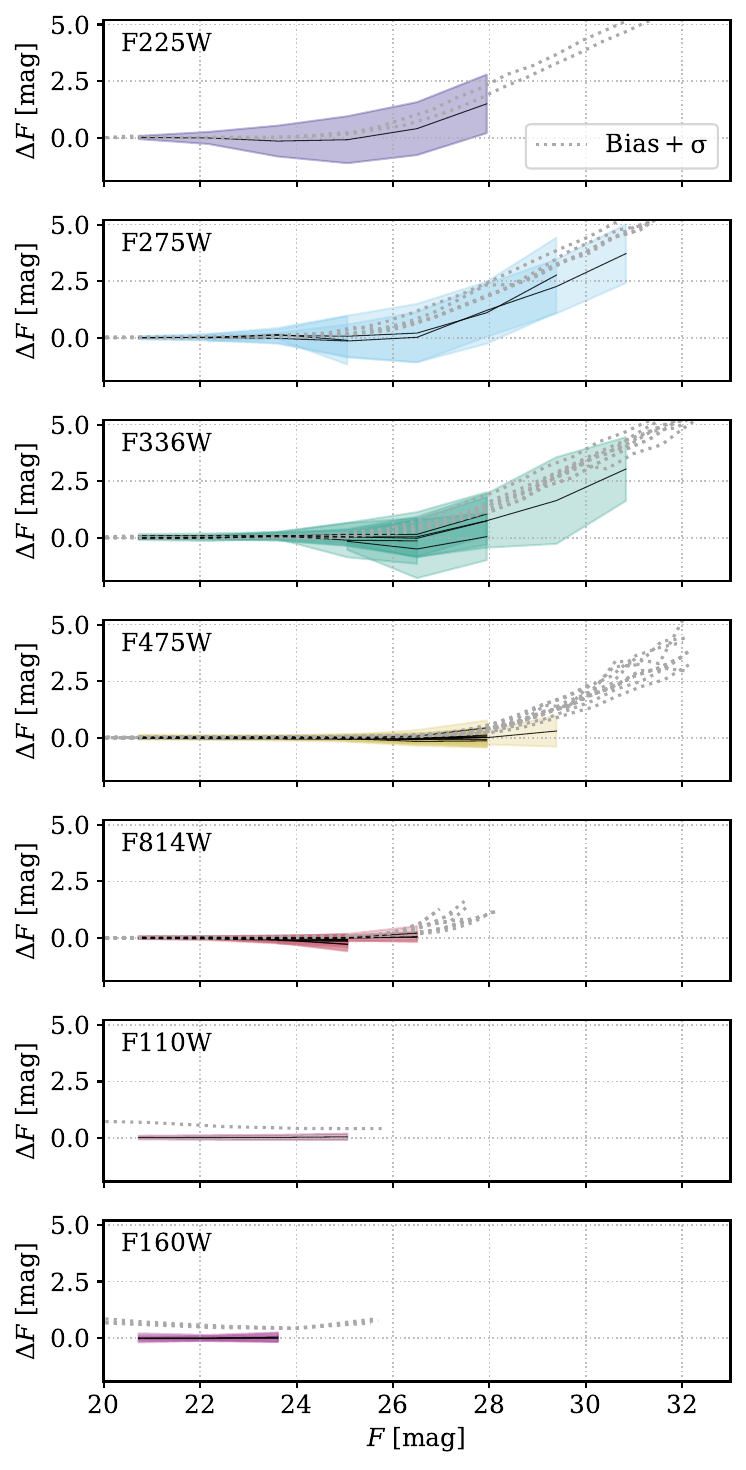}
	\caption{The difference in observed magnitude as a function of flux for fields that overlap spatially in Scylla. The black lines indicate the median difference in bins of decreasing flux (increasing magnitude) and the color-shaded regions denote $\pm1\sigma$ within each band. We overplot the results of the ASTs (bias+uncertainty) for each field in each panel and observe that the differences are all within measured uncertainties. }
    \label{fig:overlap}
\end{figure}

With a final list of ASTs for each field, we repeat the full photometry procedure for each individual artificial star in each field. Once the ASTs are processed, we compared the photometric properties of the sources to their intrinsic values and computed the completeness, uncertainty ($\sigma$), and bias ($\mu$) in each photometric band for each field. In Figure~\ref{fig:asts_all}, we plot these quantities as a function of Vega magnitude for all Scylla fields. In Appendix~\ref{a:compl}, Table~\ref{tab:summary_obs} we list the $50\%$ completeness limit in each band for each field. 

In summary, we reach average $50\%$ completeness of $m_{\rm F225W}=26.0$, $m_{\rm F275W}=26.2$, $m_{\rm F336W}=26.9$, $m_{\rm F475W}=27.8$, $m_{\rm F814W}=25.5$, $m_{\rm F110W}=24.7$, and $m_{\rm F160W}=24.0$ Vega magnitudes in the \texttt{vgst} catalogs.

As shown in Figure~\ref{fig:asts_all}, the Scylla survey is extremely deep in the optical bands (F475W and F814W, our first-priority filters), where we reach $50\%$ completeness limits of 26-28$^{\rm th}$ Vega magnitudes ($>3$ mag below the oMSTO). For a subset of fields, the primary observation setup enabled us to include short ``guard" exposures which allowed us to extract the photometry of bright sources (which would otherwise be saturated by our standard long exposures). These fields are evident in Figure~\ref{fig:asts_all} as the groups of fields whose completeness extends much brighter than $\sim15^{\rm th}$ magnitude. For the UV bands, we reach $50\%$ completeness of $<25^{\rm th}$ magnitude. In the near-IR we were limited to short, single exposures, resulting in lower completeness and higher uncertainty as a function of flux. 

\begin{figure*}
 	\centering  
    \includegraphics[width=0.95\textwidth]{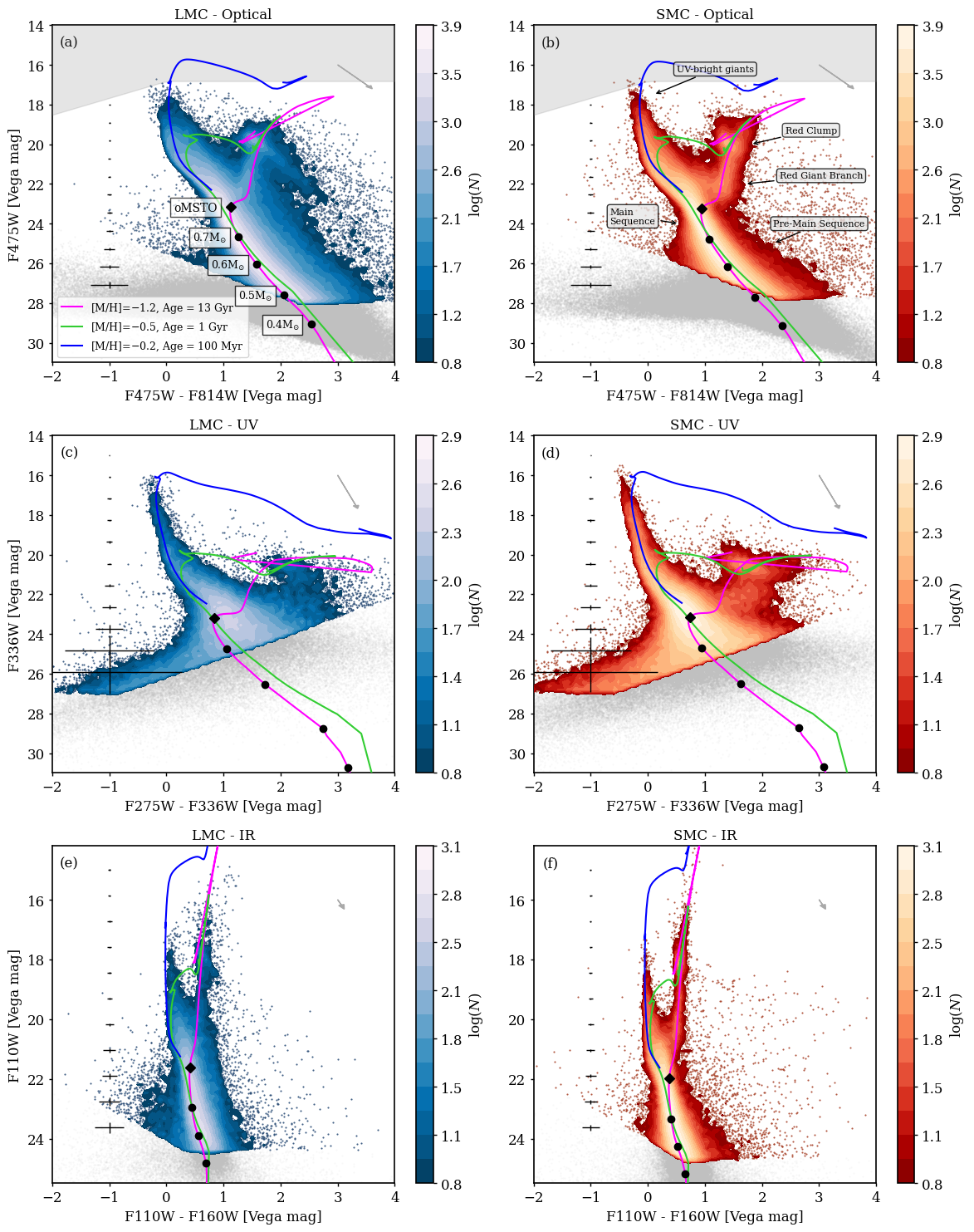}
	\caption{Stacked CMDs in the optical (top), UV (middle), and IR (bottom) bands for the LMC (left) and SMC (right) for the \texttt{vgst} catalogs. Sources brighter than the average $50\%$ completeness limits are shown in color contours, and fainter sources are grey. Average photometric uncertainties as a function of flux, computed from the ASTs are shown as error bars. Saturation limits in each band are shaded grey in the optical panels (all others are much brighter). Reddening vectors (grey, top right) and PARSEC isochrones are overlaid indicating the main sequence stellar mass range typical of our photometric catalogs.}
 \label{fig:cmds_all}
\end{figure*}

\begin{figure*}
 	\centering
\includegraphics[width=\textwidth]{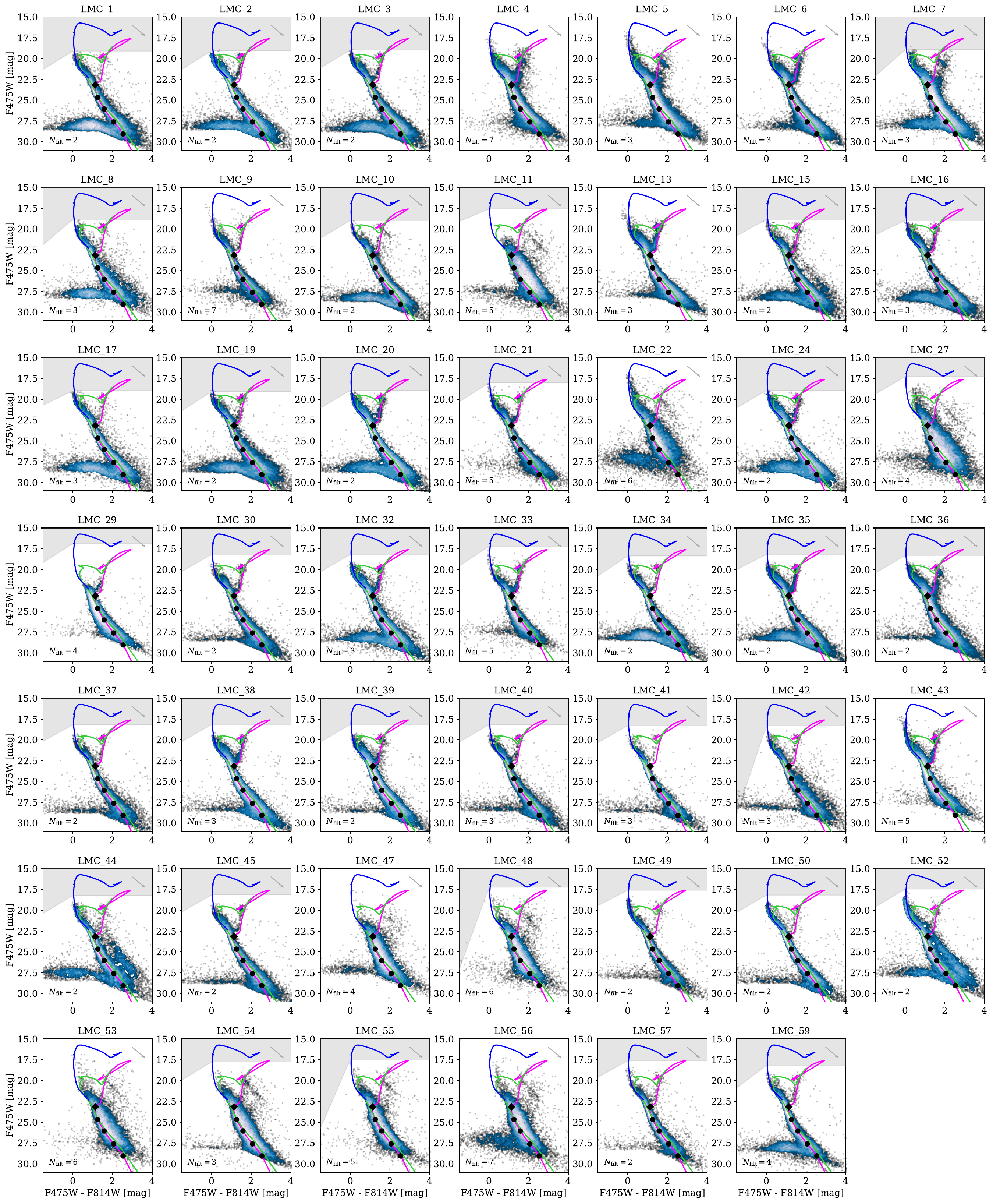}
	\caption{Optical CMDs for all LMC fields from the \texttt{vgst} catalogs. Saturation limits in each band are shaded grey in each panel. Reddening vectors (grey, top right) and PARSEC isochrones are overlaid indicating the main sequence stellar mass range typical of our photometric catalogs (same as Figure~\ref{fig:cmds_all}). We indicate the number of filters obtained for each field in the bottom right corners. }
    \label{fig:cmds_lmc}
\end{figure*}

\begin{figure*} 
 	\centering
\includegraphics[width=\textwidth]{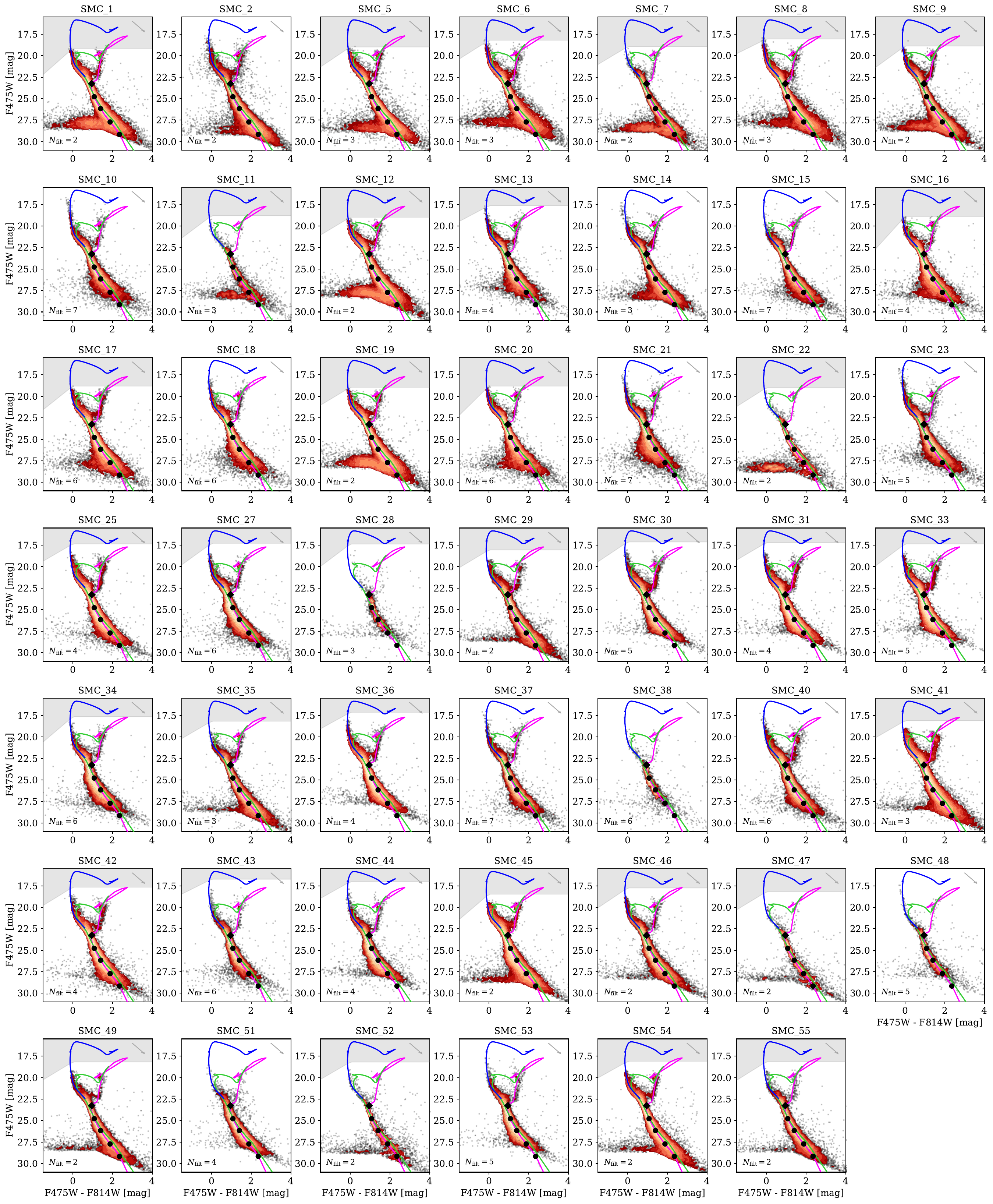}
	\caption{Optical CMDs for all SMC fields from the \texttt{vgst} catalogs. Saturation limits in each band are shaded grey in each panel. Reddening vectors (grey, top right) and PARSEC isochrones are overlaid indicating the main sequence stellar mass range typical of our photometric catalogs (same as Figure~\ref{fig:cmds_all}).We indicate the number of filters obtained for each field in the bottom right corners.}
    \label{fig:cmds_smc}
\end{figure*}

There is one Scylla field with completeness $\sim 60\%$ across the full magnitude range for F225W and F275W (LMC$\_56$). In this case, during photometric processing two observations with similar spatial coverage but different observed filters were combined into the same field. As a result, the sources with no UV observations were removed during our \texttt{vgst} quality cuts, introducing artificially low completeness in those bands. In our next round of processing, we will treat each of these fields as a separate observation and correct the completeness accordingly.

To explore how our observational completeness varies within fields, we combine all ASTs and compute the $50\%$ completeness limit in bins of source density and background density. In Figure~\ref{fig:asts_variation}, we find that the completeness remains constant within our uncertainties for all source densities probed by Scylla. The exception is the bin at $\sim0.4$ sources/deg$^2$, where residual artifacts from incompletely-subtracted diffraction spikes cause the completeness to dip slightly, especially for the shortest wavelengths. That the trend overall is flat is not surprising, given the low, uniform values of source density relative to HST's resolution limit across the survey footprint. In contrast, we observe that the completeness starts to decline with increasing background density for the optical-NIR bands. {Overall, however,} our stellar recovery for high background density fields appears to be performing well. This is likely due to the fact that {\tt DOLPHOT} successfully measures and subtracts the local background around sources. Our uncertainties are highest in the IR bands, as very few background-dominated fields also had IR exposures. For the shorter-wavelength filters, the completeness is similar to the source density-dominated cases.

\subsection{Comparison of Overlapping Fields}

In total, 15 Scylla fields have some overlap in terms of the positions of detected sources in their \texttt{vgst} catalogs. To verify that the photometry is consistent, for all filters that are the same between the overlapping fields, we compare the output fluxes in Figure~\ref{fig:overlap}. For comparison, we include the average uncertainty and bias in each filter based on the ASTs (Figure~\ref{fig:asts_all}) between the overlapping fields. We find that the average difference in the derived fluxes ($\Delta F$) falls within the uncertainties from the ASTs in all cases, and the spread in $\Delta F$ is contained within $\pm 3\sigma$ for all bands.

\subsection{Data Release}

Our catalogs are made publicly available as \texttt{st} fits files, with flags for \texttt{vgst}, at MAST as a High Level Science Product (HLSP)\footnote{https://archive.stsci.edu/hlsp/scylla}, {with a permanent DOI of \dataset[10.17909/mk54-kg51]{\doi{10.17909/mk54-kg51}}}. In this data release, we also include the drizzled .fits images, as well as the .pngs of the three-color images for all fields (two-color where only F475W and F814W are available). Artificial star tests will be made available as part of the HLSP at a later date.

\section{Results}
\label{sec:results}

The result of the Scylla survey is a deep, diverse sampling of the photometric properties of stars in the LMC and SMC. With the addition of the per-field ASTs, we also have a detailed understanding of our completeness, uncertainty and bias in each observation. In this section, we highlight these products and their potential for impact on our understanding of the ISM and stellar populations of the MCs. 

\subsection{Color-Magnitude Diagrams}
\label{sec:cmds}

To summarize our results, we create stacked CMDs for the optical, UV, and NIR bands in Scylla LMC and SMC fields (Figure~\ref{fig:cmds_all}). We observe that both galaxies include similar stellar populations, as seen by the range of color and magnitude spanned by each stacked CMD. 

In both MCs, our observations sample the unevolved main sequence down to $\gtrsim$2 mag faintward of the ancient main sequence turnoff in optical (F475W and F814W) filters, while cool lower main sequence stars preferentially evade detection in the bluest filters (F275W, F336W).  At the bright end, saturation prevents detection of massive main sequence stars younger then a few hundred Myr.  However, in all bandpasses, the breadth of the observed evolutionary sequences is due in part to age and metallicity distributions that vary across the sightlines sampled by our fields.

A key difference between the LMC and SMC in CMD space is the shape of the red clump and the main sequence in the optical bands. The red clump consists of K giants, and although its morphology in the CMD depends on the age and metallicity distribution of the underlying population, it can function as a standard candle (see \citealt{girardi16} for a review).  

In the presence of dust, however, these stars experience a shift toward fainter magnitudes and redder colors. Varying amounts of dust cause this effect to a varying degree, and the result is that the red clump appears extended in color and magnitude along a reddening vector (shown in grey in the top right corner of each panel of Figure~\ref{fig:cmds_all}). We observe from the optical and the IR CMDs that the LMC red clump is highly elongated along this reddening vector whereas the SMC red clump is much more compact along the color axis. The effect appears reversed for the UV CMD, but we note that this is due to the higher density of sources in the SMC UV CMD since in the SMC there are more Scylla fields which were observed with HST's UV filters than in the LMC (cf., Table~\ref{tab:filter_combos}). Additionally, in the optical bands the LMC main sequence appears significantly wider than the SMC main sequence. This suggests that the LMC contains more dust, which is consistent with previous analysis of IR emission in these systems \citep{gordon2014, chastenet2017, utomo2019, clark2023}. CMD-based analysis of LMC and SMC dust extinction properties similarly suggests that the amount of dust in the two galaxies causes noticeably different effects, specifically in the appearance of CMD features such as the red clump and the red giant branch \citep[RGB;][]{DeMarchi2016, ymj2017}.

To further investigate the variations between fields, we plot the individual optical CMDs (as F475W and F814W are common to all observations) for all Scylla fields in the LMC (Figure~\ref{fig:cmds_lmc}) and SMC (Figure~\ref{fig:cmds_smc}). We observe that there is a significant variation in CMD shapes between the 96 fields, between and within each MC. Due to the lack of short exposures in most fields, our saturation limits (shown in shaded grey) illustrate our lack of sensitivity to the youngest and oldest bright sources.  Even within each MC, differences in extinction among and within different sightlines are readily apparent (for example, compare LMC\_2, LMC\_53 and LMC\_54 in Fig.~\ref{fig:cmds_lmc}).  Furthermore, recent star formation is evident in some cases (e.g., LMC\_44 and LMC\_52 near the N11 star forming complex) from the presence of pre-main sequence stars redward and parallel to the main sequence \citep{gouliermis12}.  

\begin{figure*}[t!]
 	\centering  
    \includegraphics[width=0.9\textwidth]{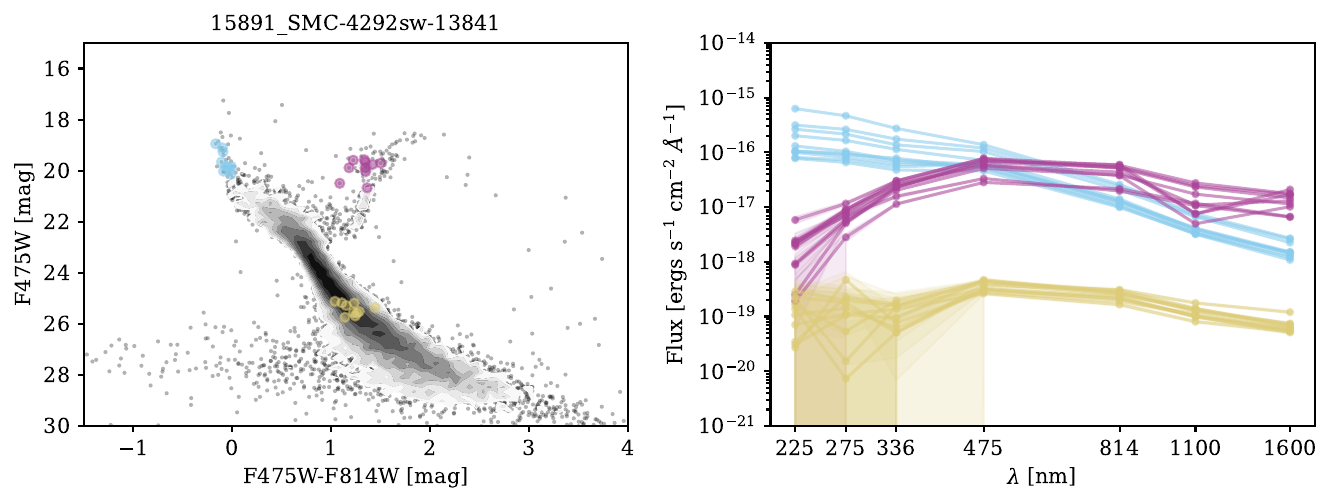}
	\caption{\textbf{Example SEDs of sources from different regions on a CMD.} Left: An optical CMD of SMC Field 15 (15891{\textunderscore}SMC-4292sw-13841) with 10 randomly selected sources in each of the three regions (Young main sequence: blue; RGB: pink; dim main sequence: yellow). Right: SED of each source, colored by region on the CMD.}
 \label{fig:sed}
\end{figure*}

\begin{figure*} 
 	\centering
\includegraphics[width=0.25\textwidth]{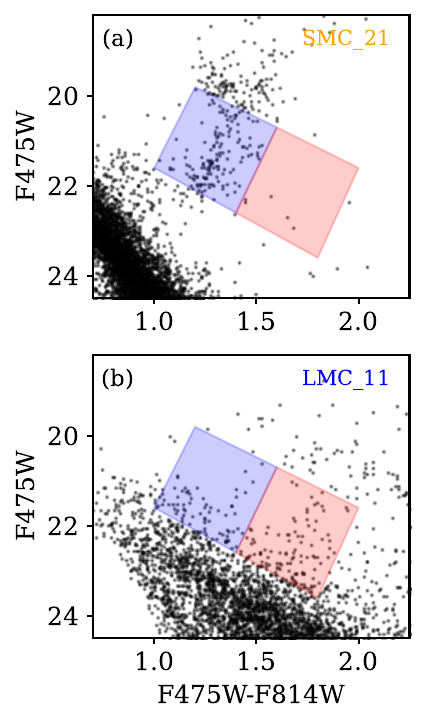}
\includegraphics[width=0.5\textwidth]{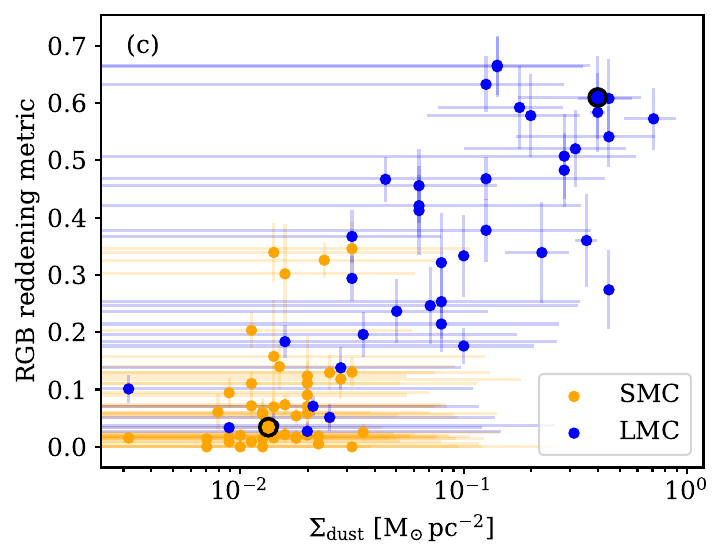}
	\caption{\textbf{Demonstration of simple reddening analysis of Scylla optical CMDs.} In panels (a) and (b), we zoom into the RGB of two example fields, SMC\_21 and LMC\_11. Two regions of the CMD are highlighted, and the reddening metric is computed by dividing the number of sources in the red box by the total number of sources in both boxes. In panel (c), we compare this RGB reddening metric with the dust mass surface density \citep{clark2023}, finding a positive correlation. The two example fields from (a) and (b) are highlighted with black outlines.}
 \label{fig:red_metric}
\end{figure*}

In Figures~\ref{fig:cmds_lmc} and~\ref{fig:cmds_smc} we observe that some fields retain a significant number of sources blueward of the main sequence with roughly constant F475W magnitude (e.g., LMC\_1, LMC\_44, SMC\_1, SMC\_22). These fields are all cases with 2 or 3 filter coverage, meaning that the {\tt DOLPHOT} source detection had access to the fewest number of images relative to the rest of the fields. In addition, by inspection of these cases, our quality cuts to remove diffraction spikes did not perform as well. We note that these sources can be removed by imposing a signal to noise cut on the photometry, but were explicitly included in the \texttt{vgst} catalogs for completeness considerations (see Section~\ref{sec:data_reduction}).

\subsection{Spectral Energy Distributions}

In addition to CMDs, we can also analyze how the observed SEDs of sources from different CMD regions compare. In Figure \ref{fig:sed}, we show the SEDs of ten different sources from three separate regions on an optical CMD in SMC\_15. This field was observed in seven bands, so the SEDs span from the NUV to the NIR. We can see that for the brightest sources on the main sequence (blue), the SEDs are brightest in the NUV as expected, as opposed to RGB sources (pink) which are brightest in the redder optical bands (F475W and F814W). For dimmer sources along the main sequence, we observe significant variability in their NUV detections within large uncertainties.

\subsection{Tracing dust reddening with a CMD}

In addition to insight into the stellar populations of the MCs, the Scylla CMDs also trace the properties of dust reddening along the line of sight. In Figure~\ref{fig:red_metric}, we demonstrate a proof of concept. In panels (a) and (b) we zoom into the RGB region of one example Scylla field from each of the SMC and LMC. We define two regions in CMD-space based on the structure of the RGB analyzed by the SMIDGE survey \citep{ymj2017}, one red and one blue. We then define an RGB reddening metric by counting the number of sources in the red box and dividing by the total number of sources in both boxes. This simple metric aims to capture, in broad strokes, the elongation of the RGB due to the effect of dust reddening.

To test this simple reddening metric, in Figure~\ref{fig:red_metric}c we plot it against the dust mass surface density ($\Sigma_{\rm dust}$) derived from IR SED modeling of the MCs \citep{clark2023}. Despite considerable scatter, we find that the two tracers are positively correlated. This suggests that we are able to recover properties of MC dust using the Scylla CMDs. In future work, we will refine this analysis to take to disentangle the effects of reddening on the structure of the RGB from the effects of LOS distance. 

\subsection{Comparison with other MC surveys}

As discussed in Section~\ref{sec:intro}, Scylla's overall spatial coverage is larger than any other HST survey of comparable depth to date. {To date, Scylla has observed 48 fields in each of the LMC and SMC. In comparison, the METAL survey covered $33$ fields \citep{romanduval2019}  dispersed throughout the LMC, the HTTP survey \citep{sabbi2016} focused on 30 Doradus (equivalent to $\sim36$ fields), and the SMIDGE survey covered the equivalent of $18$ fields in the SMC SW Bar.}

Similarly to METAL, Scylla uses HST's WFC3/UVIS (F225W, F275W, F336W, F475W, F814W) and WFC3/IR (F110W and F160W) filters. HTTP also uses the ACS/WFC in the F555W, F658N, F775W filters (without F225W), and SMIDGE in addition uses the ACS/WFC in the F550M and F658N filters. 

The Scylla survey's significant photometric depth can be compared to that of other HST MC surveys. For example, the 50\% completeness limit of METAL is on average $\sim$ 1.5 magnitudes deeper in F225W, F275W, F336W, and F475W, while it is about the same in F814W and F160W, and $\sim$ 0.5 mags shallower in F110W. The 50\% completeness limit of SMIDGE, which only targets the SMC, is 0.8 -- 1.8 mag deeper in all filters except F475W where it is only 0.4 mag deeper than Scylla's completeness. In 30 Doradus, HTTP reaches an average 50\% completeness of $\sim$ 2 mags shallower in F275W and F336W and 0.5 mag shallower in F110W, while it has the same completeness as Scylla in F160W. As for the optical bands, the two surveys observed in different filters, where the HTTP 50\% completeness in F555W is 25.8 mag, and the Scylla 50\% completeness in F475W is 27.8 mag.

\begin{figure*} 
 	\centering
\includegraphics[width=\textwidth]{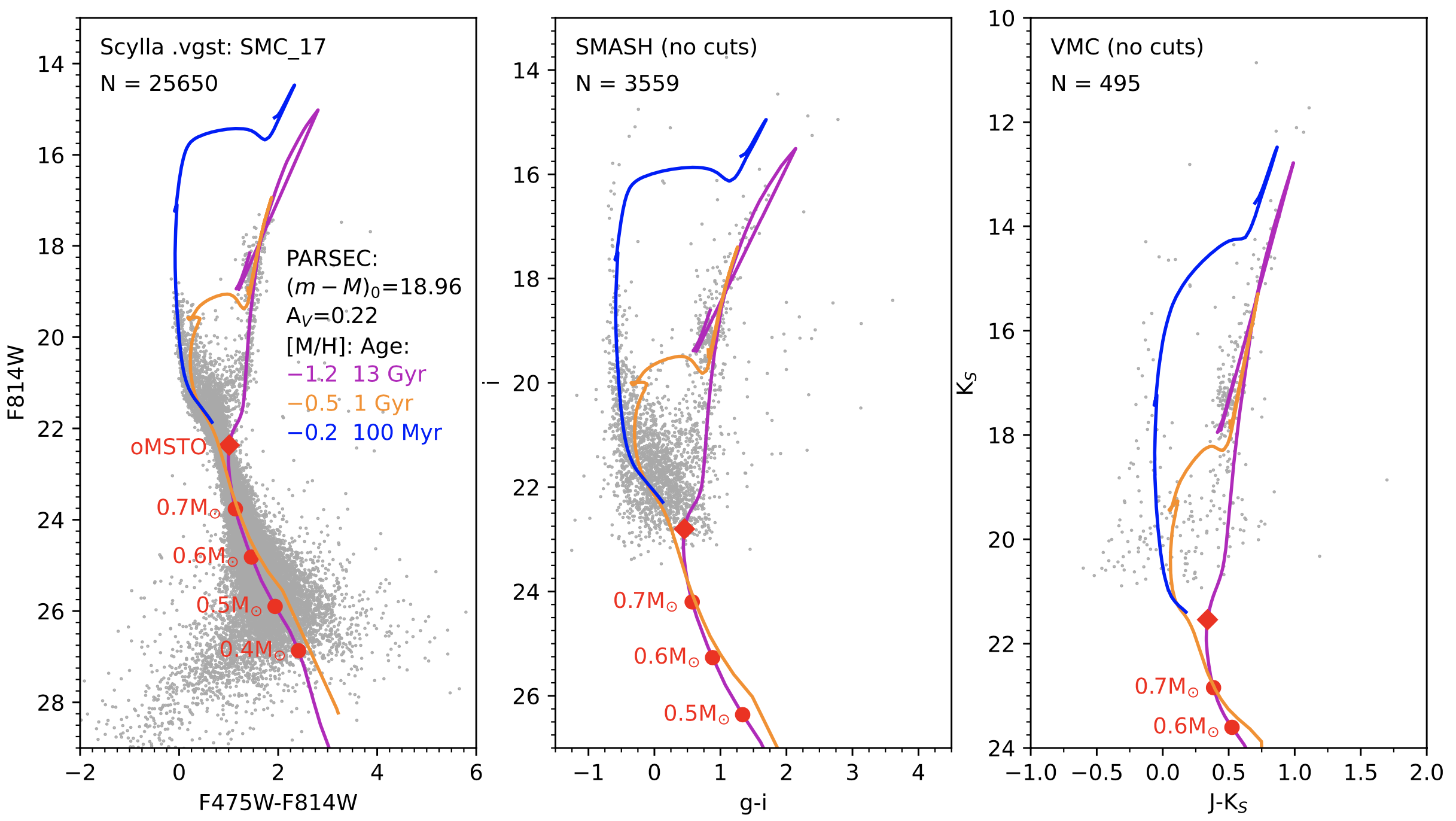}
	\caption{A comparison of optical CMDs from Scylla and ground-based surveys within: Scylla (\texttt{vgst}, left), SMASH (middle) and VMC (right). For the ground-based surveys, we extract all sources within a single Scylla field (SMC$\_$17) and apply no quality cuts. In each panel, we overlay PARSEC isochrones, indicating the main sequence stellar mass range typical of our photometric catalogs. }
 \label{fig:cmd_comparison}
\end{figure*}

While there are rich MC observations from ground-based surveys, the spatial resolution and depth of HST is unsurpassed. Due to stellar crowding, ground-based surveys show significantly shallower CMDs in the central regions of the galaxies compared to the outer parts. For example, in Figure~\ref{fig:cmd_comparison} we compare the CMD of a typical Scylla field (SMC\_17) with the results from the VISTA Survey of the MCs \citep[VISTA;][]{cioni2011} and the Survey of the MAgellanic Stellar History \citep[SMASH;][]{nidever2017, nidever2021} surveys. Within the same footprint as SMC\_17, we extracted SMASH photometry\footnote{\url{https://datalab.noirlab.edu/smash/smash.php}} in similar filters ($g,i$) to F475W and F814W, and VMC photometry\footnote{\url{http://horus.roe.ac.uk/vsa/}} in oft-used $J,K_{S}$ bandpasses.  

Based on Figure~\ref{fig:cmd_comparison}, it is clear that Scylla reaches several magnitudes faintward of the oMSTO, deeper than SMASH or VMC in crowding conditions typical of the inner SMC.  We overlay stellar isochrones from PARSEC onto the CMDs to demonstrate that the photometric depth attained by Scylla corresponds to a stellar mass sensitivity of $\sim 0.5\,M_{\odot}$. Despite our improved sensitivity, Scylla is limited to 96 HST imaging footprints (to date) across both MCs sampling only  $\sim$0.08 kpc$^{2}$ in the LMC and $\sim$0.1 kpc$^{2}$ in the SMC, whereas SMASH and VMC provide contiguous spatial coverage of both systems, as well as enhanced sensitivity to bright sources which are saturated by our typically-long exposures.

\section{Conclusions}
\label{sec:conclusions}

In this work, we present an overview of the Scylla HST survey. Scylla was designed to investigate the stellar populations, ISM, and star formation in the MCs. We describe the science goals, observing strategy, data reduction procedures, and initial results from the photometric analysis of 96 observed fields.

Although constrained by the ULLYSES primary observations, we imaged all fields in at least two filters (F475W and F814W), with 64\% of fields imaged in three or more filters spanning the UV to IR. This comprehensive approach allowed us to achieve an average 50\% completeness of $m_{\rm F225W}=26.0$, $m_{\rm F275W}=26.2$, $m_{\rm F336W}=26.9$, $m_{\rm F475W}=27.8$, $m_{\rm F814W}=25.5$, $m_{\rm F110W}=24.7$, and $m_{\rm F160W}=24.0$ Vega magnitudes in our \texttt{vgst} (quality restricted) catalogs, reaching faintward of the ancient main sequence turnoff in all filters.

Our photometric analysis revealed significant variations in CMDs between and within each MC, highlighting the diverse stellar populations and dust extinction properties across the observed fields. In addition, we demonstrated the feasibility of extracting dust reddening information from the shape of the RGB in the CMDs.

Additionally, we compared our survey statistics with previous {HST} and ground-based surveys of the MCs. We found that Scylla achieves significantly better sensitivity than ground-based surveys in the crowded central regions of the MCs, despite being limited to a smaller observing footprint.

The Scylla survey's deep, multi-wavelength imaging and photometric catalogs provide a valuable resource for characterizing the structure and properties of dust, as well as the spatially-resolved star formation and chemical enrichment histories of the MCs. These data products are available as an HLSP at MAST via \dataset[10.17909/mk54-kg51]{\doi{10.17909/mk54-kg51}}. This release will enable further investigations into the complex interplay between stellar populations and the ISM in these nearby galaxies.

In future work, we will further refine our analysis to disentangle the effects of reddening and line-of-sight distance on the structure of the MCs. We will also utilize the Bayesian Extinction and Stellar Tool (BEAST) to generate a comprehensive catalog of stellar and dust parameters, providing detailed insights into the diverse environments within the MCs.

\acknowledgements{The authors would like to thank the anonymous referee for their insightful comments which improved this work. This research is based on observations made with the NASA/ESA Hubble Space Telescope obtained from the Space Telescope Science Institute, which is operated by the Association of Universities for Research in Astronomy, Inc., under NASA contract NAS 5–26555. These observations are associated with programs 15891, 16235, and 16786. This research has made use of NASA’s Astrophysics Data System. All of the data presented in this paper were obtained from MAST at the Space Telescope Science Institute. The specific observations analyzed can be accessed via \dataset[https://doi.org/10.17909/8ads-wn75]{https://doi.org/10.17909/8ads-wn75}. Support to MAST for these data is provided by the NASA Office of Space Science via grant NAG5–7584 and by other grants and contracts.

The authors thank Benjamin Kuhn for providing insightful comments regarding WFC3 UVIS photometry variations.
}

\facilities{HST (WFC3/IR), HST (WFC3/UVIS)}
\software{DOLPHOT, BEAST \citep{gordon2016}, TOPCAT \citep{taylor2005}, astropy\citep{2018AJ....156..123A, 2013A&A...558A..33A}, numpy \citep{harris2020array}, scipy\citep{Virtanen_2020}, matplotlib \citep{Hunter:2007}, glue \citep{beaumont2015,robitaille2018}}

\bibliographystyle{aasjournal}
\bibliography{ms}

\appendix

\section{Exposure Information}
\label{a:exposures}

In this section we include a table with exposure-level information for each target field. Table~\ref{tab:detail_obs} includes the field name, file root (i.e., MAST identifier), position, position angle, detector, filter, exposure time and level of post-flash for each exposure.

\startlongtable  
\begin{deluxetable*}{llcccccccc} 
\tablecaption{\label{tab:detail_obs} Observing Parameters}  
\tablehead{  
\colhead{Name} & \colhead{Field} & \colhead{Root} & \colhead{RA}    & \colhead{Dec} &  \colhead{PA} & \colhead{Detector} & \colhead{Filter} & \colhead{$T_{\rm exp}$} & \colhead{Post-flash}     \\ 
\colhead{} & \colhead{} & \colhead{} & \colhead{($^{\circ}$)} & \colhead{($^{\circ}$)} & \colhead{($^{\circ}$)} & \colhead{}  & \colhead{}  & \colhead{(s)}   & \colhead{(s)}  \\ 
\colhead{(1)} & \colhead{(2)} & \colhead{(3)} & \colhead{(4)} & \colhead{(5)} & \colhead{(6)}  & \colhead{(7)} & \colhead{(8)} & \colhead{(9)}& \colhead{(10)} } \
\startdata  
SMC$\_ \rm $15  &  15891$\_ \rm$SMC-4292sw-13841  &  ie9m68jaq  &  9.5636155  &  -73.400765  &  138.409  &  IR-FIX  &  F160W  &  499.2  &  $\cdots$  \\ 
SMC$\_ \rm $15  &  15891$\_ \rm$SMC-4292sw-13841  &  ie9m68jdq  &  9.5604039  &  -73.403033  &  138.409  &  UVIS  &  F225W  &  382.0  &  7.3  \\ 
SMC$\_ \rm $15  &  15891$\_ \rm$SMC-4292sw-13841  &  ie9m68jfq  &  9.5604039  &  -73.403033  &  138.409  &  UVIS  &  F225W  &  436.0  &  7.3  \\ 
SMC$\_ \rm $15  &  15891$\_ \rm$SMC-4292sw-13841  &  ie9m68jjq  &  9.5604039  &  -73.403033  &  138.409  &  UVIS  &  F275W  &  433.0  &  7.3  \\ 
SMC$\_ \rm $15  &  15891$\_ \rm$SMC-4292sw-13841  &  ie9m69jnq  &  9.5636155  &  -73.400765  &  138.409  &  IR-FIX  &  F110W  &  349.2  &  $\cdots$  \\ 
SMC$\_ \rm $15  &  15891$\_ \rm$SMC-4292sw-13841  &  ie9m69joq  &  9.5604039  &  -73.403033  &  138.409  &  UVIS  &  F475W  &  710.0  &  2.3  \\ 
SMC$\_ \rm $15  &  15891$\_ \rm$SMC-4292sw-13841  &  ie9m69jqq  &  9.5604039  &  -73.403033  &  138.409  &  UVIS  &  F275W  &  400.0  &  7.3  \\ 
SMC$\_ \rm $15  &  15891$\_ \rm$SMC-4292sw-13841  &  ie9m69juq  &  9.5604039  &  -73.403033  &  138.409  &  UVIS  &  F475W  &  469.0  &  4.3  \\ 
SMC$\_ \rm $15  &  15891$\_ \rm$SMC-4292sw-13841  &  ie9m69jwq  &  9.5604039  &  -73.403033  &  138.409  &  UVIS  &  F475W  &  3.0  &  7.7  \\ 
SMC$\_ \rm $15  &  15891$\_ \rm$SMC-4292sw-13841  &  ie9m70jyq  &  9.5604039  &  -73.403033  &  138.409  &  UVIS  &  F336W  &  359.0  &  7.3  \\ 
SMC$\_ \rm $15  &  15891$\_ \rm$SMC-4292sw-13841  &  ie9m70k1q  &  9.5604039  &  -73.403033  &  138.409  &  UVIS  &  F814W  &  685.0  &  2.0  \\ 
SMC$\_ \rm $15  &  15891$\_ \rm$SMC-4292sw-13841  &  ie9m70k3q  &  9.5604039  &  -73.403033  &  138.409  &  UVIS  &  F336W  &  380.0  &  7.3  \\ 
SMC$\_ \rm $15  &  15891$\_ \rm$SMC-4292sw-13841  &  ie9m70k7q  &  9.5604039  &  -73.403033  &  138.409  &  UVIS  &  F814W  &  455.0  &  3.9  \\ 
SMC$\_ \rm $15  &  15891$\_ \rm$SMC-4292sw-13841  &  ie9m70k9q  &  9.5604039  &  -73.403033  &  138.409  &  UVIS  &  F814W  &  3.0  &  7.7  \\ 
\hline 
SMC$\_ \rm $45  &  15891$\_ \rm$SMC-641nw-12753  &  ie9m08cqq  &  12.617763  &  -72.723821  &  127.527  &  UVIS  &  F475W  &  1330.0  &  0.0  \\ 
SMC$\_ \rm $45  &  15891$\_ \rm$SMC-641nw-12753  &  ie9m08ctq  &  12.617763  &  -72.723821  &  127.527  &  UVIS  &  F814W  &  734.0  &  1.6  \\ 
SMC$\_ \rm $45  &  15891$\_ \rm$SMC-641nw-12753  &  ie9m09d0q  &  12.617763  &  -72.723821  &  127.527  &  UVIS  &  F475W  &  1720.0  &  0.0  \\ 
SMC$\_ \rm $45  &  15891$\_ \rm$SMC-641nw-12753  &  ie9m09d3q  &  12.617763  &  -72.723821  &  127.527  &  UVIS  &  F814W  &  715.0  &  2.0  \\ 
\hline 
SMC$\_ \rm $35  &  16235$\_ \rm$SMC-2773nw-32334  &  iehs77b9q  &  12.884764  &  -72.03468  &  323.343  &  UVIS  &  F814W  &  1040.0  &  0.0  \\ 
SMC$\_ \rm $35  &  16235$\_ \rm$SMC-2773nw-32334  &  iehs77bcq  &  12.884764  &  -72.03468  &  323.343  &  UVIS  &  F814W  &  1040.0  &  0.0  \\ 
SMC$\_ \rm $35  &  16235$\_ \rm$SMC-2773nw-32334  &  iehs78bhq  &  12.884764  &  -72.03468  &  323.343  &  UVIS  &  F475W  &  1183.0  &  0.0  \\ 
SMC$\_ \rm $35  &  16235$\_ \rm$SMC-2773nw-32334  &  iehs78bkq  &  12.884764  &  -72.03468  &  323.343  &  UVIS  &  F475W  &  1182.0  &  0.0  \\ 
SMC$\_ \rm $35  &  16235$\_ \rm$SMC-2773nw-32334  &  iehs79bpq  &  12.884764  &  -72.03468  &  323.343  &  UVIS  &  F336W  &  1178.0  &  7.1  \\ 
SMC$\_ \rm $35  &  16235$\_ \rm$SMC-2773nw-32334  &  iehs79bsq  &  12.884764  &  -72.03468  &  323.343  &  UVIS  &  F336W  &  1177.0  &  7.1  \\ 
\hline 
\enddata  
\tablecomments{Full table available in the online version of the paper. Observing parameters of all Scylla exposures: (1) Name, (2) Long-form name, (3) Root (aka the exposure identifier in MAST), (4) RA, (5) Dec, (6) Position Angle (PA) , (7) Detector name, (8) Filter, (9) Exposure time ($T_{\rm exp}$), (10) Post-flash level for the exposure.}  
\end{deluxetable*}  
\label{tab:detail_obs}

\section{Identifying ``vgst" Cuts}
\label{a:spikes}

\begin{figure*}[h]
 	\centering
        \includegraphics[width=\textwidth]{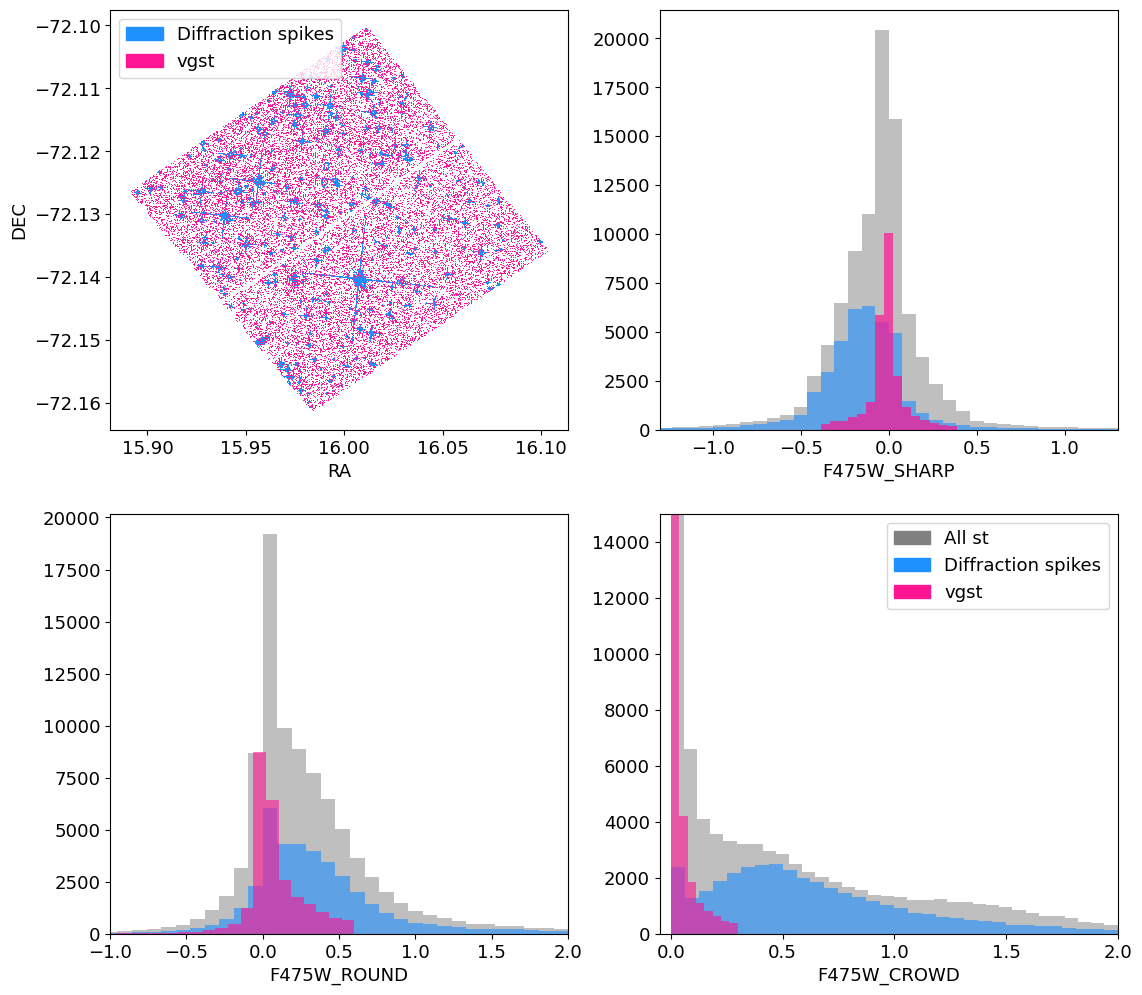}
	\caption{A sample field illustrating the selection process for the \texttt{vgst} quality cuts. (Upper left) First, we visually identify contaminant sources as indicated by diffraction spikes in the imaging footprint (blue). We then evaluate the sharpness, roundness, and crowding parameters of the full sample of stars and the diffraction spike sources, and select ranges for these photometric quality parameters which optimize the removal of diffraction spikes. }
    \label{f:vgst_cuts}
\end{figure*}

To determine photometric quality cuts for the Scylla survey, we manually inspected a test field, SMC$\_$6 (selected randomly due to its wide filter coverage). Using TOPCAT \citep{taylor2005}, we selected sources that were spatially coincident with diffraction spikes. From a total of 93,628 detected sources, 43\% (40,516) were flagged. In Figure~\ref{f:vgst_cuts}, we display the spatial distribution of the sources as well as the distributions of sharpness, roundness, and crowding for the visually-selected contaminant sample and the rest of the field in bands F475W and F814W (filters common to all Scylla fields). 

Based on this result, we selected the \texttt{vgst} quality cuts summarized in Table~\ref{tab:cuts}. 

\section{WFC3 UVIS Variations}

\begin{figure*}
    \centering
    \includegraphics[width=0.8\linewidth]{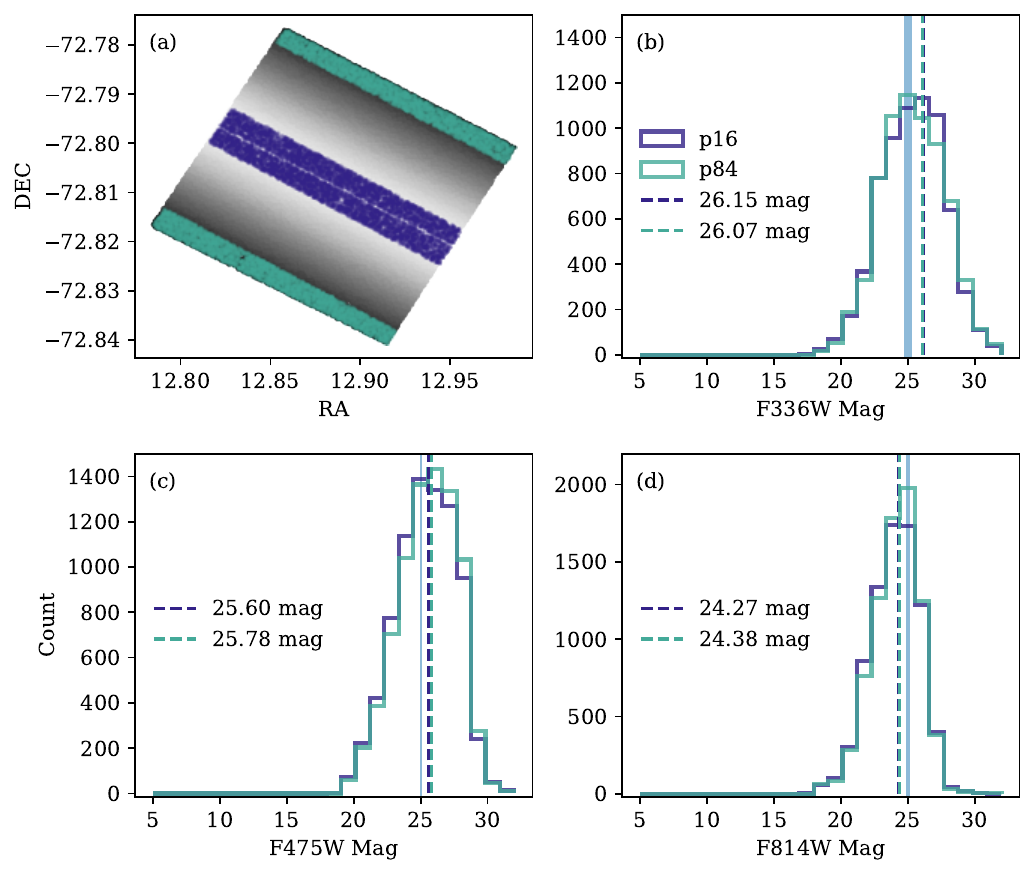}
    \caption{Observed CTE-corrected magnitude variations as a function of chip gap distance. (a) On-sky spatial distribution of all \texttt{vgst} sources in SMC Field 41 (16235{\textunderscore}SMC-286sw-34349). Sources in the lower 16th/upper 84th percentile of distances from the chip gap are highlighted in purple/green, respectively. The magnitude distributions of these sources in WFC3 UVIS filters F336W, F475W, and F814W are shown in panels b, c, and d, respectively. The median values of each distribution are shown with a dashed line, and the 3-sigma magnitude uncertainties at 25 magnitudes derived from ASTs in this particular field are shaded in blue (0.3, 0.09, 0.15, respectively).}
    \label{fig:CTE_mag}
\end{figure*}

Following photometry and quality cuts, we observed residual magnitude variations in our \texttt{vgst} catalog sources as a function of position relative to the WFC3 UVIS camera chip gap. One potential explanation for these variations could be the effects of charge-transfer efficiency (CTE) loss. %, meaning sources closer to the chip gap have underestimated fluxes. 
Space-based CCD detectors can suffer from CTE degradation when exposed to cosmic rays, impacting precision in photometric observations\footnote{CTE degradation for the WFC3 UVIS detector has been monitored since 2009 and is documented in several \hyperlink{https://www.stsci.edu/hst/instrumentation/wfc3/documentation/instrument-science-reports-isrs?keyword=2017-09&itemsPerPage=15}{Instrument Science Reports}, such as WFC3 2017-09 and WFC3 2018-15.}. WFC3 UVIS is a two-chip CCD, meaning that there is a chip gap that runs along the center of the detector, while readout amplifiers are located at the corners of the detector. As charge is transferred to the readout amplifiers and passes through pixels that may have been damaged by cosmic rays, it gets temporarily trapped in the lattice. It has been documented that charge obtained farther from the amplifiers will result in an underestimation of flux, and since regions near the chip gap are the farthest, charge from this region of the lattice is most affected. Without corrections for CTE loss, we expect sources near the chip gap to appear dimmer than the sources near the readout across all filters. 

This issue is usually remedied by including CTE corrections in the photometric reduction pipeline, increasing background levels by post-flashing the images, or obtaining dithered observations of the same field. However, since Scylla is a pure-parallel program, we were not able to obtain dithered exposures for any of our fields. However, post-flash exposures were obtained for most fields (see Table \ref{tab:detail_obs}). Despite these precautions, we still find spatial variations in the distribution of observed magnitudes. 
 
As an example, in Figure \ref{fig:CTE_mag}, we show the distributions of CTE-corrected magnitudes for the 16\% of sources closest to the chip gap (purple) and furthest from the chip gap (green). This field, SMC\_41, was selected as an example due to its homogeneous distribution of sources and lack of extended emission. For filters F475W and F814W, sources near the chip gap are systematically brighter than sources near the readouts. This is the opposite for F336W, where sources near the chip gap are dimmer than sources near the readouts. 

We test whether the two populations in each filter are statistically different by performing a Kolmogorov-Smirnov test which measures the probability that two populations were drawn from the same distribution. For the two populations in F475W, we obtain a p-value of 1.13e-4, which rejects the null hypothesis at a 95\% confidence level (p-value $<0.05$) that the two populations are drawn from the same distribution, i.e. the two populations \textit{are not drawn} from the same distribution. We report similar findings for the two populations in F814W, which have a p-value of 6.66e-4. Conversely, for the two populations in F336W, we obtain a p-value of 0.347, indicating that the two populations \textit{are drawn} from the same distribution.

We find statistically significant differences in the observed magnitudes as a function of chip gap distance in filters F475W and F814W. Based on the observed magnitude distributions, we suspect the CTE correction algorithm is over-correcting the flc images in filters F475W and F814W. However, these differences are not significantly larger than photometric uncertainties in our filters ($\sigma = 0.1, 0.03, 0.05$ mag around the 25th magnitude for F336W, F475W, and F814W, respectively). In Figure \ref{fig:CTE_mag}, we plot the median magnitude of each distribution (dashed lines) to compare with the 25th magnitude 3-sigma uncertainty derived from our ASTs in the field (shaded). From this, we see that the differences in the median magnitudes are comparable to our photometric uncertainties, indicating there may be lingering CTE effects present in our photometry. However, these effects will vary in strength depending on the conditions of each observation, and should not impact our overall science.

\section{Summary of AST Results}
\label{a:compl}

In the Table~\ref{tab:asts_sd} and Table~\ref{tab:asts_bg}, we summarize the AST results for the source-density dominated and background-density dominated fields, respectively. For bins of increasing source or background density, we compute the completeness, bias and uncertainty as a function of flux in each observed band. We include uncertainties on these parameters, which were computed by bootstrapping the binned sample with replacement over 1000 trials. 

\startlongtable
\begin{deluxetable*}{ccclll} 
\tablecaption{\label{tab:asts_sd} Summary of ASTs for source density-dominated fields}
\tablehead{\colhead{Source Density} & \colhead{Filter} &  \colhead{Magnitude}    & \colhead{Completeness} & \colhead{Bias} & \colhead{Uncertainty}   \\ 
\colhead{[sources/arcsec]} & \colhead{} & \colhead{[Vega mag]} & \colhead{[Vega mag]} & \colhead{[Vega mag]} & \colhead{[Vega mag]} \\ 
\colhead{(1)} & \colhead{(2)} & \colhead{(3)} & \colhead{(4)} & \colhead{(5)} & \colhead{(6)} }
\startdata
0.0-0.27  &  F225W  &  18.25  &  1.0$\, \pm \,$0.07  &  0.0$\, \pm \,$0.0  &  0.0$\, \pm \,$0.0  \\ 
0.0-0.27  &  F225W  &  18.75  &  0.83$\, \pm \,$0.04  &  0.0$\, \pm \,$0.0  &  0.0$\, \pm \,$0.01  \\ 
0.0-0.27  &  F225W  &  19.25  &  0.98$\, \pm \,$0.08  &  0.0$\, \pm \,$0.0  &  0.0$\, \pm \,$0.04  \\ 
0.0-0.27  &  F225W  &  19.75  &  1.0$\, \pm \,$0.1  &  0.0$\, \pm \,$0.0  &  0.0$\, \pm \,$0.01  \\ 
0.0-0.27  &  F225W  &  20.25  &  0.84$\, \pm \,$0.04  &  0.02$\, \pm \,$0.0  &  0.01$\, \pm \,$0.02  \\ 
0.0-0.27  &  F225W  &  20.75  &  1.0$\, \pm \,$0.04  &  0.02$\, \pm \,$0.01  &  0.02$\, \pm \,$0.05  \\ 
0.0-0.27  &  F225W  &  21.25  &  0.98$\, \pm \,$0.07  &  0.03$\, \pm \,$0.01  &  0.04$\, \pm \,$0.06  \\ 
0.0-0.27  &  F225W  &  21.75  &  0.99$\, \pm \,$0.13  &  0.04$\, \pm \,$0.09  &  0.04$\, \pm \,$0.24  \\ 
0.0-0.27  &  F225W  &  22.25  &  1.0$\, \pm \,$0.14  &  0.06$\, \pm \,$0.02  &  0.06$\, \pm \,$0.05  \\ 
0.0-0.27  &  F225W  &  22.75  &  0.98$\, \pm \,$0.09  &  0.07$\, \pm \,$0.02  &  0.13$\, \pm \,$0.03  \\ 
0.0-0.27  &  F225W  &  23.25  &  1.0$\, \pm \,$0.14  &  0.11$\, \pm \,$0.05  &  0.13$\, \pm \,$0.09  \\ 
0.0-0.27  &  F225W  &  23.75  &  1.0$\, \pm \,$0.06  &  0.22$\, \pm \,$0.24  &  0.17$\, \pm \,$0.16  \\ 
0.0-0.27  &  F225W  &  24.25  &  0.99$\, \pm \,$0.05  &  0.3$\, \pm \,$0.23  &  0.31$\, \pm \,$0.18  \\ 
0.0-0.27  &  F225W  &  24.75  &  0.96$\, \pm \,$0.04  &  0.41$\, \pm \,$0.33  &  0.42$\, \pm \,$0.19  \\ 
0.0-0.27  &  F225W  &  25.25  &  0.91$\, \pm \,$0.02  &  0.61$\, \pm \,$0.21  &  0.66$\, \pm \,$0.17  \\ 
0.0-0.27  &  F225W  &  25.75  &  0.73$\, \pm \,$0.01  &  0.45$\, \pm \,$0.15  &  0.92$\, \pm \,$0.11  \\ 
0.0-0.27  &  F225W  &  26.25  &  0.48$\, \pm \,0.00$  &  0.56$\, \pm \,0.00$  &  0.82$\, \pm \,0.00$  \\ 
0.0-0.27  &  F225W  &  26.75  &  0.63$\, \pm \,0.00$  &  0.05$\, \pm \,0.00$  &  0.94$\, \pm \,0.00$  \\ 
0.0-0.27  &  F225W  &  27.25  &  0.35$\, \pm \,0.00$  &  -0.77$\, \pm \,0.00$  &  1.24$\, \pm \,0.00$  \\ 
0.0-0.27  &  F225W  &  27.75  &  0.25$\, \pm \,0.00$  &  -1.3$\, \pm \,0.00$  &  0.83$\, \pm \,0.00$  \\ 
0.0-0.27  &  F225W  &  28.25  &  0.22$\, \pm \,0.00$  &  -0.87$\, \pm \,0.00$  &  1.13$\, \pm \,0.00$  \\ 
0.0-0.27  &  F275W  &  18.25  &  1.0$\, \pm \,$0.07  &  0.0$\, \pm \,$0.0  &  0.0$\, \pm \,$0.0  \\ 
0.0-0.27  &  F275W  &  18.75  &  0.97$\, \pm \,$0.04  &  0.0$\, \pm \,$0.0  &  0.0$\, \pm \,$0.01  \\ 
0.0-0.27  &  F275W  &  19.25  &  1.0$\, \pm \,$0.08  &  0.0$\, \pm \,$0.0  &  0.0$\, \pm \,$0.04  \\ 
0.0-0.27  &  F275W  &  19.75  &  0.99$\, \pm \,$0.1  &  0.0$\, \pm \,$0.0  &  0.01$\, \pm \,$0.01  \\ 
0.0-0.27  &  F275W  &  20.25  &  0.97$\, \pm \,$0.04  &  0.0$\, \pm \,$0.0  &  0.01$\, \pm \,$0.02  \\ 
0.0-0.27  &  F275W  &  20.75  &  0.96$\, \pm \,$0.04  &  0.01$\, \pm \,$0.01  &  0.02$\, \pm \,$0.05  \\ 
0.0-0.27  &  F275W  &  21.25  &  0.98$\, \pm \,$0.07  &  0.04$\, \pm \,$0.01  &  0.16$\, \pm \,$0.06  \\ 
0.0-0.27  &  F275W  &  21.75  &  0.97$\, \pm \,$0.13  &  0.03$\, \pm \,$0.09  &  0.04$\, \pm \,$0.24  \\ 
0.0-0.27  &  F275W  &  22.25  &  0.97$\, \pm \,$0.14  &  0.06$\, \pm \,$0.02  &  0.17$\, \pm \,$0.05  \\ 
0.0-0.27  &  F275W  &  22.75  &  0.98$\, \pm \,$0.09  &  0.07$\, \pm \,$0.02  &  0.08$\, \pm \,$0.03  \\ 
0.0-0.27  &  F275W  &  23.25  &  0.99$\, \pm \,$0.14  &  0.11$\, \pm \,$0.05  &  0.13$\, \pm \,$0.09  \\ 
0.0-0.27  &  F275W  &  23.75  &  0.98$\, \pm \,$0.06  &  0.18$\, \pm \,$0.24  &  0.16$\, \pm \,$0.16  \\ 
0.0-0.27  &  F275W  &  24.25  &  0.96$\, \pm \,$0.05  &  0.27$\, \pm \,$0.23  &  0.31$\, \pm \,$0.18  \\ 
0.0-0.27  &  F275W  &  24.75  &  0.94$\, \pm \,$0.04  &  0.33$\, \pm \,$0.33  &  0.44$\, \pm \,$0.19  \\ 
0.0-0.27  &  F275W  &  25.25  &  0.88$\, \pm \,$0.02  &  0.64$\, \pm \,$0.21  &  0.75$\, \pm \,$0.17  \\ 
0.0-0.27  &  F275W  &  25.75  &  0.8$\, \pm \,$0.01  &  0.61$\, \pm \,$0.15  &  0.85$\, \pm \,$0.11  \\ 
0.0-0.27  &  F275W  &  26.25  &  0.6$\, \pm \,0.00$  &  0.52$\, \pm \,0.00$  &  1.12$\, \pm \,0.00$  \\ 
0.0-0.27  &  F275W  &  26.75  &  0.46$\, \pm \,0.00$  &  -0.14$\, \pm \,0.00$  &  0.9$\, \pm \,0.00$  \\ 
0.0-0.27  &  F275W  &  27.25  &  0.35$\, \pm \,0.00$  &  0.4$\, \pm \,0.00$  &  1.4$\, \pm \,0.00$  \\ 
0.0-0.27  &  F275W  &  27.75  &  0.27$\, \pm \,0.00$  &  -0.3$\, \pm \,0.00$  &  1.12$\, \pm \,0.00$  \\ 
0.0-0.27  &  F275W  &  28.25  &  0.24$\, \pm \,0.00$  &  -0.79$\, \pm \,0.00$  &  1.06$\, \pm \,0.00$  \\ 
0.0-0.27  &  F336W  &  18.25  &  0.96$\, \pm \,$0.07  &  0.0$\, \pm \,$0.0  &  0.0$\, \pm \,$0.0  \\ 
0.0-0.27  &  F336W  &  18.75  &  1.0$\, \pm \,$0.04  &  0.0$\, \pm \,$0.0  &  0.0$\, \pm \,$0.01  \\ 
0.0-0.27  &  F336W  &  19.25  &  0.97$\, \pm \,$0.08  &  0.0$\, \pm \,$0.0  &  0.02$\, \pm \,$0.04  \\ 
0.0-0.27  &  F336W  &  19.75  &  0.98$\, \pm \,$0.1  &  0.0$\, \pm \,$0.0  &  0.0$\, \pm \,$0.01  \\ 
\enddata  
\tablecomments{The complete table will be available in the online journal}  
\end{deluxetable*}

\startlongtable
\begin{deluxetable*}{ccclll} 
\tablecaption{\label{tab:asts_bg} Summary of ASTs for background density-dominated fields}
\tablehead{\colhead{Background Density} & colhead{Filter} & \colhead{Magnitude}    & \colhead{Completeness} & \colhead{Bias} & \colhead{Uncertainty}   \\ 
\colhead{[sources/arcsec]} & \colhead{} & \colhead{[Vega mag]} & \colhead{[Vega mag]} & \colhead{[Vega mag]} & \colhead{[Vega mag]} \\ 
\colhead{(1)} & \colhead{(2)} & \colhead{(3)} & \colhead{(4)} & \colhead{(5)} & \colhead{(6)} }
\startdata
F275W  &  0.0-5.38  &  18.25  &  1.0$\, \pm \,$0.0  &  0.0$\, \pm \,$0.0  &  0.0$\, \pm \,$0.0  \\ 
F275W  &  0.0-5.38  &  18.75  &  1.0$\, \pm \,$0.0  &  0.0$\, \pm \,$0.0  &  0.0$\, \pm \,$0.0  \\ 
F275W  &  0.0-5.38  &  19.25  &  1.0$\, \pm \,$0.0  &  0.0$\, \pm \,$0.0  &  0.0$\, \pm \,$0.0  \\ 
F275W  &  0.0-5.38  &  19.75  &  1.0$\, \pm \,$0.0  &  0.0$\, \pm \,$0.0  &  0.0$\, \pm \,$0.0  \\ 
F275W  &  0.0-5.38  &  20.25  &  1.0$\, \pm \,$0.02  &  0.01$\, \pm \,$0.0  &  0.0$\, \pm \,$0.0  \\ 
F275W  &  0.0-5.38  &  20.75  &  1.0$\, \pm \,$0.0  &  0.01$\, \pm \,$0.0  &  0.02$\, \pm \,$0.0  \\ 
F275W  &  0.0-5.38  &  21.25  &  1.0$\, \pm \,$0.0  &  0.03$\, \pm \,$0.0  &  0.03$\, \pm \,$0.0  \\ 
F275W  &  0.0-5.38  &  21.75  &  1.0$\, \pm \,$0.0  &  0.03$\, \pm \,$0.01  &  0.03$\, \pm \,$0.01  \\ 
F275W  &  0.0-5.38  &  22.25  &  0.99$\, \pm \,$0.02  &  0.01$\, \pm \,$0.01  &  0.01$\, \pm \,$0.01  \\ 
F275W  &  0.0-5.38  &  22.75  &  0.93$\, \pm \,$0.03  &  0.1$\, \pm \,$0.01  &  0.07$\, \pm \,$0.0  \\ 
F275W  &  0.0-5.38  &  23.25  &  1.0$\, \pm \,$0.0  &  0.06$\, \pm \,$0.03  &  0.09$\, \pm \,$0.03  \\ 
F275W  &  0.0-5.38  &  23.75  &  1.0$\, \pm \,$0.0  &  0.3$\, \pm \,$0.05  &  0.28$\, \pm \,$0.13  \\ 
F275W  &  0.0-5.38  &  24.25  &  1.0$\, \pm \,$0.0  &  0.43$\, \pm \,$0.05  &  0.03$\, \pm \,$0.17  \\ 
F275W  &  0.0-5.38  &  24.75  &  0.81$\, \pm \,$0.14  &  0.75$\, \pm \,$0.3  &  0.41$\, \pm \,$0.1  \\ 
F275W  &  0.0-5.38  &  25.25  &  1.0$\, \pm \,$0.04  &  0.56$\, \pm \,$0.25  &  1.06$\, \pm \,$0.13  \\ 
F275W  &  0.0-5.38  &  25.75  &  1.0$\, \pm \,$0.04  &  0.91$\, \pm \,$0.26  &  1.65$\, \pm \,$0.32  \\ 
F275W  &  0.0-5.38  &  26.25  &  0.61$\, \pm \,$0.06  &  0.18$\, \pm \,$0.16  &  0.75$\, \pm \,$0.17  \\ 
F336W  &  0.0-5.38  &  18.25  &  1.0$\, \pm \,$0.01  &  0.0$\, \pm \,$0.0  &  0.0$\, \pm \,$0.0  \\ 
F336W  &  0.0-5.38  &  18.75  &  1.0$\, \pm \,$0.01  &  0.0$\, \pm \,$0.0  &  0.0$\, \pm \,$0.0  \\ 
F336W  &  0.0-5.38  &  19.25  &  0.99$\, \pm \,$0.04  &  0.0$\, \pm \,$0.0  &  0.0$\, \pm \,$0.0  \\ 
F336W  &  0.0-5.38  &  19.75  &  0.89$\, \pm \,$0.06  &  0.0$\, \pm \,$0.0  &  0.0$\, \pm \,$0.01  \\ 
F336W  &  0.0-5.38  &  20.25  &  1.0$\, \pm \,$0.0  &  0.0$\, \pm \,$0.0  &  0.0$\, \pm \,$0.0  \\ 
F336W  &  0.0-5.38  &  20.75  &  1.0$\, \pm \,$0.0  &  0.0$\, \pm \,$0.0  &  0.01$\, \pm \,$0.0  \\ 
F336W  &  0.0-5.38  &  21.25  &  1.0$\, \pm \,$0.01  &  0.01$\, \pm \,$0.0  &  0.02$\, \pm \,$0.0  \\ 
F336W  &  0.0-5.38  &  21.75  &  1.0$\, \pm \,$0.02  &  0.07$\, \pm \,$0.02  &  0.02$\, \pm \,$0.0  \\ 
F336W  &  0.0-5.38  &  22.25  &  1.0$\, \pm \,$0.0  &  0.04$\, \pm \,$0.0  &  0.02$\, \pm \,$0.03  \\ 
F336W  &  0.0-5.38  &  22.75  &  1.0$\, \pm \,$0.03  &  0.03$\, \pm \,$0.0  &  0.02$\, \pm \,$0.01  \\ 
F336W  &  0.0-5.38  &  23.25  &  0.95$\, \pm \,$0.06  &  0.1$\, \pm \,$0.01  &  0.05$\, \pm \,$0.0  \\ 
F336W  &  0.0-5.38  &  23.75  &  1.0$\, \pm \,$0.01  &  0.08$\, \pm \,$0.01  &  0.06$\, \pm \,$0.01  \\ 
F336W  &  0.0-5.38  &  24.25  &  1.0$\, \pm \,$0.0  &  0.11$\, \pm \,$0.02  &  0.08$\, \pm \,$0.02  \\ 
F336W  &  0.0-5.38  &  24.75  &  1.0$\, \pm \,$0.01  &  0.42$\, \pm \,$0.08  &  0.13$\, \pm \,$0.1  \\ 
F336W  &  0.0-5.38  &  25.25  &  1.0$\, \pm \,$0.01  &  0.35$\, \pm \,$0.27  &  0.56$\, \pm \,$0.47  \\ 
F336W  &  0.0-5.38  &  25.75  &  1.0$\, \pm \,$0.09  &  0.48$\, \pm \,$0.13  &  0.48$\, \pm \,$0.1  \\ 
F336W  &  0.0-5.38  &  26.25  &  0.92$\, \pm \,$0.06  &  0.48$\, \pm \,$0.26  &  0.84$\, \pm \,$0.25  \\ 
\enddata  
\tablecomments{The complete table will be available in the online journal}  
\end{deluxetable*}

In Table~\ref{tab:summary_obs}, we summarize the $50\%$ completeness limits as a function of observing band for each field.

\startlongtable
\begin{deluxetable*}{lcc|ccccccc} 
\tablecaption{\label{tab:summary_obs} Completeness}
\tablehead{
\colhead{Field Name} & \colhead{Avg RA} & \colhead{Avg Dec}    & \colhead{F225W} & \colhead{F275W} & \colhead{F336W} & \colhead{F475W} & \colhead{F814W} & \colhead{F110W} & \colhead{F160W} \\
\colhead{} & \colhead{[$^{\circ}$]} & \colhead{[$^{\circ}$]} & \colhead{[mag]} & \colhead{[mag]} & \colhead{[mag]}  & \colhead{[mag]} & \colhead{[mag]}   & \colhead{[mag]} & \colhead{[mag]} \\
\colhead{(1)} & \colhead{(2)} & \colhead{(3)} & \colhead{(4)} & \colhead{(5)} & \colhead{(6)}  & \colhead{(7)} & \colhead{(8)} & \colhead{(9)} & \colhead{(10)}}
\startdata  
SMC$\_ \rm $15  &  9.562  &  -73.4013  &  25.7  &  25.7  &  26.5  &  27.7  &  25.4  &  24.9  &  24.2  \\ 
SMC$\_ \rm $45  &  12.6171  &  -72.721  &  $\cdots$  &  $\cdots$  &  $\cdots$  &  27.6  &  25.3  &  $\cdots$  &  $\cdots$  \\ 
SMC$\_ \rm $35  &  12.8842  &  -72.0375  &  $\cdots$  &  $\cdots$  &  27.8  &  28.2  &  25.7  &  $\cdots$  &  $\cdots$  \\ 
SMC$\_ \rm $41  &  12.8849  &  -72.809  &  $\cdots$  &  $\cdots$  &  27.5  &  27.7  &  25.2  &  $\cdots$  &  $\cdots$  \\ 
SMC$\_ \rm $25  &  13.9209  &  -72.7082  &  $\cdots$  &  $\cdots$  &  26.9  &  27.3  &  25.0  &  $\cdots$  &  23.9  \\ 
SMC$\_ \rm $40  &  14.238  &  -72.6194  &  26.0  &  26.2  &  26.7  &  27.2  &  25.1  &  $\cdots$  &  23.9  \\ 
SMC$\_ \rm $42  &  14.2842  &  -72.5876  &  $\cdots$  &  $\cdots$  &  27.6  &  27.4  &  25.2  &  $\cdots$  &  24.0  \\ 
SMC$\_ \rm $17  &  14.4594  &  -72.5908  &  $\cdots$  &  26.4  &  26.7  &  27.8  &  25.5  &  24.9  &  24.3  \\ 
SMC$\_ \rm $7  &  14.5464  &  -71.4007  &  $\cdots$  &  $\cdots$  &  $\cdots$  &  28.5  &  25.8  &  $\cdots$  &  $\cdots$  \\ 
SMC$\_ \rm $37  &  14.6806  &  -72.0602  &  26.2  &  26.3  &  26.8  &  27.3  &  25.2  &  24.6  &  24.0  \\ 
SMC$\_ \rm $23  &  14.7218  &  -72.2607  &  $\cdots$  &  25.7  &  26.5  &  27.0  &  25.0  &  $\cdots$  &  23.8  \\ 
SMC$\_ \rm $13  &  14.9313  &  -72.1748  &  $\cdots$  &  25.8  &  26.1  &  27.1  &  25.1  &  $\cdots$  &  23.9  \\ 
SMC$\_ \rm $14  &  14.9314  &  -71.9391  &  $\cdots$  &  $\cdots$  &  26.5  &  28.1  &  25.6  &  $\cdots$  &  $\cdots$  \\ 
SMC$\_ \rm $12  &  14.9344  &  -72.1575  &  $\cdots$  &  $\cdots$  &  $\cdots$  &  27.7  &  25.3  &  $\cdots$  &  $\cdots$  \\ 
SMC$\_ \rm $1  &  14.939  &  -72.1032  &  $\cdots$  &  $\cdots$  &  $\cdots$  &  27.9  &  25.4  &  $\cdots$  &  $\cdots$  \\ 
SMC$\_ \rm $19  &  14.944  &  -72.1573  &  $\cdots$  &  $\cdots$  &  $\cdots$  &  27.6  &  25.4  &  $\cdots$  &  $\cdots$  \\ 
SMC$\_ \rm $33  &  15.0188  &  -72.5662  &  26.4  &  26.3  &  27.0  &  26.9  &  25.0  &  $\cdots$  &  $\cdots$  \\ 
SMC$\_ \rm $21  &  15.0284  &  -72.1569  &  25.9  &  26.3  &  26.9  &  27.0  &  25.0  &  24.5  &  23.8  \\ 
SMC$\_ \rm $10  &  15.0793  &  -72.1515  &  25.6  &  26.3  &  26.6  &  27.1  &  25.1  &  24.5  &  23.8  \\ 
SMC$\_ \rm $20  &  15.1156  &  -72.082  &  $\cdots$  &  26.1  &  26.7  &  27.3  &  25.3  &  25.0  &  24.4  \\ 
SMC$\_ \rm $30  &  15.1462  &  -72.5944  &  26.4  &  26.3  &  27.0  &  27.1  &  25.0  &  $\cdots$  &  $\cdots$  \\ 
SMC$\_ \rm $31  &  15.1603  &  -72.5957  &  $\cdots$  &  26.3  &  26.7  &  27.0  &  25.0  &  $\cdots$  &  $\cdots$  \\ 
SMC$\_ \rm $9  &  15.1806  &  -72.0226  &  $\cdots$  &  $\cdots$  &  $\cdots$  &  28.1  &  25.7  &  $\cdots$  &  $\cdots$  \\ 
SMC$\_ \rm $16  &  15.2425  &  -72.0912  &  $\cdots$  &  $\cdots$  &  26.7  &  27.7  &  25.4  &  $\cdots$  &  24.1  \\ 
SMC$\_ \rm $27  &  15.379  &  -72.4818  &  $\cdots$  &  26.6  &  27.0  &  26.8  &  25.7  &  24.0  &  23.5  \\ 
SMC$\_ \rm $36  &  15.403  &  -72.269  &  $\cdots$  &  25.9  &  25.1  &  27.1  &  23.5  &  $\cdots$  &  $\cdots$  \\ 
SMC$\_ \rm $2  &  15.5059  &  -72.5123  &  $\cdots$  &  $\cdots$  &  $\cdots$  &  28.4  &  25.8  &  $\cdots$  &  $\cdots$  \\ 
SMC$\_ \rm $34  &  15.706  &  -72.7421  &  26.1  &  26.2  &  26.8  &  27.2  &  25.1  &  $\cdots$  &  23.9  \\ 
SMC$\_ \rm $54  &  15.8228  &  -72.2482  &  $\cdots$  &  $\cdots$  &  $\cdots$  &  28.0  &  25.5  &  $\cdots$  &  $\cdots$  \\ 
SMC$\_ \rm $18  &  15.9413  &  -72.6589  &  $\cdots$  &  26.6  &  27.3  &  27.4  &  25.3  &  24.7  &  24.1  \\ 
SMC$\_ \rm $6  &  15.9954  &  -72.1305  &  $\cdots$  &  $\cdots$  &  26.7  &  27.7  &  25.4  &  $\cdots$  &  $\cdots$  \\ 
SMC$\_ \rm $8  &  16.0039  &  -72.1341  &  $\cdots$  &  $\cdots$  &  26.9  &  27.9  &  25.6  &  $\cdots$  &  $\cdots$  \\ 
SMC$\_ \rm $5  &  16.4523  &  -72.835  &  $\cdots$  &  $\cdots$  &  27.6  &  28.2  &  25.7  &  $\cdots$  &  $\cdots$  \\ 
SMC$\_ \rm $29  &  16.5543  &  -72.4672  &  $\cdots$  &  $\cdots$  &  $\cdots$  &  28.0  &  25.6  &  $\cdots$  &  $\cdots$  \\ 
SMC$\_ \rm $51  &  16.5598  &  -72.1312  &  $\cdots$  &  25.9  &  26.8  &  27.5  &  25.2  &  $\cdots$  &  $\cdots$  \\ 
SMC$\_ \rm $49  &  16.9656  &  -72.089  &  $\cdots$  &  $\cdots$  &  $\cdots$  &  27.9  &  25.6  &  $\cdots$  &  $\cdots$  \\ 
SMC$\_ \rm $46  &  17.0188  &  -72.4006  &  $\cdots$  &  $\cdots$  &  $\cdots$  &  27.6  &  25.3  &  $\cdots$  &  $\cdots$  \\ 
SMC$\_ \rm $55  &  17.3518  &  -72.0303  &  $\cdots$  &  $\cdots$  &  $\cdots$  &  28.1  &  25.7  &  $\cdots$  &  $\cdots$  \\ 
SMC$\_ \rm $44  &  17.4098  &  -73.1726  &  $\cdots$  &  26.2  &  27.2  &  27.3  &  25.0  &  $\cdots$  &  $\cdots$  \\ 
SMC$\_ \rm $53  &  17.5022  &  -73.1831  &  26.0  &  26.4  &  27.2  &  27.4  &  25.2  &  $\cdots$  &  $\cdots$  \\ 
SMC$\_ \rm $43  &  17.7638  &  -72.633  &  $\cdots$  &  26.8  &  26.8  &  27.1  &  25.1  &  24.5  &  23.8  \\ 
SMC$\_ \rm $38  &  20.82  &  -73.3444  &  $\cdots$  &  26.8  &  26.8  &  27.3  &  25.2  &  24.6  &  24.0  \\ 
SMC$\_ \rm $11  &  21.4361  &  -73.1212  &  $\cdots$  &  $\cdots$  &  26.7  &  28.2  &  25.7  &  $\cdots$  &  $\cdots$  \\ 
SMC$\_ \rm $48  &  21.5707  &  -73.3374  &  26.0  &  26.3  &  27.2  &  27.6  &  25.5  &  $\cdots$  &  $\cdots$  \\ 
SMC$\_ \rm $22  &  21.723  &  -73.127  &  $\cdots$  &  $\cdots$  &  $\cdots$  &  28.6  &  26.0  &  $\cdots$  &  $\cdots$  \\ 
SMC$\_ \rm $47  &  21.9543  &  -73.2765  &  $\cdots$  &  $\cdots$  &  $\cdots$  &  28.2  &  25.7  &  $\cdots$  &  $\cdots$  \\ 
SMC$\_ \rm $52  &  22.5766  &  -73.4592  &  $\cdots$  &  $\cdots$  &  $\cdots$  &  28.3  &  25.9  &  $\cdots$  &  $\cdots$  \\ 
SMC$\_ \rm $28  &  22.9725  &  -73.4318  &  $\cdots$  &  $\cdots$  &  27.3  &  27.6  &  25.3  &  $\cdots$  &  $\cdots$  \\ 
LMC$\_ \rm $59  &  73.0619  &  -68.0239  &  $\cdots$  &  $\cdots$  &  27.7  &  28.1  &  25.6  &  $\cdots$  &  24.4  \\ 
LMC$\_ \rm $38  &  73.9133  &  -69.1883  &  $\cdots$  &  $\cdots$  &  27.7  &  28.0  &  25.5  &  $\cdots$  &  $\cdots$  \\ 
LMC$\_ \rm $9  &  73.9739  &  -67.58  &  25.6  &  25.8  &  26.9  &  27.5  &  25.2  &  24.7  &  24.1  \\ 
LMC$\_ \rm $21  &  74.0085  &  -70.0449  &  $\cdots$  &  25.7  &  26.9  &  27.8  &  25.5  &  $\cdots$  &  24.4  \\ 
LMC$\_ \rm $1  &  74.0669  &  -66.3823  &  $\cdots$  &  $\cdots$  &  $\cdots$  &  28.1  &  25.6  &  $\cdots$  &  $\cdots$  \\ 
LMC$\_ \rm $52  &  74.1457  &  -66.4882  &  $\cdots$  &  $\cdots$  &  $\cdots$  &  27.2  &  24.9  &  $\cdots$  &  $\cdots$  \\ 
LMC$\_ \rm $44  &  74.2426  &  -66.4259  &  $\cdots$  &  $\cdots$  &  $\cdots$  &  27.4  &  25.1  &  $\cdots$  &  $\cdots$  \\ 
LMC$\_ \rm $41  &  74.3489  &  -65.5843  &  $\cdots$  &  $\cdots$  &  27.9  &  28.4  &  26.0  &  $\cdots$  &  $\cdots$  \\ 
LMC$\_ \rm $8  &  74.3669  &  -66.3639  &  $\cdots$  &  $\cdots$  &  26.9  &  28.1  &  25.5  &  $\cdots$  &  $\cdots$  \\ 
LMC$\_ \rm $42  &  74.3789  &  -66.3868  &  $\cdots$  &  $\cdots$  &  27.7  &  28.1  &  25.7  &  $\cdots$  &  $\cdots$  \\ 
LMC$\_ \rm $16  &  74.4019  &  -67.735  &  $\cdots$  &  $\cdots$  &  27.4  &  28.3  &  25.8  &  $\cdots$  &  $\cdots$  \\ 
LMC$\_ \rm $24  &  74.4057  &  -68.4953  &  $\cdots$  &  $\cdots$  &  $\cdots$  &  28.1  &  25.5  &  $\cdots$  &  $\cdots$  \\ 
LMC$\_ \rm $20  &  74.4149  &  -68.4953  &  $\cdots$  &  $\cdots$  &  $\cdots$  &  28.1  &  25.6  &  $\cdots$  &  $\cdots$  \\ 
LMC$\_ \rm $32  &  74.7905  &  -70.1869  &  $\cdots$  &  $\cdots$  &  27.6  &  28.1  &  25.5  &  $\cdots$  &  $\cdots$  \\ 
LMC$\_ \rm $39  &  74.9718  &  -68.067  &  $\cdots$  &  $\cdots$  &  $\cdots$  &  28.0  &  25.5  &  $\cdots$  &  $\cdots$  \\ 
LMC$\_ \rm $35  &  76.1349  &  -68.1583  &  $\cdots$  &  $\cdots$  &  $\cdots$  &  27.8  &  25.4  &  $\cdots$  &  $\cdots$  \\ 
LMC$\_ \rm $43  &  76.2446  &  -70.5113  &  25.8  &  26.1  &  26.8  &  27.1  &  25.0  &  $\cdots$  &  $\cdots$  \\ 
LMC$\_ \rm $29  &  76.3845  &  -70.3046  &  $\cdots$  &  26.1  &  26.9  &  27.3  &  25.2  &  $\cdots$  &  $\cdots$  \\ 
LMC$\_ \rm $33  &  76.5819  &  -71.2035  &  26.2  &  26.3  &  27.0  &  27.2  &  25.1  &  $\cdots$  &  $\cdots$  \\ 
LMC$\_ \rm $30  &  78.7752  &  -67.1761  &  $\cdots$  &  $\cdots$  &  $\cdots$  &  28.1  &  25.6  &  $\cdots$  &  $\cdots$  \\ 
LMC$\_ \rm $37  &  80.3377  &  -65.7595  &  $\cdots$  &  $\cdots$  &  $\cdots$  &  28.3  &  25.7  &  $\cdots$  &  $\cdots$  \\ 
LMC$\_ \rm $15  &  81.675  &  -67.7037  &  $\cdots$  &  $\cdots$  &  $\cdots$  &  28.2  &  25.7  &  $\cdots$  &  $\cdots$  \\ 
LMC$\_ \rm $19  &  81.9348  &  -67.3754  &  $\cdots$  &  $\cdots$  &  $\cdots$  &  28.3  &  25.7  &  $\cdots$  &  $\cdots$  \\ 
LMC$\_ \rm $45  &  82.1444  &  -66.9497  &  $\cdots$  &  $\cdots$  &  $\cdots$  &  28.2  &  25.7  &  $\cdots$  &  $\cdots$  \\ 
LMC$\_ \rm $2  &  82.6984  &  -67.1787  &  $\cdots$  &  $\cdots$  &  $\cdots$  &  28.0  &  25.7  &  $\cdots$  &  $\cdots$  \\ 
LMC$\_ \rm $3  &  82.8737  &  -67.3005  &  $\cdots$  &  $\cdots$  &  $\cdots$  &  28.2  &  25.8  &  $\cdots$  &  $\cdots$  \\ 
LMC$\_ \rm $13  &  83.0047  &  -70.7874  &  $\cdots$  &  $\cdots$  &  26.5  &  27.6  &  25.3  &  $\cdots$  &  $\cdots$  \\ 
LMC$\_ \rm $6  &  83.06  &  -70.9753  &  $\cdots$  &  $\cdots$  &  26.5  &  27.8  &  25.5  &  $\cdots$  &  $\cdots$  \\ 
LMC$\_ \rm $4  &  83.7896  &  -69.8  &  25.6  &  25.9  &  26.5  &  27.1  &  24.9  &  24.5  &  24.0  \\ 
LMC$\_ \rm $57  &  83.793  &  -66.1738  &  $\cdots$  &  $\cdots$  &  $\cdots$  &  27.7  &  25.4  &  $\cdots$  &  $\cdots$  \\ 
LMC$\_ \rm $34  &  84.0445  &  -66.6476  &  $\cdots$  &  $\cdots$  &  $\cdots$  &  28.0  &  25.5  &  $\cdots$  &  $\cdots$  \\ 
LMC$\_ \rm $40  &  84.1042  &  -67.0506  &  $\cdots$  &  $\cdots$  &  27.7  &  28.1  &  25.7  &  $\cdots$  &  $\cdots$  \\ 
LMC$\_ \rm $27  &  84.3877  &  -69.1643  &  $\cdots$  &  $\cdots$  &  26.8  &  27.1  &  24.6  &  $\cdots$  &  23.1  \\ 
LMC$\_ \rm $48  &  84.4053  &  -69.0844  &  26.0  &  26.3  &  26.8  &  26.9  &  24.9  &  $\cdots$  &  23.8  \\ 
LMC$\_ \rm $47  &  84.4514  &  -69.0828  &  $\cdots$  &  26.3  &  26.8  &  27.1  &  25.0  &  $\cdots$  &  $\cdots$  \\ 
LMC$\_ \rm $36  &  84.55  &  -69.4356  &  $\cdots$  &  $\cdots$  &  $\cdots$  &  27.7  &  25.3  &  $\cdots$  &  $\cdots$  \\ 
LMC$\_ \rm $56  &  84.5993  &  -69.1631  &  25.3  &  25.1  &  26.8  &  26.9  &  24.8  &  24.2  &  23.7  \\ 
LMC$\_ \rm $53  &  84.6333  &  -69.2891  &  26.0  &  26.0  &  26.8  &  26.9  &  24.9  &  $\cdots$  &  23.9  \\ 
LMC$\_ \rm $22  &  84.6425  &  -69.0671  &  $\cdots$  &  26.0  &  26.3  &  26.0  &  23.7  &  23.1  &  22.6  \\ 
LMC$\_ \rm $11  &  84.7783  &  -69.1768  &  $\cdots$  &  26.1  &  26.7  &  27.3  &  23.9  &  $\cdots$  &  29.4  \\ 
LMC$\_ \rm $54  &  84.8145  &  -68.9016  &  $\cdots$  &  $\cdots$  &  27.6  &  27.8  &  25.4  &  $\cdots$  &  $\cdots$  \\ 
LMC$\_ \rm $55  &  84.9219  &  -69.0085  &  $\cdots$  &  26.1  &  26.8  &  27.3  &  25.1  &  $\cdots$  &  24.1  \\ 
LMC$\_ \rm $7  &  85.1607  &  -69.3493  &  $\cdots$  &  $\cdots$  &  26.9  &  27.8  &  25.5  &  $\cdots$  &  $\cdots$  \\ 
LMC$\_ \rm $5  &  85.1885  &  -69.5809  &  $\cdots$  &  $\cdots$  &  26.7  &  27.7  &  25.3  &  $\cdots$  &  $\cdots$  \\ 
LMC$\_ \rm $10  &  85.9116  &  -67.9299  &  $\cdots$  &  $\cdots$  &  $\cdots$  &  28.3  &  25.8  &  $\cdots$  &  $\cdots$  \\ 
LMC$\_ \rm $17  &  86.1452  &  -67.325  &  $\cdots$  &  $\cdots$  &  27.5  &  28.3  &  25.8  &  $\cdots$  &  $\cdots$  \\ 
LMC$\_ \rm $49  &  86.3622  &  -67.0819  &  $\cdots$  &  $\cdots$  &  $\cdots$  &  27.7  &  25.4  &  $\cdots$  &  $\cdots$  \\ 
LMC$\_ \rm $50  &  86.5299  &  -67.1893  &  $\cdots$  &  $\cdots$  &  $\cdots$  &  28.1  &  25.7  &  $\cdots$  &  $\cdots$  
\enddata  
\tablecomments{Summary of $50\%$ completeness limits for each field. We include: (1) Name, (2) Average catalog RA, (3) Average catalog Dec, (4-10) $50\%$ completeness limits in F225W, F275W, F336W, F475W, F814W, F110W and F160W.}  
\end{deluxetable*}

\end{document}